\documentclass[superscriptaddress,nofootinbib]{revtex4-1}
\usepackage{graphicx}
\usepackage{subfigure}
\usepackage{ulem}
\usepackage{color}
\usepackage{slashed}
\usepackage[]{hyperref}
\usepackage{amsmath}
\usepackage{amsfonts}
\usepackage{amssymb}

\numberwithin{equation}{section}

\providecommand{\U}[1]{\protect\rule{.1in}{.1in}}

\def\avg#1{\left\langle#1\right\rangle}
\def\bra#1{\left\langle#1\right|}
\def\ket#1{\left|#1\right\rangle}

\def\kc#1{\left(#1\right)}
\def\kd#1{\left[#1\right]}
\def\ke#1{\left\{#1\right\}}
\def\Re{{\rm Re}}
\def\Im{{\rm Im}}
\def\sgn{{\rm sgn}}

\def\be{\begin{equation}}       \def\ee{\end{equation}}
\def\bea{\begin{eqnarray}}      \def\eea{\end{eqnarray}}
\def\ba{\begin{array}}
	\def\ea{\end{array}}
\def\bnum{\begin{enumerate} }
	\def\enum{\end{enumerate}}

\def\nn{\nonumber}

\def\=>{\Rightarrow}
\def\>{\rightarrow}

\def\eye2{Fathbb{I}}

\def\Tr{\mathrm{Tr}}

\DeclareMathOperator\csch{csch}

\newcommand{\UCAS}{\affiliation{School of Physics, University of Chinese Academy of Sciences, Beijing 100049, China}}
\newcommand{\ITP}{\affiliation{CAS Key Laboratory of Theoretical Physics, Institute of Theoretical Physics, Chinese Academy of Sciences, Beijing, 100190, China}}

\begin{document}
\title{Wormholes and the Thermodynamic Arrow of Time}
\author{Zhuo-Yu Xian}
\email{xianzy@itp.ac.cn}
\ITP
\author{Long Zhao}
\email{zhaolong@itp.ac.cn}
\ITP
\UCAS

\begin{abstract}
	In classical thermodynamics, heat cannot spontaneously pass from a colder system to a hotter system, which is called the thermodynamic arrow of time. However, if the initial states are entangled, the direction of the thermodynamic arrow of time may not be guaranteed. Here we take the thermofield double state at $0+1$ dimension as the initial state and assume its gravity duality to be the eternal black hole in AdS$_2$ space. We make the temperature difference between the two sides by changing the Hamiltonian. We turn on proper interactions between the two sides and calculate the changes in energy and entropy. The energy transfer, as well as the thermodynamic arrow of time, are mainly determined by the competition between two channels: thermal diffusion and anomalous heat flow. The former is not related to the wormhole and obeys the thermodynamic arrow of time; the latter is related to the wormhole and reverses the thermodynamic arrow of time, {\it i.e.} transferring energy from the colder side to the hotter side at the cost of entanglement consumption. Finally, we find that the thermal diffusion wins the competition, and the whole thermodynamic arrow of time has not been reversed. 
\end{abstract}

\maketitle

\tableofcontents

\section{Introduction}

With the experimental advances in the last two decades, it is possible to prepare and control systems with an increasing number of atoms, whose scale stretches across the gap between macroscopic thermodynamics and microcosmic quantum mechanics. Taking these former two theories as cornerstones, quantum thermodynamics tries to extend the framework of thermodynamics to finite size quantum systems, nonequilibrium dynamics, strong coupling, and quantum information processes \cite{QT:book}.

The laws of thermodynamics constrain the evolution of a system coupled to heat baths. The second law can be derived when the initial correlation between the system and its heat baths is absent \cite{SecondLaw}. The correlation generated during the interaction between the system and the heat baths leads to non-negative entropy production. The thermodynamic arrow of time also plays a crucial role in the second law of thermodynamics, which refers to the phenomenon that heat will spontaneously pass from a hotter system to a colder system. Similarly, it can be derived if the two systems are uncorrelated initially \cite{Partovi:2008}.

\begin{figure}
	\includegraphics[height=0.1\linewidth]{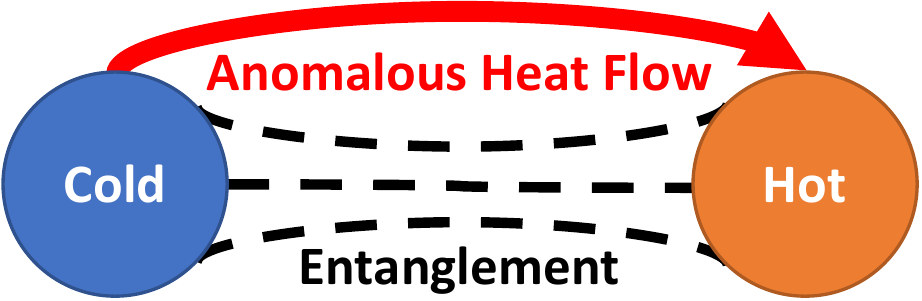}
	\caption{Anomalous heat flow.}
	\label{FigAHF}
\end{figure}

The author of Ref.~\cite{Partovi:2008} argued that the second law, and the thermodynamic arrow of time, are emergent phenomena in the low-entropy and low-correlation environment. Furthermore, for the two systems with different temperatures and initial correlation, the thermodynamic arrow of time can be reversed for some proper interactions: heat passes from colder systems to hotter systems by consuming their correlation \cite{Partovi:2008,Jennings:2010}, which is called anomalous heat flow, as illustrated in Fig.~\ref{FigAHF}. The authors of Ref.~\cite{Lloyd:2015} discussed the conditions for its presence. Such a phenomenon was observed in the experiment on two initially correlated spins \cite{Micadei:2017} and the quantum computer \cite{Lesovik:2019klx}. The reversion of the thermodynamic arrow of time may conflict with the original second law and urges a generalized second law in the presence of correlations \cite{Bera:2017}. 

This traces back to the relationship between work and correlation \cite{Vitagliano:2018}. By acting unitary transformation on a state, one can lower the energy and extract work from the system. A state from which no work can be extracted is called the passive state. All the thermal states are passive \cite{PassiveState}. But work can be extracted from the correlation between two systems, each of which is locally in a thermal state with the same temperature \cite{Llobet:2015}. Then, the anomalous heat flow is a refrigeration driven by the work stored in correlations \cite{Bera:2017}. Furthermore,  the authors of Refs.~\cite{Bera:2017,Brandao:2013,Alipour:2016} discussed the generalizations of the second law in the presence of correlation. 

Similar issues have been being studied in gravity since two decades ago as well, while the famous AdS/CFT correspondence was discovered. The AdS/CFT correspondence is a realization of the holography principle, proposed by Susskind according to the surface law of black hole entropy. It conjectures that the degree of freedom of a gravitational system can be mapped to its boundary \cite{Susskind:1994vu}. Furthermore, the AdS/CFT correspondence offers the dictionary between the quantum gravity in asymptotic anti-de Sitter (AdS) space and the conformal field theory (CFT) on its boundary \cite{AdSCFT}. Notably, a large AdS-Schwarzschild black hole is dual to the CFT at finite temperature. The correspondence reveals that black hole thermodynamics is not only an analogy of the classical thermodynamics but also an equivalent description. 

Another significant development in the AdS/CFT correspondence is the Ryu-Takayanagi (RT) formula, which conjectures that the von Neumann entropy of a subregion on the boundary is proportional to the area of the minimal homological surface in the bulk \cite{HEE}. Furthermore, the black hole entropy is interpreted as the entanglement entropy between the two field theories on the two boundaries of the eternal black hole. The spatial region behind the two horizons is called an Einstein-Rosen bridge or a wormhole.

The AdS/CFT correspondence also sheds light on the black hole information problem by utilizing the unitary of dual boundary theories \cite{Polchinski:2016hrw,Harlow:2014yka}. Before the AdS/CFT correspondence was proposed, Page embedded the unitary in the process of black hole evaporation and study the entanglement between the remainder black hole and its Hawking radiation \cite{Page:1993df,Page:2013dx}. Motivated by the Hartle-Hawking-Israel state \cite{Israel:1976ur} and the black hole complementarity \cite{Susskind:1993if}, Maldacena conjectured that the eternal black hole is described by the thermofield-double (TFD) state and tried to resolve part of the information loss paradox \cite{Maldacena:2001kr}. Motivated by the RT formula, Van Raamsdonk conjectured the correspondence between geometry and entanglement \cite{VanRaamsdonk}. To resolve the firewall paradox \cite{Firewall}, Maldacena and Susskind applied the idea to general entangled states and raised the ER=EPR conjecture where ER refers to the Einstein-Roson bridge, and EPR (Einstein-Podolsky-Rosen) refers to EPR pair, {\it i.e.} quantum entanglement \cite{Maldacena:2013xja}. 

The connection between the TFD state and the eternal black hole helps us to understand the complicated dynamics of boundary field theories from the gravity side. Quantum chaos, which is diagonalized by out-of-time-order correlators (OTOCs), can be understood from the shock wave on black holes \cite{Chaos,Shenker:2014cwa}. The teleportation protocol or Hayden-Preskill protocol corresponds to constructing a traversable wormhole in an eternal black hole \cite{Hayden:2007cs,Susskind:2017nto,Gao:2016bin,Maldacena:2017axo,Maldacena:2018lmt,Chen:2019qqe,Garcia-Garcia:2019poj,Gao:2018yzk}. 
To make the wormhole traversable, one should turn on proper interactions between the two sides of the eternal black hole, where the interactions are equivalent to the measurements in the teleportation protocol. The evolution of the bi-systems with interactions and entanglement is an object of research in quantum thermodynamics.

We raise the following questions in the intersection between quantum thermodynamics and the AdS/CFT correspondence.
\begin{itemize}
	\item Does anomalous heat flow exist in the systems which have holographic duality?
	\item If yes, will it challenge the laws of black hole thermodynamics?
\end{itemize}
In previous works, anomalous heat flow was realized in weakly coupled or weakly chaotic systems. In this paper, we try to find similar phenomena in strongly coupled systems based on the AdS$_2$/CFT$_1$ correspondence and the ER=EPR conjecture. In Sec. \ref{SectionEnergyCorrelation}, we review the relation between correlation and anomalous heat flow. In Sec.~\ref{SectionTFD}, we prepare a TFD state and the dual eternal black hole in Jackiw-Teitelboim (JT) gravity with different temperatures on each side. In Sec.~\ref{SectionWork}, we show that work can be extracted from the wormhole. In Sec.~\ref{SectionAnomalous}, with proper interactions, we find two channels which mainly contribute to the energy transfer: the thermal diffusion via the boundary and the anomalous heat flow via the wormhole in the eternal black hole. The former is numerically larger than the latter. So the total thermodynamic arrow of time has not been reversed. In Sec.~\ref{SectionEntropy}, we study the concomitant consumption of entanglement. Although we focus on the eternal black hole in the main text, in Appendixes \ref{SectionProductBH} and \ref{SectionSYK}, similar issues are investigated for a product state in JT gravity and a TFD or product state in the Sachdev-Ye-Kitaev (SYK) model for comparison.

\section{Energy transfer with correlation}\label{SectionEnergyCorrelation}

\subsection{Thermodynamics for bi-systems}

Consider bi-systems labeled by $L$ and $R$. The state of the bi-systems is described by the density matrix $\rho$; then the local density matrices of systems $L$ and $R$ follow $\rho_{\gamma}=\Tr_{\bar\gamma}\rho$, where $\gamma=L,R$ and $\bar\gamma$ is the complement of $\gamma$. The entropy of the bi-systems and local systems are defined as the von Neumann entropy $S_{LR}=-\Tr[\rho\ln\rho]$ and $S_\gamma=-\Tr[\rho_\gamma\ln\rho_\gamma]$. 
The two systems may be correlated. We use mutual information $I_{L;R}=S_L+S_R-S_{LR}$ to measure their correlation, where $I_{L;R}\geq0$ because of the sub-additivity of entanglement entropy. 

The local Hamiltonians of systems $L$ and $R$ are $H_L$ and $H_R$. Then the local energies are $E_{\gamma}=\Tr[\rho_{\gamma}H_{\gamma}]$.
We will consider the following process. Prepare an initial state $\rho(t_i)$ of the bi-systems at time $t_i$. Then we turn on an interaction $H_I$ between them. After time $t_i$, the total Hamiltonian $H_{tot}=H_L+H_R+H_I$ is time independent. We define $E_I=\Tr[\rho H_I]$. In this paper, we will consider a weak interaction $H_I$ compared to the local Hamiltonian $H_\gamma$. So the local energies of each local systems are well defined by the expectational valued of their local Hamiltonians. At time $t_f>t_i$, the state of the bi-systems becomes $\rho(t_f)$ through the unitary evolution with total Hamiltonian $H_{tot}$. So the entropy of the bi-systems $S_{LR}$ does not change. The changes in the local entropies and energies are $\Delta S_{\gamma}=S_{\gamma}(t_f)-S_{\gamma}(t_i)$, $\Delta E_{\gamma}=E_{\gamma}(t_f)-E_{\gamma}(t_i)$, and $\Delta E_I=E_I(t_f)-E_I(t_i)$.

After time $t_i$, the total energy is conserved,
\begin{align}\label{Conserve}
\Delta E_L+\Delta E_R+\Delta E_I=0.
\end{align}
According to the first law of thermodynamics, the change in local energy can be divided into work $W$ and heat $Q$. We will adopt the definitions of work and heat in Ref.~\cite{Alipour:2016}, as reviewed in Appendix \ref{SectionFirstLaw}. For a specific model, we can calculate them accordingly.

\subsection{Anomalous heat flow}

We review some theoretical observations about the heat transfer in the presence of correlation in Refs.~\cite{Partovi:2008,Jennings:2010}. 
Given an inverse temperature $\beta_\gamma=1/T_\gamma$, one can define a local thermal state for a local Hamiltonian, namely $\tau_\gamma(\beta_\gamma)=Z_{\gamma}^{-1} \exp(-\beta_{\gamma} H_{\gamma})$. 
Now we specify the initial state $\rho(t_i)$ of the bi-systems at time $t_i$ such that the two local systems are individually thermal, {\it i.e.} $\rho_{\gamma}(t_i)=\tau_\gamma(\beta_\gamma)$.

Note that, given a constant $\beta$, free energy $F=E-S/\beta$ is minimized by thermal state. As each local system is thermal initially, we have
\begin{align}\label{EnergyEntropy}
\beta_L \Delta E_L \geq \Delta S_L,    \quad
\beta_R \Delta E_R \geq \Delta S_R.
\end{align}
where $\beta_{L,R}$ are taken to be constant during the process. We will call (\ref{EnergyEntropy}) as energy-entropy inequalities. When the inequalities are saturated, they are just the first law of entropy of local systems. Combining the two inequalities, we have
\begin{align}\label{EnergyInformation}
(\Delta E_L+\Delta E_R)(\beta_L+\beta_R) +     (\Delta E_L-\Delta E_R)(\beta_L-\beta_R) \geq 2\Delta I_{L;R} = 2(\Delta S_L+\Delta S_R),
\end{align}
where the last equality follows from the unitary evolution.

For an initial state which is a product state $\rho(t_i)=\rho_L(t_i)\otimes\rho_R(t_i)$, we have $I_{L;R}(t_i)=0$. So $\Delta I_{L;R}\geq0$, meaning that the interaction may induce entanglement. If we further consider the case of heat contact, {\it i.e.} $\Delta E_L+\Delta E_R=0$, one can easily obtain the thermodynamic arrow of time from inequality (\ref{EnergyInformation}),
\begin{align}\label{Obey}
(\Delta E_L-\Delta E_R)(\beta_L-\beta_R)\geq0.
\end{align}
For example, we assume that system $R$ is hotter than system $L$, {\it i.e.} $\beta_L>\beta_R$. From the above inequality, we know $\Delta E_L=-\Delta E_R\geq0$, which just means that heat can only pass from a hotter system to a colder system. The above inequality also holds once $\Delta E_L+\Delta E_R\leq0$, which means that one cannot transfer energy from a colder system to a hotter system without positive work done on the bi-systems. 

However, the above argument fails if the two systems are entangled initially. It is possible that the mutual information $I_{L;R}$ decreases. Then the left-hand side of inequality (\ref{EnergyInformation}) is not required to be non-negative anymore. Especially for a pure initial state, $S_L=S_R$ always holds. Then $\Delta S_L=\Delta S_R$. Combining it with inequalities (\ref{EnergyEntropy}), we have $\Delta E_L+\Delta E_R\geq\Delta S_L(T_L+T_R)$. If $\Delta E_L+\Delta E_R\leq0$ during the process, we find $\Delta I_{L;R}=2\Delta S_L\leq0$. So the reversal of inequality (\ref{Obey}),
\begin{align}
(\Delta E_L-\Delta E_R)(\beta_L-\beta_R)<0,
\end{align}
is not forbidden anymore. Energy may pass from a colder system to a hotter system by consuming (quantum) correlation, which is called anomalous heat flow. No doubt that such a phenomenon highly depends on the state and the interaction. The reversal of the thermodynamic arrow of time during heat contact was observed in the recent experiment on the two-spin system \cite{Micadei:2017}. The time-reversal algorithms on a quantum computer were developed in Ref.~\cite{Lesovik:2019klx}.

In a word, we say that the thermodynamic arrow of time is reversed if
\begin{align}\label{ReversedArrow}
\Delta E_L+\Delta E_R\leq0 \text{~~and~~} (\Delta E_L-\Delta E_R)(\beta_L-\beta_R)<0.
\end{align}

\subsection{Extract work from correlation}

No doubt that, the anomalous heat flow stems from (quantum) correlations, which plays an important role in quantum thermodynamics. One can extract work from correlation or create correlation by costing work \cite{Llobet:2015,Huber:2015,Bruschi:2015,Friis:2016,Vitagliano:2018}. The trade-off between works and correlations explains the presence of anomalous heat flow. For a state $\rho$ and its Hamiltonian $H$, the work extracted from the system after a unitary transformation $U$ is
\begin{align}
W=\Tr[\rho H]-\Tr[U\rho U^\dagger H].
\end{align}
For a Hamiltonian $H$, a state is called passive if and only if no positive work can be extracted from it, namely 
\begin{align}
W\leq 0, \quad \forall U.
\end{align}
Furthermore, it is proved in Refs.~\cite{PassiveState} that a state $\pi$ is passive if and only if it is diagonal in the energy eigenbasis $\ke{\ket{n}}$ and its eigenvalues $\ke{p_n}$ are monotonically decreasing with increasing energy $\ke{E_n}$, namely
\begin{align}\label{Passive}
\pi=\sum_n p_n \ket n \bra n \quad \text{where} \quad p_n\geq p_m \text{ if } E_n\leq E_m.
\end{align}
Obviously, a thermal state $\tau=Z^{-1}e^{-\beta H}$ is a passive state of its Hamiltonian $H$. Furthermore, the outer product of thermal state $\tau^{\otimes n}$ is also a passive state of the $n$ copies of $H$ for any $n$, while an outer product of a passive state $\pi^{\otimes n}$ may not be a passive state of the $n$ copies of $H$.

The authors of Ref.~\cite{Bera:2017} argued that, for entropy-preserving evolution, the extractable work solely from the correlation between system and heat bath is bounded by the product of mutual information and temperature. Here we take the entangled pure state of bi-systems as an example. From inequalities (\ref{EnergyEntropy}), the work extracted is bounded by
\begin{align}\label{WorkCorrelation}
W\leq I_{L;R}(T_L+T_R)/2.
\end{align}
The authors in Ref.~\cite{Bera:2017} further explained the anomalous heat flow as the refrigeration driven by the work stored in correlations. Finally, they also modified the Clausius statement of the second law: no process is possible whose sole result is the transfer of heat from a colder system to a hotter system, where the work potential stored in the correlations does not decrease.

\section{A Thermo field double state as an initial state}\label{SectionTFD}

\subsection{Balanced bi-systems and holography}
Consider a local Hamiltonians $H$ of a local system, with eigenbasis $H\ket{n}=E_n\ket{n}$. Further, consider two copies of the local system labeled by $L$ and $R$ with Hamiltonian
\begin{align}\label{BalancedHamiltonian}
H_{0}=H_L+H_R, \quad H_L=H\otimes 1,\quad H_R=1\otimes H.
\end{align}
We call it the balanced Hamiltonian since the local Hamiltonians are the same.
We consider a state $\ket{I}=\sum_n \ket{n}_L\ket{n}_R$, where $\ket{n}_\gamma$ is the energy eigenstate in system $\gamma$. We prepare a TFD state at $t=0$,
\begin{align}\label{TFD}
\ket{\beta}=\frac1{\sqrt Z} e^{-\beta H_0/4}\ket{I}=\frac1{\sqrt Z} \sum_n \ket{n}_L\ket{n}_R e^{-\beta E_n/2}    
\end{align}
where $Z=\Tr e^{-\beta H}$. The TFD state is not an eigenstate of $H_0$. We define
$
\ket{\beta(t)}=e^{-iH_0t}\ket\beta.
$
Tracing out a local system on one side, we obtain a thermal state on another side,
$
\rho_\gamma=\Tr_{\bar\gamma} \ket{\beta(t)}\bra{\beta(t)}=Z^{-1} e^{-\beta H}
$.
Both of the inverse temperatures of local systems are $\beta$.

Throughout this paper, we consider $H$ to be a conformal field theory at $0+1$ dimension. We can also consider that the local Hamiltonian $H$ in Eq.~(\ref{BalancedHamiltonian}) is the SYK model \cite{SYK,Maldacena:2016hyu,Kitaev:2017awl} in Appendix \ref{SectionSYK}. The TFD state in the two-site SYK model was constructed in Refs.~\cite{Maldacena:2018lmt,Gu:2017njx,Garcia-Garcia:2019poj}. The ways of preparing TFD states were discussed in Refs.~\cite{Cottrell:2018ash,Maldacena:2018lmt}.

Given an operator $O$ on the local system, we define
\begin{align}\label{OLOR}
O_L=O^T\otimes 1,\quad O_R= 1 \otimes O,
\end{align}
where $O^T$ is the transpose of $O$ on the basis of energy eigenstates. The interaction picture for both sides is
\begin{align}\begin{split}
O_{L}(t)=e^{iH_0t}O_{L}e^{-iH_0t}=e^{iHt}O^Te^{-iHt}\otimes 1,\\
O_{R}(t)=e^{iH_0t}O_{R}e^{-iH_0t}=1\otimes e^{iH t}Oe^{-iH t},
\end{split}\end{align}
Due to the entanglement in the TFD state, the operators on both side are related by
\begin{align}
O_L(t)\ket{\beta}=O_R\kc{-t+i\frac\beta2}\ket{\beta}.
\end{align}

We assume that the systems have a holographic duality of nearly AdS$_2$. The bulk description of the TFD state is the Rindler patch in AdS$_2$ space \cite{Maldacena:2016upp,JT}. For higher-dimensional CFT, one can further consider higher-dimensional generalization in the eternal black hole in AdS space.
We consider that the AdS$_2$ holography is described by dilaton gravity
\begin{align}\label{DilatonGravity}\begin{split}
I_0=&
-\frac{\phi_0}{16\pi G_N}\kd{\int dx^2 \sqrt{g}R+\int dx \sqrt{h}K}
-\frac1{16\pi G_N}\kd{\int d^2x \sqrt{g}\phi (R+2)+2\int_\partial dx \sqrt{h}\phi_\partial K}+...     \\
&- \sum_{i=1}^M \int d^2x \sqrt{g}\kd{(\partial\chi_i)^2+m^2\chi_i^2}
\end{split}\end{align}
The first term is a topological term. It does not control the dynamic and only contributes to extremal entropy $S_0=\phi_0/4G_N$.
The second term is called Jackiw-Teitelboim (JT) theory. After integrating out dilaton field $\phi$, one can fix the metric to be AdS$_2$ (EAdS$_2$). The remaining degrees of freedom are the boundary trajectory and the matter fields \cite{Maldacena:2016upp}. The dots refer to the expansion of the dilaton field from higher-dimensional reduction, which is neglectable for $\phi\ll \phi_0$. We will consider $M$ free scalar fields with the same mass $\ke{\chi_i}$ in the bulk. So the theory has $SO(M)$ symmetry of rotation $\chi_i\to \sum_j G_{ij}\chi_j$ where $G\in SO(M)$. In this paper, we will frequently exchange these scalar fields, which is a subgroup of the $SO(M)$ symmetry.

The coordinates of EAdS$_2$ and AdS$_2$ used in this paper are
\begin{align}
ds^2=\frac{d\mu^2+dz^2}{z^2}=\sinh^2\rho d\varphi ^2+d\rho^2.\\
ds^2=\frac{-d\nu^2+dz^2}{z^2}=-\sinh^2\rho d\psi^2+d\rho^2=\frac{-d\theta^2+d\sigma^2}{\sin^2\sigma}.
\end{align}
where $(\mu,z)$ and $(\nu,z)$ are Poincare coordinates, $(\phi,\rho)$ and $(\psi,\rho)$ are Rindler coordinates and $(\theta,\sigma)$ are conformal coordinates. We will first work on Euclidean time. One can impose boundary conditions
\begin{align}\label{BCtau}
ds_{\partial}^2=\frac1{\epsilon^2}d\tau^2,\quad \phi_{\partial}=\frac{\bar\phi}{\epsilon}.
\end{align}
The cutoff is imposed to prevent dilaton field $\phi$ from exceeding $\phi_0$. So $\bar\phi/\phi_0$ is related to the UV cutoff $\epsilon$ of the dual boundary theory, namely $\bar\phi/\phi_0=\alpha_\epsilon\epsilon$ with an order $1$ constant $\alpha_\epsilon$.
One can parametrize the boundary as $(\mu(\tau),z(\tau))$, where $z(\tau)$ is determined by the boundary condition 
\begin{align}
z=\epsilon\mu'(\tau)+O(\epsilon^3).
\end{align}
From the JT term in EAdS$_2$, one can obtain the effective action of reparametrization $\mu(\tau)$
\begin{align}\label{SchwarzianAction}\begin{split}
I_{eff}=- C \int d\tau \ke{\mu,\tau}=\frac C2 \int_0^{2\pi} d\tau (\varepsilon''^2-\varepsilon'^2)+O(\varepsilon^3),\\
\ke{\mu,\tau}=\kc{\frac{\mu''(\tau)}{\mu'(\tau)}}'-\frac{\mu''(\tau)^2}{2 \mu'(\tau)^2},\quad 
\mu=\tan\frac{\varphi}{2},\quad \varphi=\tau+\varepsilon(\tau) ,\quad  
C=\frac{\bar\phi}{8\pi G_N}, \quad \beta=2\pi.
\end{split}\end{align}
where $\ke{\mu,\tau}$ is the Schwarzian derivative. Coefficient $C$ has the dimension of time. The dependence on $\beta$ can be recovered by dimensional analysis. Those $SL(2,R)$ zero modes $\varepsilon=1,e^{i\tau},e^{-i\tau}$ should be excluded in the functional integrate on $\varepsilon$. One can obtain the correlation of reparametrization modes
\begin{align}\label{ReparametrizationCorrelation}
\avg{\varepsilon(\tau)\varepsilon(0)}=\frac1{2\pi C}\kd{-\frac12(|\tau|-\pi)^2+(|\tau|-\pi)\sin|\tau|+c_1 + c_2\cos \tau },
\end{align}
where arbitrary coefficients $\ke{c_1,c_2}$ come from the redundancy of $SL(2,R)$ zero modes.

For a free scalar field $\chi$ in the bulk, the correlation of its dual operator is modulated by reparametrization modes. The sources $\hat J(\mu)$ are related to the source $J(\tau)$ by
\begin{align}\label{HolographicDictionary}
\chi\sim z^{1-\varDelta}\hat J(\mu)\sim (\epsilon\mu'(\tau))^{1-\varDelta}\hat J(\mu)=\epsilon^{1-\varDelta}J(\tau)        
\quad\Rightarrow\quad \hat J(\mu)=\mu'(\tau)^{\varDelta-1}J(\tau).
\end{align} 
After integrating out the bulk field $\chi$, we obtain
\begin{align}
-I_M=& \int d\mu_1d\mu_2 \frac{\hat J(\mu_1)\hat J(\mu_2)}{(\mu_1-\mu_2)^{2\varDelta}}
=\int d\tau_1d\tau_2 J(\tau_1)J(\tau_2)\kd{\frac{\mu_1'(\tau_1)\mu_2'(\tau_2)}{(\mu_1(\tau_1)-\mu_2(\tau_2))^2}}^{\varDelta}\nn\\
=&\int d\tau_1d\tau_2 \frac{J(\tau_1)J(\tau_2)}{(2\sin\frac{\tau_{12}}{2})^{2\varDelta}}(1+\mathcal B(\tau_1,\tau_2)+\mathcal C(\tau_1,\tau_2)+O(\varepsilon^3)),\\
\mathcal B(\tau_1,\tau_2)=&\varDelta \left(\varepsilon_1 '+\varepsilon_2 '+\frac{\varepsilon_2-\varepsilon_1}{\tan\frac{\tau_{12}}{2}}\right),\\
\mathcal C(\tau_1,\tau_2)=&\frac{\varDelta ^2}{2}  \left(\varepsilon _1'+\varepsilon _2'+\frac{\varepsilon _2-\varepsilon _1}{\tan\frac{\tau_{12}}{2}}\right){}^2
+\varDelta \left(\left(\frac{\varepsilon _1-\varepsilon _2}{2\sin\frac{\tau_{12}}{2}}\right){}^2 -\frac12\left(\varepsilon _1'{}^2+\varepsilon _2'{}^2\right)\right).
\end{align}
where $\tau_{12}=\tau_1-\tau_2$, $\varepsilon_1=\varepsilon(\tau_1)$. The operator $O$ has been normalized such that the common factor  $D_\varDelta=\frac{(\varDelta-1/2)\Gamma(\varDelta)}{\sqrt{\pi}\Gamma(\varDelta-1/2)}$ does not appear in the correlation function. After integrating out $\varepsilon$, we obtain the generating functional
\begin{align}\label{GeneratingFunctional}
\ln \avg{e^{-I_M}}=&\int d\tau_1d\tau_2 \frac{J(\tau_1)J(\tau_2)}{(2\sin\frac{\tau_{12}}{2})^{2\varDelta}}(1+\avg{\mathcal C(\tau_1,\tau_2)}) \nn    \\
&+\frac{1}{2}\int d\tau_1d\tau_2d\tau_3d\tau_4 \frac{J(\tau_1)J(\tau_2)J(\tau_3)J(\tau_4)}{(2\sin\frac{\tau_{12}}{2})^{2\varDelta}(2\sin\frac{\tau_{34}}{2})^{2\varDelta}}\avg{\mathcal B(\tau_1,\tau_2)\mathcal B(\tau_3,\tau_4)}+O(C^{-2}),\\
\avg{\mathcal C(\tau_1,\tau_2)}=&-\frac{\varDelta}{8\pi C}  \left(\sin\frac{\tau_{12}}{2}\right)^{-2} \left(-\tau_{12}^2+2 \pi  |\tau_{12}|+2 \left(|\tau_{12}|-\pi\right) \sin |\tau_{12}|+2 \cos \tau_{12}-2\right)\nn\\
&+\frac{\varDelta ^2}{4\pi C} \left(\frac{\tau_{12}}{\tan\frac{\tau_{12}}{2}}-2\right) \left(\frac{\tau_{12}^2-2 \pi  |\tau_{12}|} {\tau_{12}\tan \frac{\tau_{12}}{2}}-2\right),    \label{Loop}
\end{align}
where the bracket $\avg{...}=\int \frac{D\varepsilon}{SL(2,R)} ... e^{-I_{eff}}$ and Eq.~(\ref{ReparametrizationCorrelation}) is used. The specific expression of $\avg{\mathcal B(\tau_1,\tau_2)\mathcal B(\tau_3,\tau_4)}$ is too tedious to be shown here, which does not only depend on $\tau_{12}$ and $\tau_{34}$ for ordering $\tau_1>\tau_3>\tau_2>\tau_4$.

The real-time correlator can be obtained by Wick rotation $\tau\to i t$. The two boundaries of AdS$_2$ space satisfy the boundary conditions
\begin{align}\begin{split}
ds^2_{\partial,L}=&-\frac1{\epsilon^2}d t^2,\quad \phi_{\partial,L}=\frac{\bar\phi}{\epsilon},\\
ds^2_{\partial,R}=&-\frac1{\epsilon^2}d t^2,\quad \phi_{\partial,R}=\frac{\bar\phi}{\epsilon}.
\end{split}\end{align}
The boundaries are parametrized by $\ke{\nu_L(t),\nu_R(t)}$ in Poincare coordinates or $\ke{\psi_L(t),\psi_R(t)}$ in Rindler coordinates,
whose effective action is
\begin{align}
I_{eff}=-C\int dt_L \ke{\nu_L, t_L}-C \int d t_R \ke{\nu_R, t_R}
=-C\int dt_L \ke{-\coth\frac{\psi_L}{2}, t_L}-C \int d t_R \ke{\tanh\frac{\psi_R}2, t_R}.
\end{align}
The $SL(2,R)$ symmetry of parametrization $\ke{\psi_L( t),\psi_R( t)}$ is
\begin{align}
\psi_L \to \psi_L + \epsilon_0 + \epsilon_1 e^{\psi_L} + \epsilon_2 e^{-\psi_L},    \quad 
\psi_R \to \psi_R - \epsilon_0 + \epsilon_1 e^{-\psi_R} + \epsilon_2 e^{\psi_R}.
\end{align}
Without the excitation of scalar fields, those three $SL(2,R)$ charges are required to vanish for the consistency of the effective action. Then the equations of motion are automatically satisfied. Corresponding to the TFD state $\ket{\beta(t)}$, the solution is 
\begin{align}
\psi_L= \frac{2\pi}{\beta}  t_L ,\quad     \psi_R= \frac{2\pi}{\beta}  t_R,\quad t_L=t_R=t.
\end{align}
The correlator in bi-systems can be obtained by doing analytical continuation $t\to t\pm i\frac{\beta}{2}$. For example, the two-point functions at tree level in real time are
\begin{align}
\avg{O_L(t_1)O_L(t_2)}&=\avg{O_R(t_1)O_R(t_2)}
=\Tr[O(t_1)O(t_2)y^2]
=2 \left(2 i \sinh\frac{t_{12}}{2}\right)^{-2 \varDelta },\\
\label{2Pt}
\avg{O_L(t_1)O_R(t_2)}&
=\Tr[O(-t_1)yO(t_2)y]
=2\left(2\cosh\frac{t_1+t_2}{2}\right)^{-2 \varDelta },
\end{align}
where, within the trace,
\begin{gather}
O(t)=e^{iHt}Oe^{-iHt},\\
y=Z^{-1/2}e^{-\beta H/2}.
\end{gather}

\subsection{Unbalanced bi-systems and holography}

Now we make the Hamiltonian of the bi-systems unbalanced
\begin{align}\label{UnbalancedHamiltonian}
\tilde H_{0}=H_L+\tilde H_R, \quad H_L=H\otimes 1,\quad \tilde H_R=1\otimes \lambda H,
\end{align}
where the Hamiltonian of system $R$ is multiplied by a constant $\lambda$. We still consider the TFD state $\ket{\beta}$ in Eq.~(\ref{TFD}) at time $t=0$, which is prepared by imaginary-time evolution with $H_0$ rather than $\tilde H_0$, while $\ket\beta$ evolves with $\tilde H_0$ in real time, $\ket{\tilde\beta(t)}=e^{-i\tilde H_0t}\ket\beta=\ket{\beta(\frac{1+\lambda}{2}t)}$. Tracing it by part, we have
$
\rho_\gamma=\Tr_{\bar\gamma} \ket{\tilde\beta(t)}\bra{\tilde\beta(t)}=\frac1Z e^{-\beta H}=\frac1Z e^{-(\beta/\lambda) \lambda H}
$.
Thus $\beta_L=\beta$ and $\beta_R=\beta/\lambda$. Without loss of generality, we set $\lambda\geq1$ through this paper. So system $R$ is hotter than system $L$ or they have the same temperature, {\it i.e.} $\beta_L\geq\beta_R$. Effectively, multiplying the Hamiltonian of system $R$ by $\lambda$ is equivalent to redefining the time of system $R$ as $\tilde t=t/\lambda$.

We still consider the same operator $O$ at $t=0$. Since we have changed the Hamiltonian, the time evolution of $O$ is changed as well. Through this paper, $\tilde O(t)$ refers to the time evolution of $O$ with $\tilde H_0$ and $O(t)$ refers to the time evolution of $O$ with $H_0$. They are related by
\begin{align}\begin{split}\label{InteractionPicture}
\tilde O_{L}(t)=e^{i\tilde H_0t}O_{L}e^{-i\tilde H_0t}=e^{iHt}O^Te^{-iHt}\otimes 1=O_L(t),\\
\tilde O_{R}(t)=e^{i\tilde H_0t}O_{R}e^{-i\tilde H_0t}=1\otimes e^{iH\lambda t}Oe^{-iH\lambda t}=O_R(\lambda t),
\end{split}\end{align}
For system $R$, the evolution with $\tilde H_0$ by time $t$ is just the evolution with $H_0$ by time $\lambda t$. Take the correlator as an example, 
\begin{align}\label{CorrelationRelation}
\avg{\tilde O_L(t_1)\tilde O_R(t_2)}=\avg{e^{iH t_1}O^Te^{-iH t_1} \otimes e^{i\lambda Ht_2} O e^{-i\lambda Ht_2}}=\avg{O_L(t_1) O_R(\lambda t_2)},
\end{align}
where $\avg{...}\equiv\bra\beta...\ket\beta$ for short. Then we can map the observables under the evolution with $\tilde H_0$ to those that under the evolution with $H_0$. This trick will be frequently used in later calculations.

\begin{figure}
	\centering
	\includegraphics[height=0.25\linewidth]{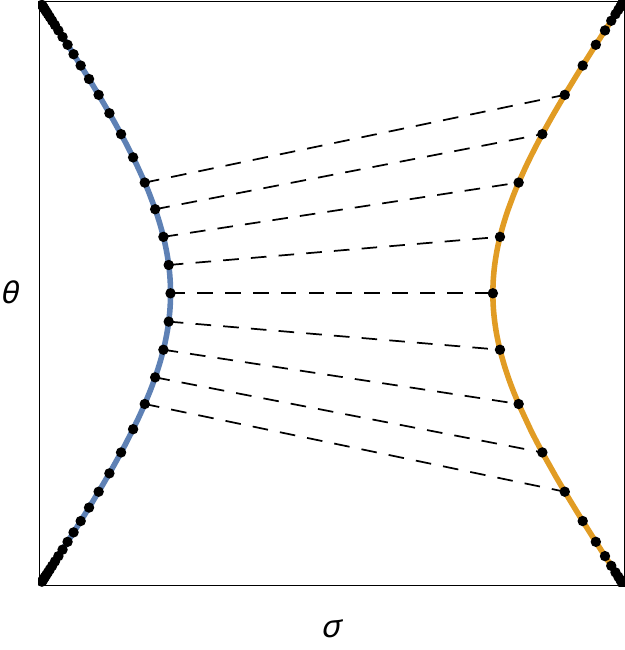}
	\caption{Two boundaries $\tilde\nu_L(t)$ and $\tilde\nu_R(t)$ with $\lambda=2$ in conformal coordinates $(\sigma,\theta)$. Systems $L,R$ are denoted by the blue curve and the orange curve. The time parametrizations are shown by the graduation line on the curves. The dashed lines indicate the insertion of interactions.}
	\label{FigUnbalance}
\end{figure}

We can further discuss the effect on the gravity side under the change of Hamiltonian.
Corresponding to the evolution with unbalanced $\tilde H_0$ by time $t$, the conditions of the two boundaries in AdS$_2$ become
\begin{align}\begin{split}
ds^2_{\partial,L}=&-\frac1{\epsilon^2}dt^2,\quad \phi_{\partial,L}=\frac{\bar\phi}{\epsilon}\\
ds^2_{\partial,R}=&-\frac{\lambda^2}{\epsilon^2}dt^2,\quad \phi_{\partial,R}=\frac{\bar\phi/\lambda}{\epsilon/\lambda}
\end{split}\end{align}
where the UV cutoff of system $R$ becomes $\epsilon/\lambda$. Let the boundaries corresponding to the evolution with unbalanced $\tilde H_0$ be parametrized by $\ke{\tilde\nu_L(t),\tilde\nu_R(t)}$ or $\ke{\tilde\psi_L(t),\tilde\psi_R(t)}$.
The evolution with unbalanced $\tilde H_0$ can inherit the bulk description from the evolution with $H_0$ according to Eq.~(\ref{InteractionPicture}) and
\begin{align}\begin{split} 
\tilde \nu_L(t)&=\nu_L(t),\quad \tilde \nu_R(t)=\nu_R(\lambda t),\\
\tilde \psi_L(t)&=\psi_L(t),\quad \tilde \psi_R(t)=\psi_R(\lambda t).
\end{split}\end{align}
Then the effective action becomes
\begin{align}
I_{eff}=-C\int d t \ke{\tilde\nu_L, t}-\frac{C}{\lambda} \int d t \ke{\tilde\nu_R, t}
=-C\int d t \ke{-\coth\frac{\tilde\psi_L}{2}, t} - \frac{C}{\lambda} \int d t \ke{\tanh\frac{\tilde\psi_R}2, t}.
\end{align}
Corresponding to the TFD state $\ket{\tilde\beta(t)}$, the saddle point solution of an eternal black hole with unbalanced $\tilde H_0$ is
\begin{align}
\tilde\psi_L= \frac{2\pi}{\beta}  t ,\quad     \tilde\psi_R= \frac{2\pi\lambda}{\beta}  t.
\end{align}
As expected, the black hole $R$ is hotter than the black hole $L$.
Although the parametrizations are unbalanced, the classical location of the boundary is balanced, as shown in Fig.~\ref{FigUnbalance}.

The source $\tilde J_R$ of system $R$ can be identified as follows
\begin{align}
O_R(\lambda t)=\tilde O_R(t)\quad \text{and}\quad \int dt   J_R( t)O_R( t)=\int dt  \tilde J_R(t)\tilde O_R(t) 
\quad\Rightarrow\quad \lambda J_R(\lambda t)=\tilde J_R(t)
\end{align}
which is consistent with the correlation (\ref{CorrelationRelation}), since
\begin{align}\label{CorrelationRelationMatch}
\int dt_1 dt_2 J_L(t_1)J_R(t_2)\kd{\frac{\nu_L'(t_1)\nu_R'(t_2)}{(\nu_L(t_1)-\nu_R(t_2))^2}}^{\varDelta}
=\int d t_1d t_2 \tilde J_L(t_1)\tilde J_R(t_2)\kd{\frac{\nu_L'( t_1)\nu_R'(\lambda t_2)}{(\nu_L(t_1)-\nu_R(\lambda t_2))^2}}^{\varDelta}.
\end{align}
Compared with dictionary (\ref{HolographicDictionary}), the dictionary for $\tilde J_R$ is
\begin{align}
&\chi\sim z^{1-\varDelta}\hat J_R(\nu_R)\sim(\epsilon\tilde\nu_R'(t)/\lambda)^{1-\varDelta}\hat J_R(\nu_R)=\epsilon^{1-\varDelta}J_R(\lambda t)=\epsilon^{1-\varDelta}\lambda^{-1}\tilde J_R(t)
\, \nn\\
\Rightarrow \, & 
\hat J_R(\nu_R)=\tilde\nu_R'(t)^{\varDelta-1}\lambda^{-\varDelta} \tilde J_R(t)
\end{align}
rather than $\chi\sim (\epsilon/\lambda)^{1-\varDelta} \tilde J_R(\tilde \tau_R)$ and $\hat J_R(\nu_R)=\tilde\nu_R'(t)^{\varDelta-1} \tilde J_R(t)$. This means that the holographic dictionary of $\tilde O_R(t)$ is different from the one of $O_R(t)$.

One may wonder whether any physics will be changed by redefining the time. Indeed, no physics has been changed so far. However, physics will be changed if we couple the two systems.

\subsection{Product states as a comparison}\label{SubSectionProductState}

As a comparison, we still adopt the unbalanced Hamiltonian $\tilde H_0$ in Eq.~(\ref{UnbalancedHamiltonian}) but replaced the initial state by the product state
\begin{align}\label{ProductState}
\rho=\tau(\beta)\otimes\tau(\beta/\lambda),
\end{align}
which is the same as the TFD state (\ref{TFD}) locally. The interaction picture (\ref{InteractionPicture}) also holds.

The Hamiltonian $H$ of conformal field theory is gapless. Considering a non-trivial Hamiltonian $\tilde H_0\not\equiv0$, we find that product state (\ref{ProductState}) is a passive state if and only if $\lambda=1$. The analysis is as follows. The probability and energy of the energy eigenstate $\ket{n}_L\ket{m}_R$ are $p_{nm}=(Z_LZ_R)^{-1} \exp(-\beta(E_n+E_m))$ and $E_{nm}=E_n+\lambda E_m$. 
When $\lambda=1$, $p_{nm}\propto \exp(-\beta E_{nm})$, then $\rho$ satisfies condition (\ref{Passive}). Recall that local Hamiltonian $H_\gamma$ here is gapless. When $\lambda>1$, we can choose $m$ and $m'$ such that $E_m-E_{m'}=\delta>0$, and then choose $n$ and $n'$ such that $\lambda\delta>E_{n'}-E_{n}>\delta$. We find $E_{nm}>E_{n'm'}$ but $p_{nm}>p_{n'm'}$ and then $\rho$ does not satisfy condition (\ref{Passive}).

The product of two thermal states is dual to two black holes without a wormhole connecting them \cite{VanRaamsdonk,Product}. We will call them disconnected black holes. In AdS$_2$ holography, we should consider the union of two AdS$_2$ spaces, each of which is controlled by a theory of JT gravity. The disconnected black holes are described by two Rindler patches of the two AdS$_2$ spaces, as discussed in Appendix \ref{SectionProductBH}. The wormhole in each Rindler path is caused by the purification of the corresponding thermal state and has nothing to do with the entanglement between system $L$ and system $R$. 

\section{Extract work from wormhole}\label{SectionWork}

The wormhole in the eternal black hole geometrically represents the entanglement between system $L$ and system $R$, each of which is in a thermal state.
From quantum thermodynamics, as work can be extracted from correlation, it is natural to expect that work can be extracted from the wormhole as well. 

We take the TFD state as the initial state and turn on the interaction $H_I$ at time $t_i$. After time $t_i$, the total Hamiltonian is $H_{tot}=\tilde H_0+H_I$. In the interaction picture, the state at time $t_f>t_i$ is
\begin{align}\label{TFDInteraction}
\ket{\tilde\beta_I(t_f)}=\mathcal T \exp\kc{-i\int_{t_i}^{t_f}dt\, \tilde H_I(t)}\ket{\beta},
\end{align}
where $\tilde H_I(t)$ is the interaction picture of $H_I$ under the evolution with $\tilde H_0$ and  $\mathcal T$ is time ordering.

To study the work extracted from the eternal black hole, in this section, we consider the interaction
\begin{align}\label{HIOO}
H_I= g  O_L  O_R.
\end{align}
Such an interaction is also introduced in the literature to couple two spacetimes \cite{Product} or construct a traversable wormhole for $g<0$ \cite{Gao:2016bin,Maldacena:2017axo,Maldacena:2018lmt}. For balanced $H_0$, the energy changes at the first-order perturbation of the interaction (\ref{HIOO}) have been calculated in Refs.~\cite{Gao:2016bin,Maldacena:2017axo}. Here we revisit them from the viewpoint of quantum thermodynamics. Due to the consideration of the weak interaction $H_I$, we should consider a small $g$, such that the perturbation theory holds. For unbalanced $\tilde H_0$, the energy changes are
\begin{subequations}\begin{align}
	\Delta E_L=&\bra{\tilde\beta_I(t_f)}H_L\ket{\tilde\beta_I(t_f)}-\avg{H_L}=\Delta E_L^{(1)}+\Delta E_L^{(2)}+O(g^3),    \\
	\Delta E_R=&\bra{\tilde\beta_I(t_f)}\tilde H_R\ket{\tilde\beta_I(t_f)}-\avg{\tilde H_R}=\Delta E_R^{(1)}+\Delta E_R^{(2)}+O(g^3),    \\
	\Delta E_I=&\bra{\tilde\beta_I(t_f)}\tilde H_I(t_f)\ket{\tilde\beta_I(t_f)}-\avg{\tilde H_I(t_i)}=\Delta E_I^{(1)}+\Delta E_I^{(2)}+O(g^3),
	\end{align}\end{subequations}
where the superscript in $\Delta E^{(i)}$ denotes the energy change at $O(g^i)$. After $t_i$, energy conservation holds at each order of $g$.
By virtue of the entanglement, the work extracted at $O(g)$ is
\begin{align}\label{Work}
W^{(1)}=\Delta E_I^{(1)}=g\kc{\avg{O_L(t_f)O_R(\lambda t_f)}-\avg{O_L(t_i)O_R(\lambda t_i)}}.
\end{align}
According to Eq.~(\ref{2Pt}) at tree level, $\sgn(W^{(1)})=-\sgn(g)\sgn(t_f+t_i)$, which can be positive. As explained in Ref.~\cite{Maldacena:2017axo}, the interaction (\ref{HIOO}) imposes a potential energy $\avg{\tilde H_I(t)}=g\avg{O_L(t)O_R(\lambda t)}\sim ge^{-ml(t)}$ and does work on each system, where $l(t)$ is the distance between the location of the two operators $O_L(t)$ and $O_R(\lambda t)$ in AdS$_2$ space, and $m$ is the mass of the dual scalar field $\chi$.

The first law of entropy holds at first-order perturbation, then
\begin{align}\label{EnergyRelation}
\beta\Delta E_L^{(1)}=\Delta S_L^{(1)}=\Delta S_R^{(1)}=(\beta/\lambda)\Delta E_R^{(1)}.
\end{align}
Combining it with Eqs.~(\ref{Conserve}) and (\ref{Work}), we have
\begin{align}
\Delta E_L^{(1)}=-\frac{1}{1+\lambda}W^{(1)}, \quad 
\Delta E_R^{(1)}=-\frac{\lambda}{1+\lambda}W^{(1)}, \quad 
\Delta S_L^{(1)}=\Delta S_R^{(1)}=-\frac{1}{1+\lambda}\beta W^{(1)}.
\end{align}
Then we find such work is extracted from correlation
\begin{align}\label{MI2Work}
W^{(1)}=-\frac12\kc{\frac1{\beta_L}+\frac1{\beta_R}}\Delta I_{L;R}^{(1)}.
\end{align}
We can alternatively calculate the change in the von Neumann entropy from the replica trick in Sec.~\ref{SectionEntropy}. 

According to the definitions of the work and heat of the local system in Appendix \ref{SectionFirstLaw}, we find that $\Delta E_I^{(1)}$ is the change in the binding energy during the process. The work $W^{(1)}$ is extracted through the binding energy in Eq.~(\ref{Work}). The local energy change in the way of heat, $\Delta E_\gamma^{(1)}=\Delta Q_\gamma^{(1)}$, and the work done on the local system vanishes, $\Delta W_\gamma^{(1)}=0$.

As a comparison, we try to extract work from the disconnected black holes, as calculated in Appendix \ref{SectionProductBH}. But, without a wormhole, $\Delta E^{(1)}\propto\avg{\dot O(t)}=0$, so no work can be extracted at $O(g)$. We further go to $O(g^2)$ and find that $\Delta E_L^{(2)}+\Delta E_R^{(2)}\geq0$, so no positive work can be extracted via this interaction. It is expectable when $\lambda=1$, since the disconnected black holes form a passive state and the eternal black hole does not. However, when $\lambda>1$, the disconnected black holes do not form a passive state anymore, as explained below Eq.~(\ref{ProductState}). The failure of extracting positive work results from the unsatisfactory form of the interaction.

\section{Anomalous heat flow via wormhole}\label{SectionAnomalous}

\subsection{Interaction}

In this section, we continue to consider the unbalanced $\tilde H_0$ and the TFD state. As explained in the previous section, the appearance of anomalous heat flow is highly dependent on the form of the interaction term \cite{Lloyd:2015}. To look for the anomalous heat flow in the eternal black hole, we will alternatively consider the interaction
\begin{align}\label{HIVW-WV}
H_I=\frac1{\sqrt2}g( V_L W_R - W_L V_R),
\end{align}
where two scalar operators $V$ and $W$ are dual to two scalar fields $\chi_V=\chi_1$ and $\chi_W=\chi_2$ separately in the bulk for $M=2$. The scaling dimensions of $V$ and $W$ are the same. Due to the $SO(M)$ symmetry of Eq.~(\ref{DilatonGravity}), any Green's functions of $V$ and $W$ are invariant under exchange $V\leftrightarrow W$. Similar interactions was appeared in the thermofield dynamics \cite{Alfinito:2007zz} \footnote{We thank Sang Pyo Kim for pointing this point out.}.
One may consider other kinds of interactions, such as Eq.~(\ref{HIOO}) or
\begin{align}\label{HIVW}
H_I=g V_L W_R
\end{align}
or
\begin{align}\label{HIVW+WV}
H_I=\frac1{\sqrt2}g( V_L W_R + W_L V_R).
\end{align}

Since $V$ and $W$ are dual to different scalar fields, $\avg{V_L(t_1)W_R(t_2)}=\avg{W_L(t_1)V_R(t_2)}=0$. So the effect of each of Eqs.~(\ref{HIVW-WV}), (\ref{HIVW}), and (\ref{HIVW+WV}) at $O(g)$ vanishes.
We have to look at the contribution at $O(g^2)$.
Actually, even if it does not vanish, {\it e.g.} for the interaction (\ref{HIOO}), the first law of entropy leads to Eq.~(\ref{EnergyRelation}). Then we have
\begin{align}
\sgn((\Delta E_L^{(1)}-\Delta E_R^{(1)})(\beta_L-\beta_R)) = -\sgn(\Delta E_L^{(1)} + \Delta E_R^{(1)}).
\end{align}
Comparing it with criterion (\ref{ReversedArrow}), we find that the arrow of time will not be reversed at first order. This statement holds for other forms of interaction and higher dimensional generalization, while it does not hold for the mixed initial state. So we design the interaction (\ref{HIVW-WV}) to kill the effect of first-order perturbation.

At second-order perturbation, different interactions will generate different sets of Feynman diagrams. As we will see, the sets of diagrams generated by interactions (\ref{HIVW}) and (\ref{HIVW+WV}) are contained in the set of diagrams generated by the interaction (\ref{HIVW-WV}). Furthermore, the set of diagrams generated by the interaction (\ref{HIVW-WV}) is contained in the set of diagrams generated by the interaction (\ref{HIOO}). 
We do not observe any anomalous heat flow with interactions (\ref{HIOO}), (\ref{HIVW}), and (\ref{HIVW+WV}).

\subsection{Early time}

We will consider the interaction (\ref{HIVW-WV}) and go to $O(g^2)$. The changes in the local energies $\Delta E_\gamma$ and interaction energy $\Delta E_I$ from time $t_i$ to time $t_f$ at $O(g^2)$ are
\begin{subequations}\label{PerturbationA}\begin{align}
	&\Delta E_L\approx\int_{t_i}^{t_f}dt\int_{t_i}^{t}dt' K_L(t,t'),\nn\\
	&K_L(t,t')=(-i)^3g^2\kc{
		\avg{[\dot V_L(t)W_R(\lambda t),V_L(t')W_R(\lambda t')]}
		-\avg{[\dot V_L(t)W_R(\lambda t),W_L(t')V_R(\lambda t')]}    },    
	\label{PerturbationAL}        \\
	&\Delta E_R\approx\int_{t_i}^{t_f}dt\int_{t_i}^{t}dt' K_R(t,t'),\nn\\
	&K_R(t,t')=(-i)^3g^2\lambda\kc{
		\avg{[V_L(t)\dot W_R(\lambda t),V_L(t')W_R(\lambda t')]}
		-\avg{[V_L(t)\dot W_R(\lambda t),W_L(t')V_R(\lambda t')]}    },
	\label{PerturbationAR}    \\
	&\Delta E_I\approx\int_{t_i}^{t_f}dt K_I(t_f,t),\nn\\
	&K_I(t,t')=(-i)g^2\kc{
		\avg{[V_L(t)W_R(\lambda t),V_L(t')W_R(\lambda t')]}
		-\avg{[V_L(t)W_R(\lambda t),W_L(t')V_R(\lambda t')]}    }.
	\label{PerturbationAI}
	\end{align}\end{subequations}
We have used the $SO(M)$ symmetry to exchange $V\leftrightarrow W$. Since all the following results are of $O(g^2)$, we have omitted the superscript in the $\Delta E^{(2)}$s. The first (second) term in each integral kernel of Eq.~(\ref{PerturbationA}) comes from the diagonal (cross) terms in $H_I^2$. If we choose the interaction (\ref{HIVW}), the second term in each integral kernel of Eq.~(\ref{PerturbationA}) disappears. If we choose interaction (\ref{HIVW+WV}), the second term in each integral kernel of Eq.~(\ref{PerturbationA}) has an inverse sign. The following discussion corresponding to the interaction (\ref{HIVW-WV}) can be easily translated into the case of the interactions (\ref{HIVW}) and (\ref{HIVW+WV}).

The two terms in each integral kernel of Eq.~(\ref{PerturbationA}) are denoted as $(K_\gamma(t,t'))_{VWVW}$ and $(K_\gamma(t,t'))_{VWWV}$ where $\gamma=L,R,I$. They are related to the following two commutators
\begin{subequations}\label{4PtC}\begin{align}
	C_{VWVW}=&\avg{[V_L(t_1)W_R(t_2),V_L(t_3)W_R(t_4)]}
	=2i\Im\Tr[V(-t_3)V(-t_1)yW(t_2)W(t_4)y]\nn\\
	\approx& (C_{VWVW})_{discon.} + (C_{VWVW})_{t-channel} + O(C^{-2}),    \label{VWVW}    \\
	C_{VWWV}=&\avg{[V_L(t_1)W_R(t_2),W_L(t_3)V_R(t_4)]}
	=2i\Im\Tr[W(-t_3)V(-t_1)yW(t_2)V(t_4)y]\nn\\
	\approx& (C_{VWWV})_{u-channel}    + O(C^{-2}), \label{VWWV} \\
	&\ke{t_1,t_2,t_3,t_4}\approx\ke{t,\lambda t,t',\lambda t'},\quad t'<t\nn
	\end{align}\end{subequations}
and their derivative. $C_{VWVW}$ is transformed into $VVWW$-order four-point functions and $C_{VWWV}$ is transformed into $VWVW$-order four-point functions at finite temperature.

\begin{figure}
	\newcommand{\figsize}{100pt}
	\centering
	\subfigure[]{
		\includegraphics[height=\figsize]{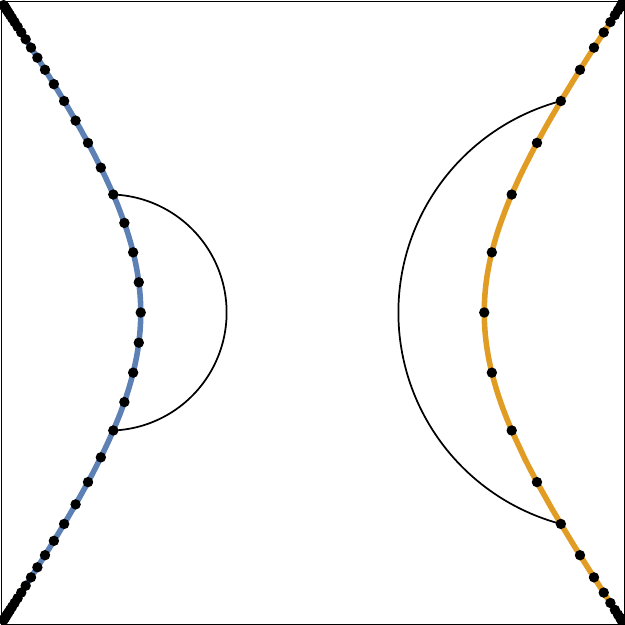}
		\label{Fig4Pt_d1}
	}    
	\subfigure[]{
		\centering
		\includegraphics[height=\figsize]{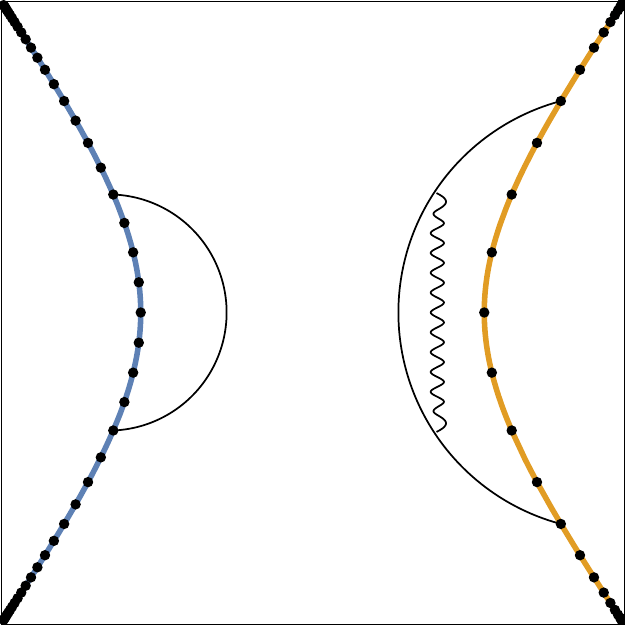}
		\label{Fig4Pt_d2}    
	}    
	\subfigure[]{
		\centering
		\includegraphics[height=\figsize]{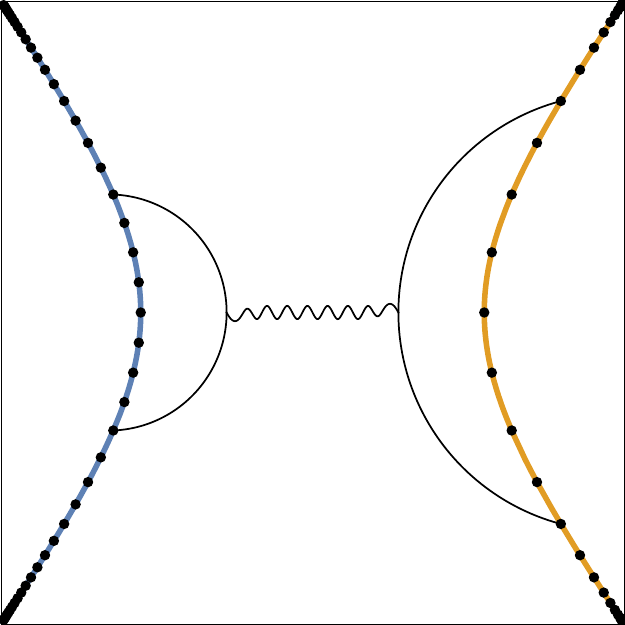}
		\label{Fig4Pt_t}
	}
	\subfigure[]{
		\includegraphics[height=\figsize]{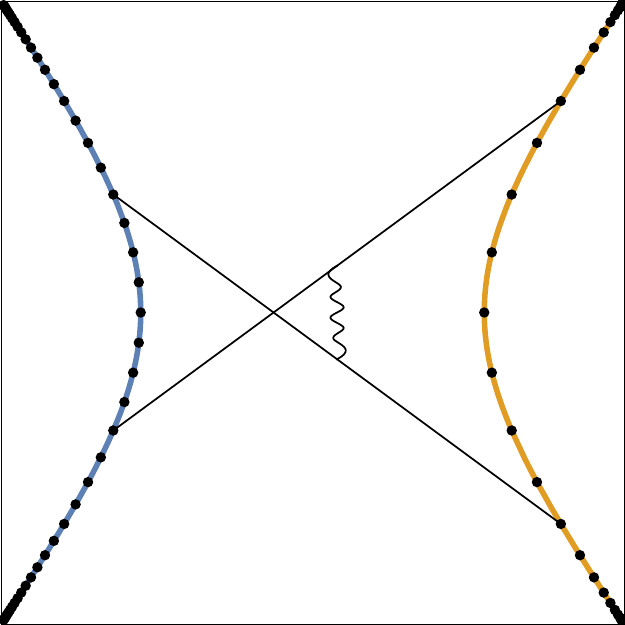}
		\label{Fig4Pt_u}
	}
	\caption{Four diagrams contribute to Eq.~(\ref{PerturbationA}): (a) disconnected-tree diagram, (b) disconnected-loop diagram, (c) t-channel, and (d) u-channel. The loop in (b) can appear on either side. The smooth (wavy) curve is the correlator of the scalar field (graviton). The commutator $C_{VWVW}$ in Eq.~(\ref{VWVW}) consists of (a)(b)(c) . The commutator $C_{VWWV}$ in Eq.~(\ref{VWWV}) consists of (d).}
	\label{Fig4PtA}
\end{figure}

Those diagrams contributing to the commutators (\ref{PerturbationA}) stem from the causal connection between time $t$ and time $t'$, which is shown in Fig.~\ref{Fig4PtA}. Since there are two operators on each side, (dis)connected diagrams will (not) go across the wormhole.

\begin{figure}
	\centering
	\includegraphics[height=100pt]{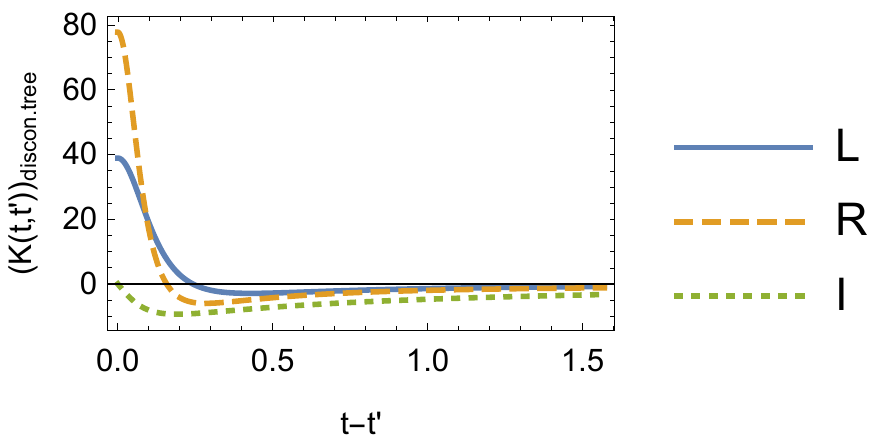}~~~~
	\includegraphics[height=100pt]{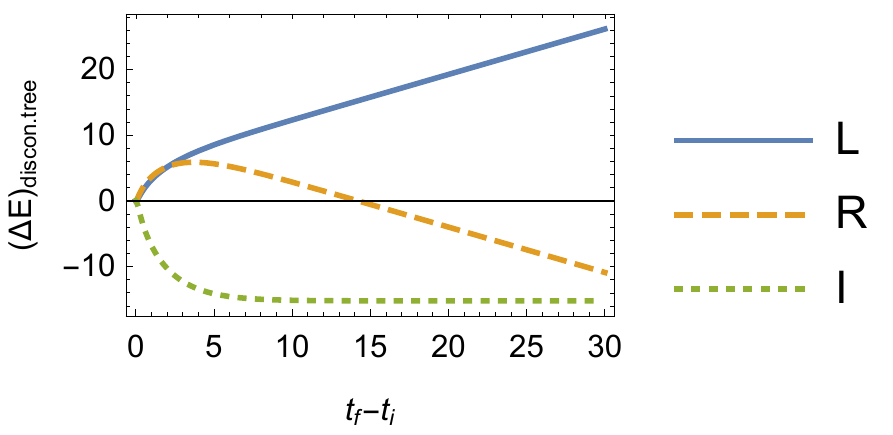}
	\caption{Kernels $(K_L(t,t'))_{discon.},\,(K_R(t,t'))_{discon.},\,(K_I(t,t'))_{discon.}$ as functions of $t-t'$ and energy changes $(\Delta E_L)_{discon.},\,(\Delta E_R)_{discon.},\,(\Delta E_I)_{discon.}$ as functions of $t_f-t_i$, contributed by the disconnected-tree diagram given in Fig.~\ref{Fig4Pt_d1}, where $\varDelta=1/6,\, \beta=2\pi,\,\lambda=2,\, C=10^5$ and $\epsilon=0.2$. Actually, the contribution of the loop diagram is neglectable here.}
	\label{FigE_tree}
\end{figure}

The first commutator (\ref{VWVW}) contain the diagrams in Figs.~\ref{Fig4Pt_d1}, \ref{Fig4Pt_d2}, and \ref{Fig4Pt_t} up to $O(C^{-2})$. The tree diagram in Figs.~\ref{Fig4Pt_d1} and its one-loop correction in Fig.~\ref{Fig4Pt_d2} are disconnected and do not go across the wormhole. They are due to the propagation of the two scalar fields $\chi_V,\chi_W$ in each side separately. So, they are functions of $t_{13}$ and $t_{24}$, which are summed into
\begin{align}
(C_{VWVW})_{discon.}=-i 2^{3} \sin (2 \pi  \varDelta ) \left(4\sinh\frac{t_{13}}{2} \sinh\frac{t_{24}}{2}\right)^{-2 \varDelta }(1+O(C^{-1}))
\end{align}
This has an important contribution on the kernels at short time $t-t'<t_d/2$, where $t_d=\beta/2\pi$ is the dissipation time. 
It suffers from UV divergence once $0<\varDelta$. However, when $\varDelta<1/4$, the integral $(\Delta E_I)_{discon.}$ contributed by $(C_{VWVW})_{discon.}$ is finite. The kernel $(K_L(t,t')-K_R(t,t'))_{discon.}$ turns out to be $O((t-t')^{1-4\varDelta})$ so that the integral $(\Delta E_L-\Delta E_R)_{discon.}$ converges when $\varDelta<1/4$. We will focus on the case of $0<\varDelta<1/4$. Of course, one can also introduce a small separation $\epsilon$ on imaginary time to regularize the UV behavior, such as $\avg{V(t_1)V(t_2)}\to\avg{V(t_1-i\epsilon)V(t_2)}$. It acts as the UV cutoff of the boundary theory which is dual to JT gravity and satisfies $\epsilon\ll\beta$. However, the integrals $(\Delta E_L-\Delta E_R)_{discon.}$ and $(\Delta E_I)_{discon.}$ are not UV sensitive. They only have some shifts if we change $\epsilon$ and remain finite if we set $\epsilon\to0$.
We illustrate the energy changes contributed by the disconnected diagrams shown in Figs.~\ref{Fig4Pt_d1} and \ref{Fig4Pt_d2} in Fig.~\ref{FigE_tree}.
After the dissipation time, $(\Delta E_I(t,t'))_{discon.}$ is saturated, while energy starts to diffuse from hotter system $R$ to colder system $L$ via the channel of disconnected diagrams.
Further numerical calculation shows that the energy changes contributed by these disconnected diagrams never satisfy criterion (\ref{ReversedArrow}). This is expected since these disconnected diagrams are not related to the wormhole. They can also appear in the interaction between two disconnected black holes in Appendix \ref{SectionProductBH}, where the thermodynamic arrow of time should not be reversed. 

The connected diagram in Fig.~\ref{Fig4Pt_t} is called the t-channel, which goes across the wormhole. 
The interpretation of the t-channel is that the energy-momentum tensors of the two scalar fields $\chi_V,\chi_W$ on each side are correlated with each others caused by the propagation of the virtual graviton through the wormhole. The propagation of the virtual graviton just reflects the energy correlation between the two sides of TFD state, $H_L\ket\beta=H_R\ket\beta$. The t-channel contributes to the commutator $C_{VWVW}$ with
\begin{align}
(C_{VWVW})_{t-channel}
=-i  2^{3} \sin (2 \pi  \varDelta )  \left(4\sinh\frac{t_{13}}{2} \sinh\frac{t_{24}}{2}\right)^{-2 \varDelta }  \frac{\varDelta ^2}{2\pi C} \left(t_{13} \coth \frac{t_{13}}{2}-2\right) \left(t_{24} \coth\frac{t_{24}}{2}-2\right).
\end{align}
It also depends on $t_{13}$ and $t_{24}$ only, but it is free from UV divergence once $\varDelta<1$. It is always $O(C^{-1})$, so it is smaller than the contribution of the tree diagram in Fig.~\ref{Fig4Pt_d1}. Since $t_{13}\geq0,\,t_{24}\geq0$ and $0<\varDelta<1/4$, we find $(-i)(C_{VWVW})_{t-channel}\leq0$; then $(\Delta E_L+\Delta E_R)_{t-channel}=-(\Delta E_I)_{t-channel}\geq0$, which does not satisfy criterion (\ref{ReversedArrow}).

Thus, the first commutator (\ref{VWVW}) is a function of $t_{13}$ and $t_{24}$, which describes the process without the signal passing the wormhole since the original wormhole in the eternal black hole is not traversable. The energy change contributed by the first commutator (\ref{VWVW}) obeys the thermodynamic arrow of time.

\begin{figure}
	\centering
	\includegraphics[height=100pt]{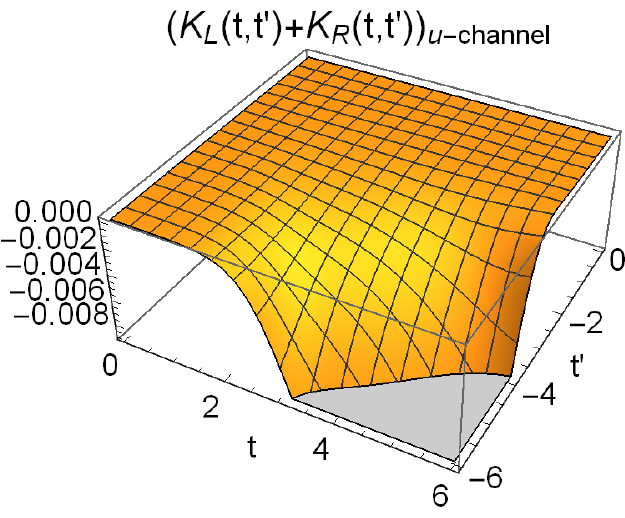}~~~~
	\includegraphics[height=100pt]{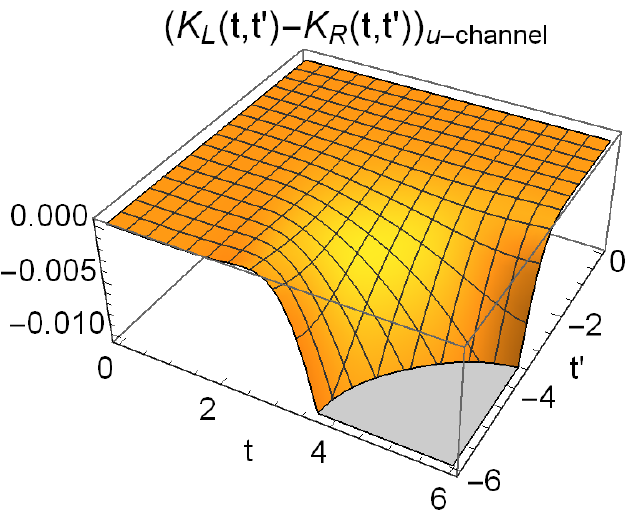}
	\caption{Kernels $(K_L(t,t')+K_R(t,t'))_{u-channel}$ and $(K_L(t,t')-K_R(t,t'))_{u-channel}$ contributed by the u-channel at early time, where $\varDelta=1/6,\,C=10^5,\, \beta=2\pi,\,\lambda=2$, and then $t_*/2=(\ln C)/2\approx 5.7$.}
	\label{FigKLKR_u}
\end{figure}

The second commutator (\ref{VWWV}) is an OTOC. It leads to the diagram in Fig.~\ref{Fig4Pt_u} at $O(C^{-1})$, which is called the u-channel. It plays an important role in quantum chaos and traversable wormholes \cite{Gao:2016bin,Maldacena:2017axo}. Its corresponding process in the bulk is that two particles scatter near the horizon by exchanging gravitons and then reach the opposite boundaries. In contrast, without the exchange of gravitons, the diagram does not contribute to Eq.~(\ref{4PtC}) since the wormhole is not traversable. The u-channel contributes to the commutator $C_{VWWV}$ with
\begin{align}\label{u-Channel}
&(C_{VWWV})_{u-channel}\nn\\
=&-i 2^{3} \left(4\cosh\frac{t_2+t_3}{2}\cosh\frac{t_1+t_4}{2}\right)^{-2\varDelta}        \frac{\varDelta ^2}{C}
\left(\frac{\sinh\frac{t_{14}+t_{23}}{2}}{\cosh\frac{t_2+t_3}{2}\cosh\frac{t_1+t_4}{2}}+\frac{t_{14}+t_{23}}{2}\tanh\frac{t_2+t_3}{2}\tanh\frac{t_1+t_4}{2}\right) 
\end{align}
Due to the suppression of $C^{-1}$, it is visible only within the chaos region, $t'\approx-t$ and $t_d/2\ll t<t_*/2$, where $t_*=\frac{\beta}{2\pi}\ln\frac{2\pi C}{\beta}$ is the scrambling time.
Specifically, let $\lambda=1$, $t_1=t_2=t>0$ and $t_3=t_4=t'=-t$, this becomes
\begin{align}
(C_{VWWV})_{u-channel}= -i 2^{3-4 \varDelta } \varDelta ^2 C^{-1}\sinh (2 t),
\end{align}
which will exponentially grow at early time $t_d/2\ll t<t_*/2$. At late time $t\gg t_*/2$, the approximation in Eq.~(\ref{u-Channel}) at $O(C^{-1})$ is unreliable, because of the strong backreaction on the metric, which will be investigated in the next section.
If we let $t_i=-t_f$ and $\lambda\approx1$ in Eq.~(\ref{PerturbationA}), the u-channel will contribute exponential growth on the energy changes at early time $t_d/2\ll t_f<t_*/2$, whose leading behaviors are
\begin{align}\label{Energyu-Channel}
(\Delta E_L+\Delta E_R)_{u-channel}=&-(\Delta E_I)_{u-channel}\sim -g^2C^{-1} e^{2 t_f}+ O((\lambda-1)^1), \\
(\Delta E_L-\Delta E_R)_{u-channel}    \sim & -(\lambda-1)g^2C^{-1} t_f e^{2 t_f}+ O((\lambda-1)^2),
\end{align}
both of which are negative and satisfy criterion (\ref{ReversedArrow}). This means that the u-channel with the interaction (\ref{HIVW-WV}) contributes to anomalous heat flow, at least at early time, when $\lambda\approx1$. Our numerical calculation shows that, for general $\lambda>1$, such anomalous heat flow also appears. We illustrate kernels contributed by the u-channel in Fig.~\ref{FigKLKR_u}.

\begin{figure}
	\centering
	\includegraphics[height=0.25\linewidth]{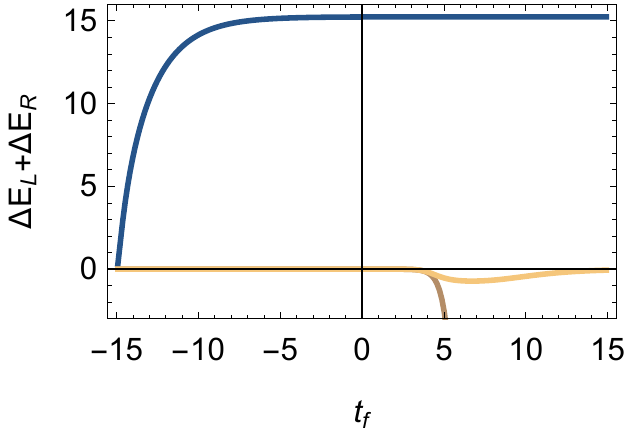}
	\includegraphics[height=0.25\linewidth]{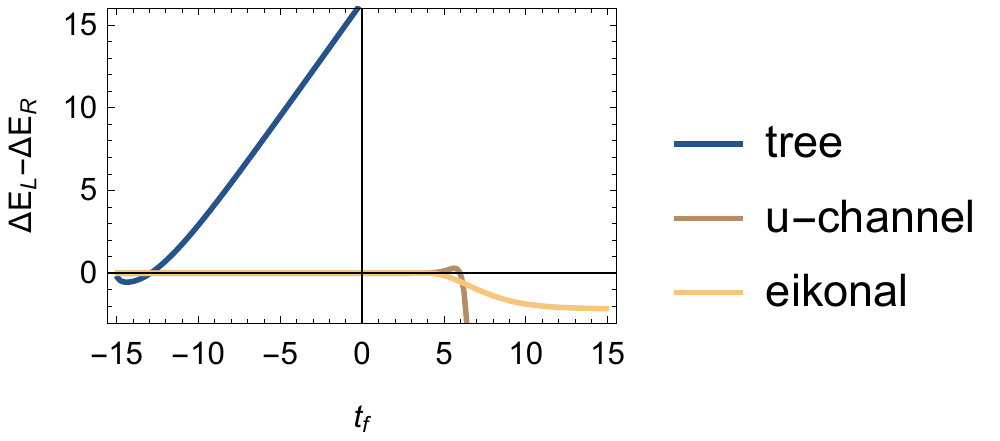}
	\caption{$\Delta E_L\pm\Delta E_R$ contributed by different diagrams in Table \ref{TabA} as functions of $t_f$, where the fifth line ``eikonal'' refers to the contribution of $C_{VWWV}$ in Eq.~(\ref{Modifiedu-Channel}). Parameters are chosen to be $\varDelta=1/6,\, C = 10^5,\, \beta= 2\pi,\,\lambda=2, \, \epsilon=0.2$, and $t_i=-15$. So $t_*\approx11.5$ and the early time is $t_f<5.8$. The contributions of the one-loop correction and the t-channel are too small to be shown.}
	\label{FigL2R2pm}
\end{figure}

However, in Eq.~(\ref{PerturbationA}), we should sum over all the diagrams. We estimate their scales, as concluded in Tab.~\ref{TabA}.
Specially, the energy change contributed by the u-channel is order $C^{-1}$ before the dissipation time $t_d$ and it reaches order $C^{-1}e^{t_*}\sim1$ near the scrambling time $t_*$. 
First, the wormhole supports two channels, the t-channel and the u-channel. Near the scrambling time, the contribution of the u-channel surpasses the contribution of the t-channel. So, with the interaction (\ref{HIVW-WV}), the wormhole supports an anomalous heat flow near the scrambling time. 
Second, since the contributions of the tree diagram and the u-channel are in the same order, to determine the whole thermodynamic arrow of time, we should compare the contribution of the two diagrams numerically, as shown in Fig.~\ref{FigL2R2pm}. We find that the contribution of the tree diagram is larger than the contribution of the u-channel at early time. So we conclude that the whole thermodynamic arrow of time has not been reversed at early time.

\begin{table}
	\centering
	\begin{tabular}{|c|c|c|c|}
		\hline 
		diagrams    &    (dis)connected & the thermodynamic arrow of time  & scale of energy change \\ 
		\hline 
		tree    &    disconnected  & obey & $1$  \\ 
		\hline 
		one-loop    &    disconnected  & obey & $C^{-1}$  \\ 
		\hline 
		t-channel    &    connected  & obey & $C^{-1}$  \\ 
		\hline 
		u-channel    &    connected  & reverse at early time & from $C^{-1}$ to $C^{-1}e^{t_*}\sim1$  \\ 
		\hline 
	\end{tabular} 
	\caption{Comparing the diagrams in Fig.~\ref{Fig4PtA}.}
	\label{TabA}
\end{table}

\subsection{late time}

\begin{figure}
	\centering
	\includegraphics[height=100pt]{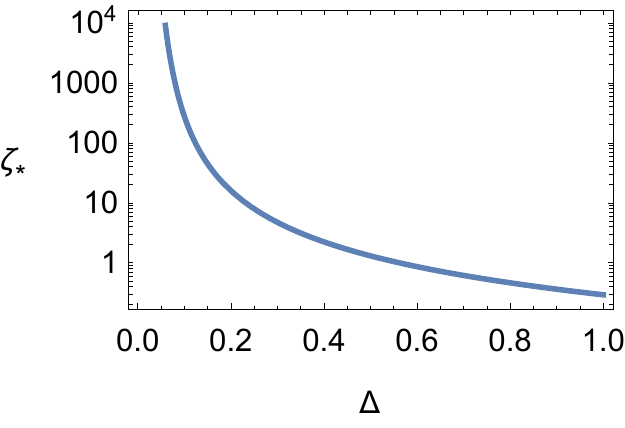}
	\caption{$\zeta_*$ as a function of $\varDelta$.}
	\label{FigMaxPoint}
\end{figure}

At late time, one should consider the higher-order $C^{-1}$ correction on $C_{VWWV}$ beyond the u-channel. By taking the limit $t\to\infty,\,C\to\infty$ while keeping $C^{-1}e^{t}$ finite, one can calculate $C_{VWWV}$ in the eikonal approximation \cite{Maldacena:2016upp,Shenker:2014cwa}
\begin{align}\label{Modifiedu-Channel}
C_{VWWV}
=-2i\avg{V_L(t_1)V_R(t_4)}\avg{W_R(t_2)W_L(t_3)} \Im \mathcal F    \nn    \\
\mathcal F=\kc{\frac i\zeta}^{2 \varDelta } U\left(2 \varDelta ,1,\frac i\zeta\right),\quad
\zeta=\frac{1}{8C}\frac{e^{\left(t_1+t_2-t_3-t_4\right)/2}}{\cosh\frac{t_2+t_3}{2} \cosh\frac{t_1+t_4}{2}},\\
t_1\approx t_2/\lambda\approx t,\quad t_3\approx t_4/\lambda\approx t',\quad t'<0<t    \nn
\end{align}
where the two-point functions are given in Eq.~(\ref{2Pt}) and the confluent hypergeometric function $U(a,1,x)=\Gamma(a)^{-1}\int_0^\infty ds e^{-sx}\frac{s^{a-1}}{(1+s)^a} $ is used. In real time, $\zeta>0$ always holds; then $\Im\mathcal F>0$ always holds. Equation~(\ref{Modifiedu-Channel}) is negligible before the early time. It agrees with the u-channel in Eq.~(\ref{u-Channel}) well at early time, while it decays to zero at late time. So its amplitude is maximized at the time scale of the scrambling time. More precisely, $\Im\mathcal F$ is maximal at $\zeta=\zeta_*$, whose dependence on $\varDelta$ is shown in Fig.~\ref{FigMaxPoint}. It corresponds to the time scale on $\kc{t,t'}$ as
\begin{align}\label{Max}
\frac{e^{(\lambda +1)t}}{ 2 C \left(e^{t'+\lambda  t}+1\right) \left(e^{\lambda  t'+t}+1\right)}\approx \zeta_*.
\end{align}  
The time dependence of the two-point functions in Eq.~(\ref{2Pt}) also affects the time of the maximal amplitude of $C_{VWWV}$. Basically, when $\lambda\approx1$, it is maximized at the time scale (\ref{Max}); when $\lambda\gg1$, it is maximized at the time scale which is smaller than (\ref{Max}).

Besides, there are two lines in the plane of $\kc{t,t'}$ where the two-point functions in Eq.~(\ref{Modifiedu-Channel}) are maximized respectively
\begin{align}
\lambda t+t'=0,    \label{T1}  \\
t+\lambda t'=0.        \label{T2}
\end{align}
The intersections of Eqs.~(\ref{Max}) and (\ref{T1}) and Eqs.~(\ref{Max}) and (\ref{T2}) are approximately
\begin{align}
\kc{t,t'}=\kc{t_a,-\lambda t_a},\quad t_a\approx \frac1{1+\lambda}\frac{t_*}{2},\\
\kc{t,t'}=\kc{t_b,-t_b/\lambda},\quad t_b\approx \frac{\lambda}{1+\lambda}\frac{t_*}{2}.
\end{align}
The two time scales $t_a$ and $t_b$ can characterize the process of energy change.

\begin{figure}
	\newcommand{\figheight}{105pt}
	\centering
	\subfigure[]{
		\includegraphics[height=\figheight]{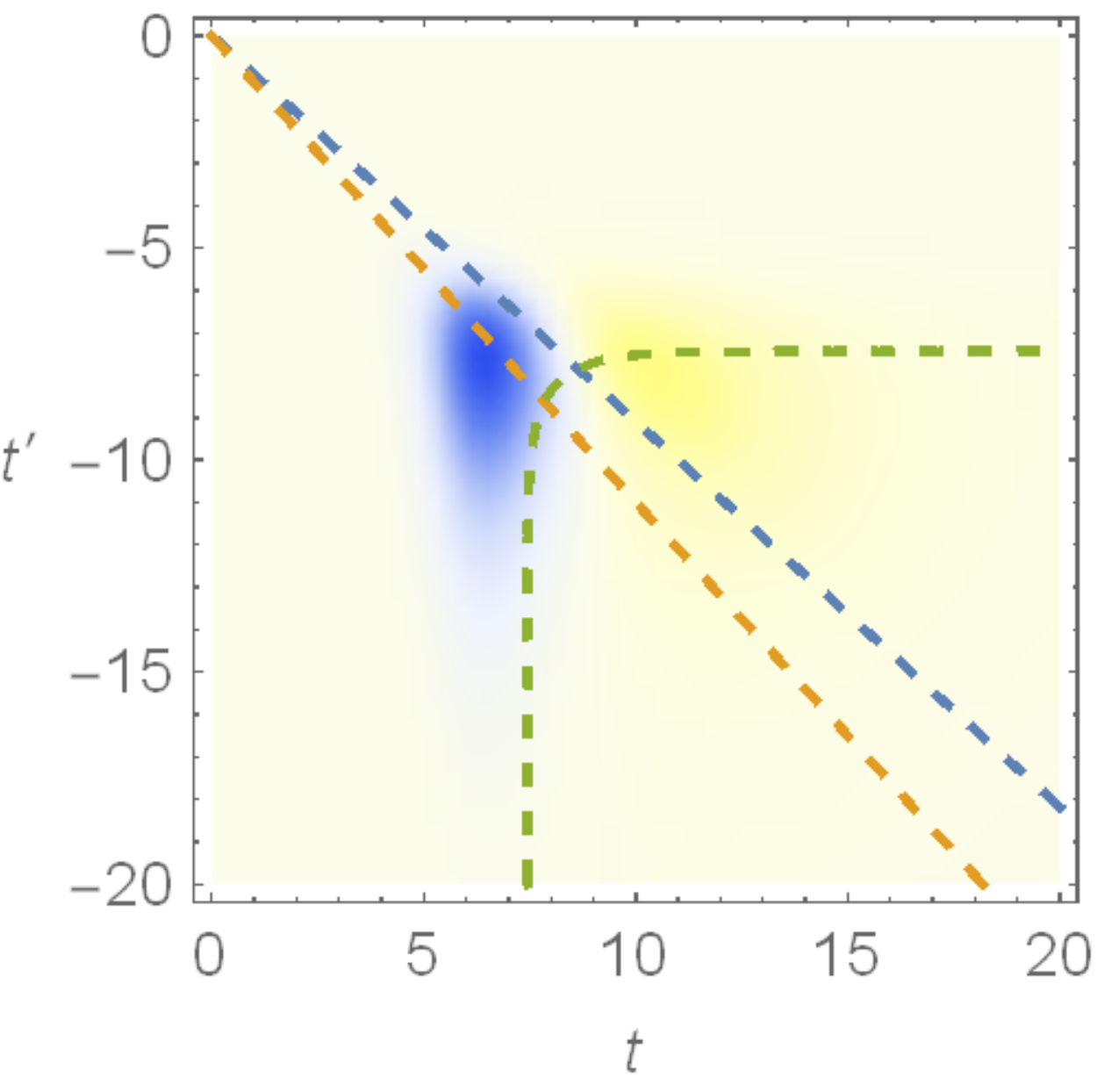}
		\includegraphics[height=\figheight]{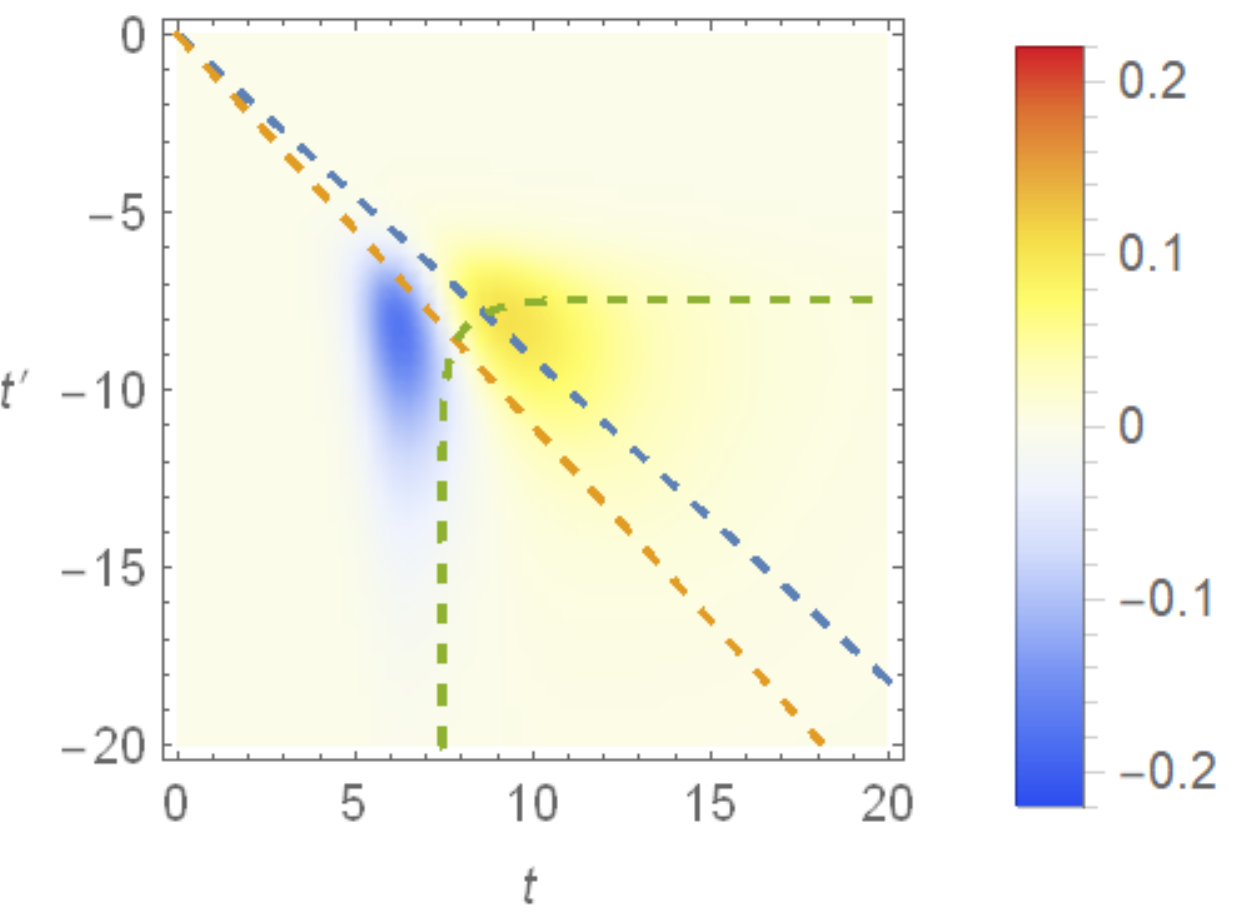}
		\includegraphics[height=\figheight]{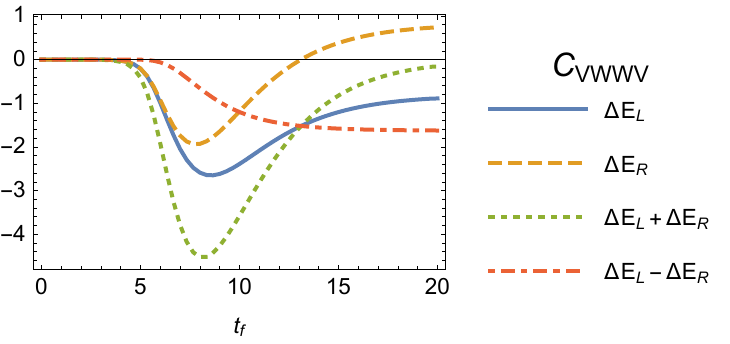}
		\label{FigKL2R2_1p1}
	}    
	\subfigure[]{
		\centering
		\includegraphics[height=\figheight]{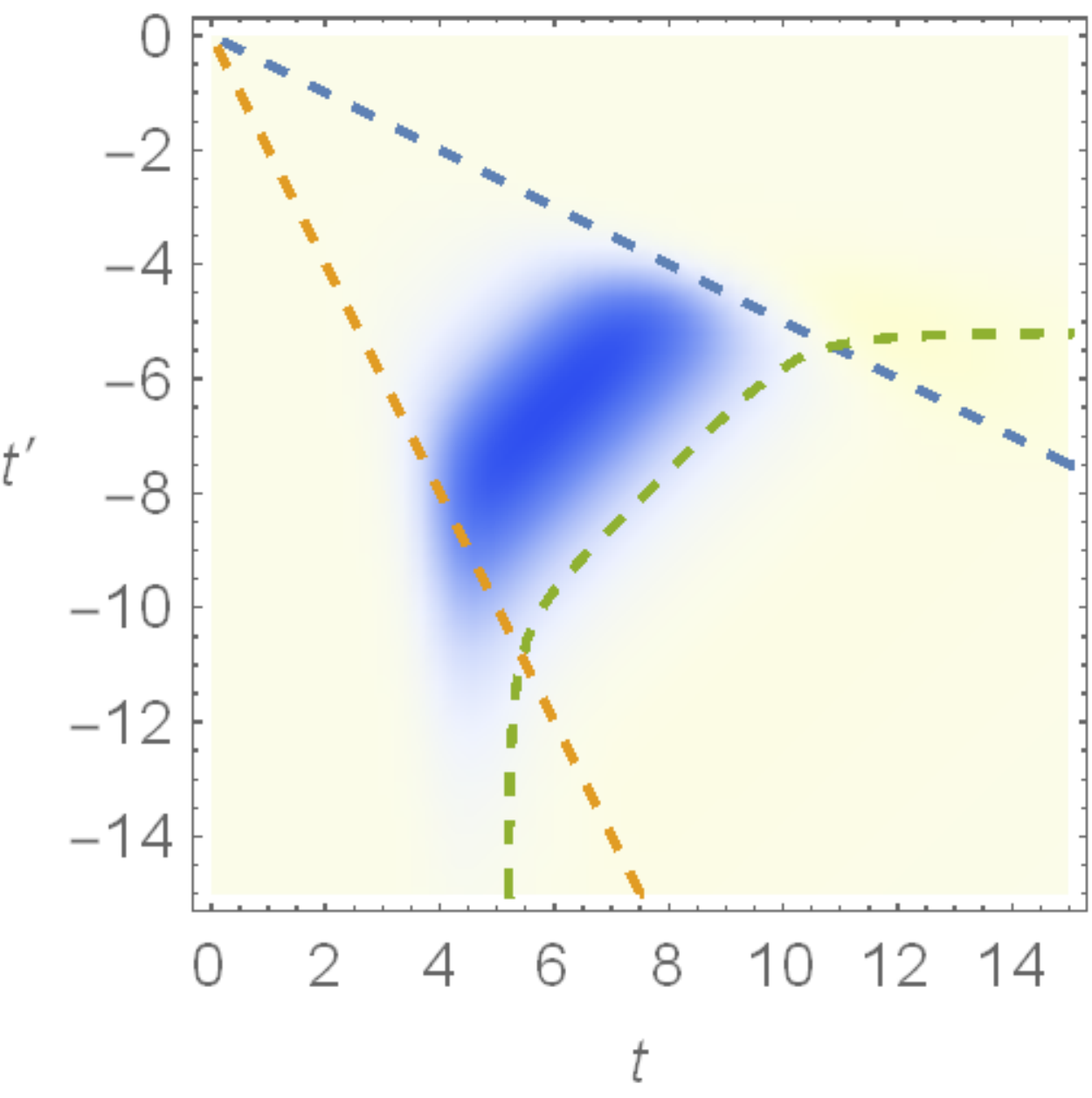}
		\includegraphics[height=\figheight]{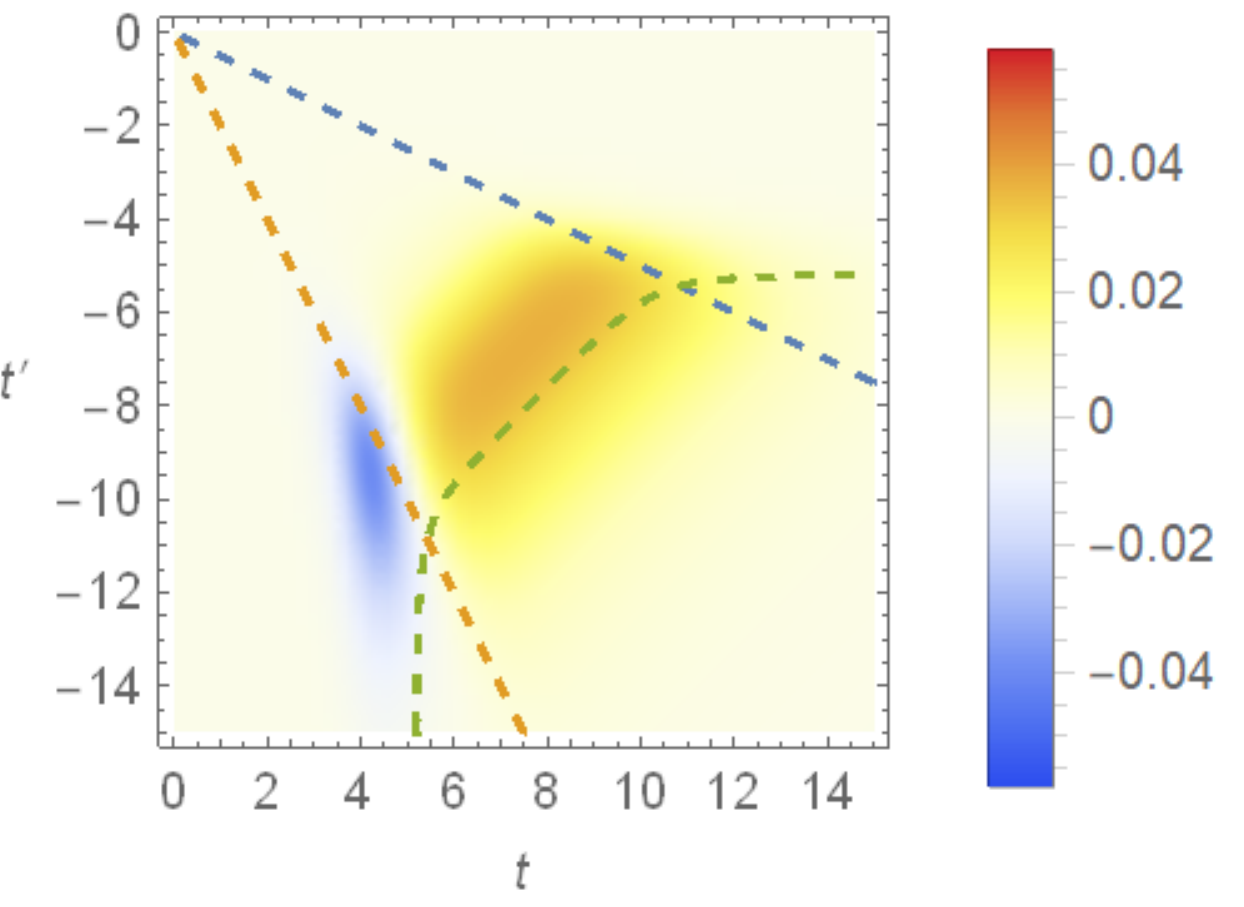}
		\includegraphics[height=\figheight]{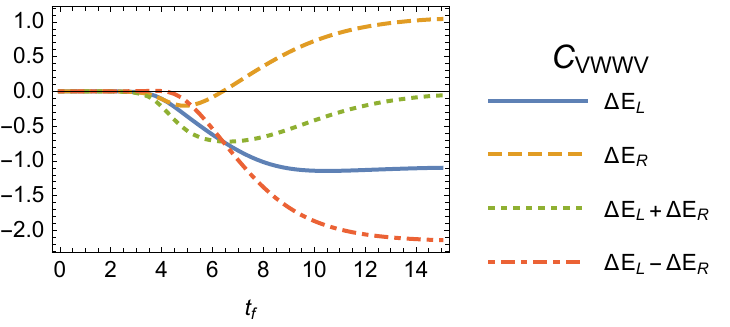}
		\label{FigKL2R2_2}    
	}    
	\subfigure[]{
		\centering
		\includegraphics[height=\figheight]{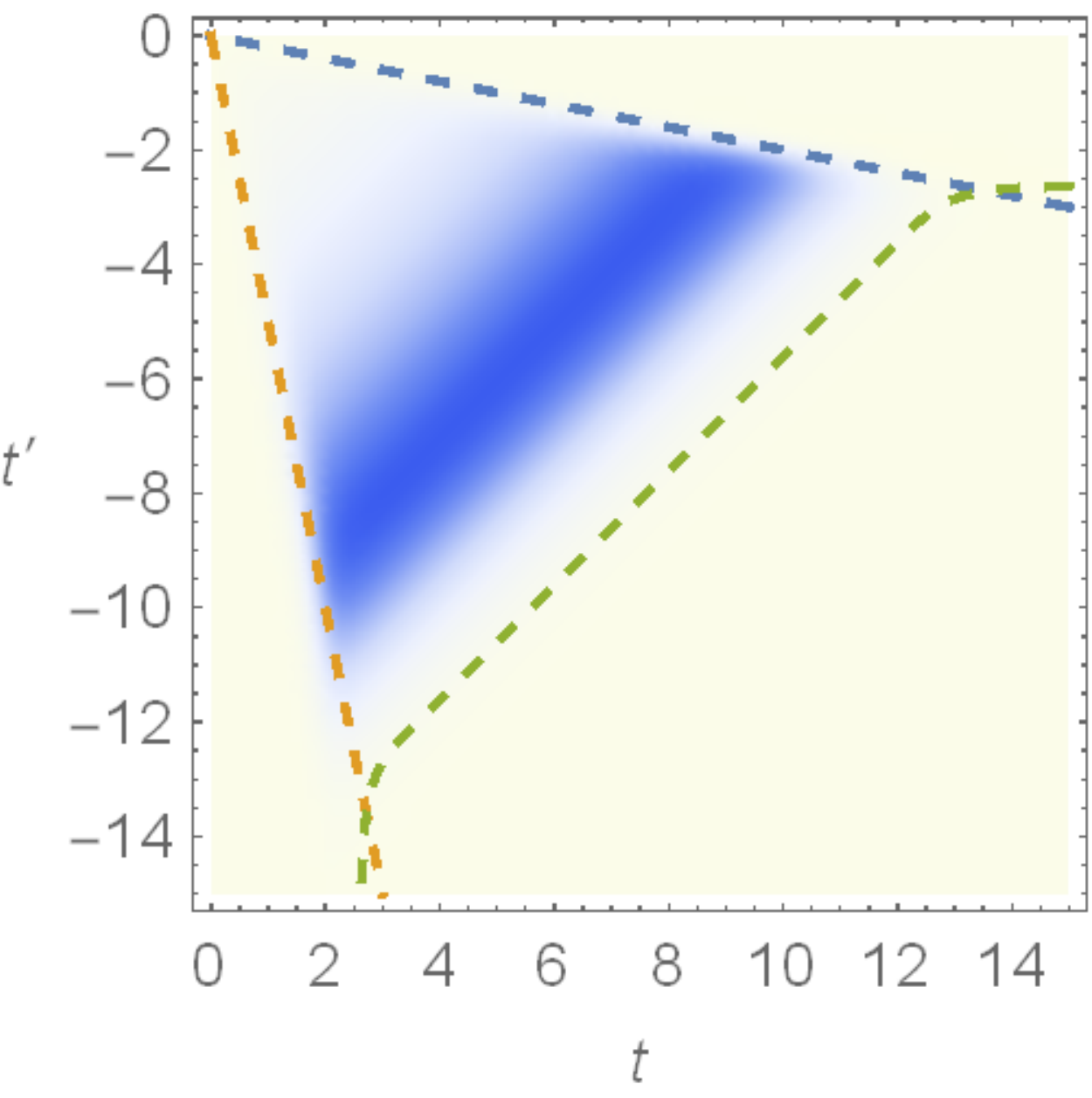}
		\includegraphics[height=\figheight]{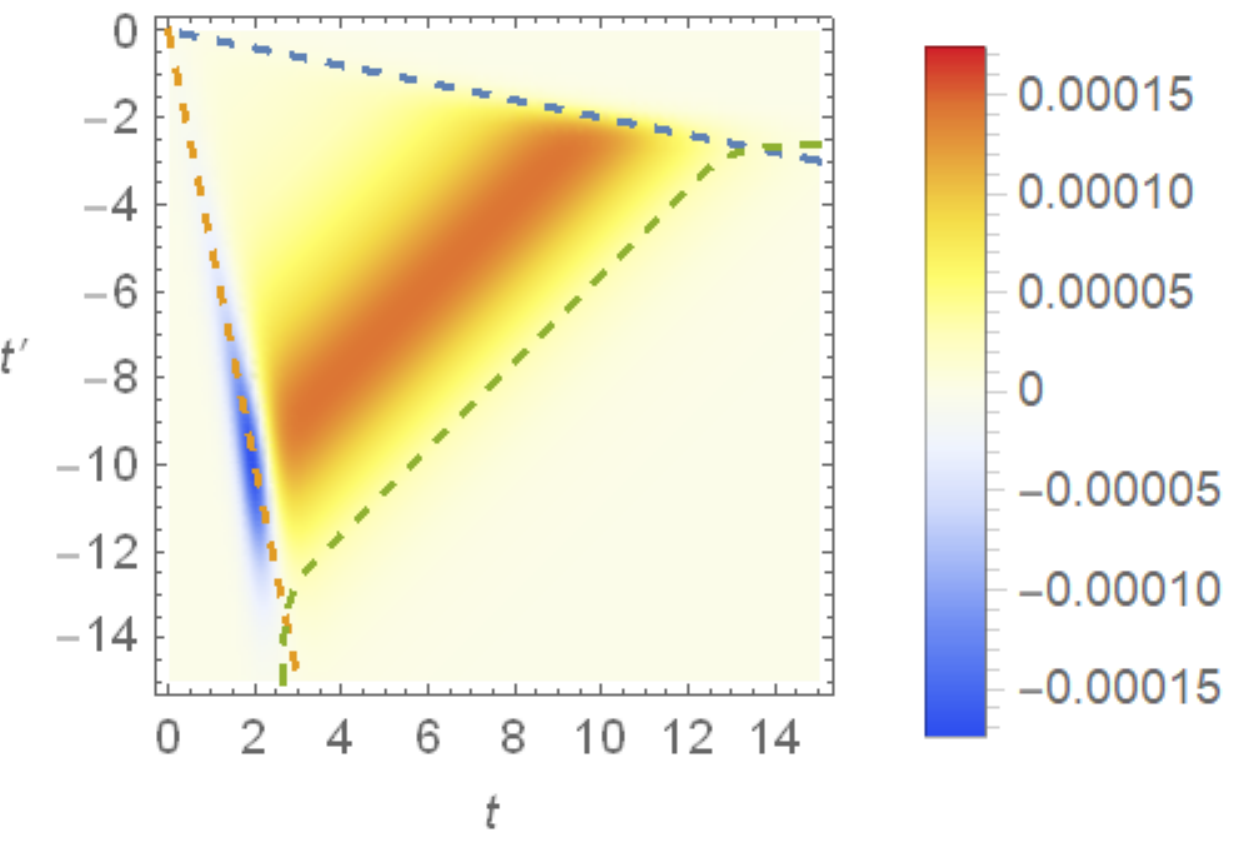}
		\includegraphics[height=\figheight]{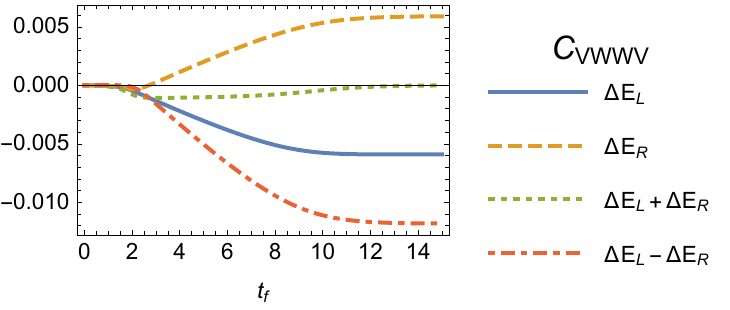}
		\label{FigKL2R2_5}
	}
	\caption{Left and middle: The integral kernels (left) $(K_L(t,t'))_{VWWV}$ and (middle) $(K_R(t,t'))_{VWWV}$ contributed by $C_{VWWV}$ as functions of $\ke{t,t'}$ under the interaction (\ref{HIVW-WV}). The dashed curves denote (blue) $t+\lambda t'=0$, (orange) $\lambda t+t'=0$, and (green) $\zeta=\zeta_*$ in Eq.~(\ref{Max}). Right: The energy changes contributed by $C_{VWWV}$ as functions of $t_f$. Parameters are $\varDelta=1/6,\,C=10^5,\,\beta=2\pi$, $g=1$, and (a) $\lambda=1.1$, $t_i=-20$, (b) $\lambda=2$, $t_i=-15$, or (c) $\lambda=5$, $t_i=-15$.}
	\label{FigKL2R2}
\end{figure}

In Fig.~\ref{FigKL2R2}, we show the energy changes $(\Delta E_\gamma)_{VWWV}$ and their kernels $(K_\gamma(t,t'))_{VWWV}$ contributed by $C_{VWWV}$ in Eq.~(\ref{Modifiedu-Channel}) including the late time. We can consider the process with initially time $t_i\ll-t_*$ and observe the energy changes contributed by $C_{VWWV}$ when $t_f$ goes from $t_i$ to $+\infty$. 
We find several periods. All the energy changes in the following list refer to the contribution by $C_{VWWV}$ only.
\begin{enumerate}
	\item When $t_f\ll t_a$, these energy changes are negligible, since the system hardly scrambles.
	\item When $t_f\sim t_a$, the amplitude of commutator $C_{VWWV}$ exponentially grows. Both $(\Delta E_L)_{VWWV}$ and $(\Delta E_R)_{VWWV}$ synchronously decrease and $(\Delta E_I)_{VWWV}$ increases. During this time, $|(\Delta E_L-\Delta E_R)_{VWWV}|$ is still small.
	\item When $t_a\ll t_f\ll t_b$, $(\Delta E_L)_{VWWV}$ keeps dropping, while $(\Delta E_R)_{VWWV}$ in turn increases. Then $(\Delta E_L-\Delta E_R)_{VWWV}$ starts to drop and system $R$ keeps absorbing the energy from system $L$, which is quite apparent in Figs.~\ref{FigKL2R2_2} and \ref{FigKL2R2_5}. $(\Delta E_I)_{VWWV}$ reaches its maximum in this period.
	\item When $t_f\sim t_b$, $(\Delta E_L)_{VWWV}$ in turn increases and $(\Delta E_R)_{VWWV}$ increases, whose speeds are low when $\lambda\gg1$. $E_I$ decreases.
	\item When $t_b\ll t_f$, all the energies go to some constants, because of the quasinormal decay of commutator $C_{VWWV}$ at late time \cite{Maldacena:2016upp}. Especially, $(\Delta E_I)_{VWWV}$ is forced to vanish, while the net energy transfer $(\Delta E_L-\Delta E_R)_{VWWV}$ reaches a negative constant.
\end{enumerate}

During the whole process contributed by $C_{VWWV}$, energy transfers from colder system $L$ to hotter system $R$, which exhibits the anomalous heat flow via the wormhole. Note that the final magnitude of the anomalous heat flow decreases when $\lambda$ increases from $1$, which agrees with the analysis of Renyi entropy in the next section.

Similar to the u-channel in the previous subsection, the amplitude of $C_{VWWV}$ is order $1$. Numerically, it is usually smaller than the contribution of the disconnected-tree diagram, as shown in Fig.~\ref{FigL2R2pm}. In other words, the thermal diffusion is stronger than the anomalous heat flow. So the total thermodynamic arrow of time cannot be reversed even at late time.

According to Appendix \ref{SectionFirstLaw}, we find that the work done on the local systems vanishes as well because one point functions $\avg{V}$ and three points functions $\avg{VVW}$ vanish. So the energies change in the way of heat. This agrees with our terminologies, ``thermal diffusion'' and  ``anomalous heat flow''.

Finally, in Appendix \ref{SectionSYK}, we show that such anomalous heat flow via wormhole can be constructed in the SYK model as well since its breaking of conformal symmetry at the large-$N$ limit is also described by Schwarzian action at low energy.

\subsection{An interpretation from boundary theories}

According to the AdS/CFT correspondence, the dual system on the boundary is strongly coupled and highly chaotic. It goes beyond the ordinary realization of anomalous heat flow in weakly coupled or weakly chaotic models \cite{Micadei:2017,Partovi:2008,Jennings:2010} and is relatively difficult to be understood from boundary theories. Here we try to explain. Recall that energy changes highly rely on the form of interaction, which must be taken into consideration. For simplicity, we assume $g>0$.

We first consider the interaction $H_I=gV_LW_R$, so only the disconnected diagrams are presented. We will explain the behaviors of kernels $(K_\gamma(t_2,t_1))_{discon.}$ for $\gamma=L,R,I$, as shown in Fig.~\ref{FigE_tree}, which corresponds to the energy change after the instantaneous deformation $\Delta H(t)=H_I(\delta(t-t_1)+\delta(t-t_2))$. At time $t_1$, we turn on a potential $gV_LW_R$ and then intermediately turn it off. This creates an EPR pair, which is a superposition of all the combinations tending to lower the value of the potential $g\avg{V_LW_R}$. However, it will not change $E_L+E_R$, since $\avg{V_LW_R}$ initially vanishes on quantum average. In other words, the first-order perturbation vanishes. The EPR pair will interact with other particles on their own sides and then dissipate when time goes forward, characterized by a dissipation time $t_d$ which scales as the inverse temperature $\beta_\gamma$ on each side. At time $t_2>t_1$, we instantaneously turn on the potential $gV_LW_R$ again. The accumulation of the tendency of the EPR pair has lowered $g\avg{V_LW_R}$ and leads to $(K_I(t_2,t_1))_{discon.}<0$. However, when $t_2-t_1\gg t_d$, the dying EPR pair is unable to lower $g\avg{V_LW_R}$, so that $g\avg{V_LW_R}$ as well as $|(K_I(t_2,t_1))_{discon.}|$ returns to zero. Meanwhile $(K_\gamma(t_2,t_1))_{discon.}$ is proportional to the power of $gV_LW_R$ done on system $\gamma$ at time $t_2$. When $t_2$ goes from $t_1$ to infinity, the value $g\avg{V_LW_R}(t_2)$ decreases first and then returns to zero, such that its power done on system $\gamma$ is positive first and is negative later, which agrees with the tendency of $(K_\gamma(t_2,t_1))_{discon.}$. The hotter the system is, the faster the EPR pair decays. So $(K_R(t_2,t_1))_{discon.}$ becomes negative earlier than $(K_L(t_2,t_1))_{discon.}$. For the sustained interaction $H_I=gV_LW_R$, since $(\Delta E_I)_{discon.}$ does not change after $t_d$, the difference between the temperatures leads to a net energy transfer from system $R$ to system $L$.

Now we consider the interaction $H_I=g(V_LW_R-W_RV_L)/\sqrt2$. The analysis of the contribution of the disconnected diagrams is the same as before. We will focus on kernels $(K_\gamma(t_2,t_1))_{VWWV}$ for $\gamma=L,R,I$, as shown in Fig.~\ref{FigKL2R2}, whose main contributions are near $t_2\sim -t_1=t$. We first analyze $(K_I(t,-t))_{VWWV}$ when $\lambda=1$. We find
\begin{align}
C_{VWWV}(t,t,-t,-t)=&\Tr\kd{y\ke{W(t),V(-t)}y\kd{W(t),V(-t)}},\\
(K_I(t,-t))_{VWWV}=&\frac i\hbar g^2 C_{VWWV}(t,t,-t,-t)
\xrightarrow{\hbar\to0} -\frac{g^2}2\avg{\avg{\ke{W(t)^2,V(-t)^2}_P}}, \label{ClassicalLimit}
\end{align}
where we have restored $\hbar$ and taken the classical limit $\hbar\to0$. So $C_{VWWV}(t,t,-t,-t)$ is just the cross-correlation between fluctuation and dissipation \cite{Tsuji:2016kep}. $\ke{...,...}_P$ is the Poisson bracket. $\avg{\avg{...}}$ is the classical ensemble average. From the classical limit, we know that $(K_I(t,-t))_{VWWV}$ characterizes the dissipation of fluctuation, which can be understood from linear response. 

Consider a classical and highly chaotic system containing $N$ canonical coordinates $\ke{x_i}(i=1,...,N)$ with zero average value $\avg{\avg{x_i}}=0$. The classical model in Ref.~\cite{Maldacena:2017axo} is a good example, as discussed in Appendix \ref{SectionClassical}. At time $t_1$, we add an instantaneous potential $\Delta H(t) = gx_1^2\delta(t-t_1) $, which gives the system a momentum along the direction of reducing the fluctuation $\avg{\avg{x_1^2}}$. At $t_2>t_1$, we measure the fluctuation $\avg{\avg{x_2^2}}$. $g\avg{\avg{\ke{x_2(t_2)^2,x_1(t_1)^2}_P}}$ characterizes the change of the fluctuation $\avg{\avg{x_2^2}}$ at time $t_2$ due to the instantaneous potential at time $t_1$, which is the classical interpretation of Eq.~(\ref{ClassicalLimit}) if we identify $x_1=V, x_2=W$.
Since the system is highly chaotic, the fluctuation of $x_i$ is sourced from the noises given by the other $(N-1)$ canonical coordinates. When the fluctuation of $x_1$ is suppressed by the instantaneous potential at time $t_1$, we expect that the fluctuation of other $x_{i\neq1}$ will be suppressed later due to their interactions with $x_1$. If we only focus on one of them, such as $x_2$, the response is split into $N$ parts. So $g\avg{\avg{\ke{x_2(t_2)^2,x_1(t_1)^2}_P}}$ is negative and is suppressed by an $N^{-1}$ factor. Furthermore, since the system is chaotic, such respond should exponentially grow along time $t_2-t_1$. Finally, when $t_2$ goes to infinity, the system should reach a new equilibration and lose its memory. The final value of fluctuation $\avg{\avg{x_2^2}}$ only depends on the final equilibrium state which is determined by the energy of the system after the perturbation $\Delta H(t)$, while the energy at $t_2$ does not change at linear order, {\it i.e.} $\avg{\avg{\ke{H,gx_1(t)^2]}_P}}\propto \partial_t\avg{\avg{x_1(t)^2}}=0$. So linear response $g\avg{\avg{\ke{x_2(t_2)^2,x_1(t_1)^2}_P}}$ should decay to zero with a time scale near the inverse temperature. We summarize the behavior from the classical approximation
\begin{align}\label{ClassicalApproxKI}
(K_I(t,-t))_{VWWV}\approx
\left\{\begin{array}{ll}
g^2N^{-1} e^{2\lambda_L t}, & \text{early time}\\
g^2\alpha_1 e^{-c_1t/\beta}, & \text{later time}
\end{array}\right.,
\end{align}
where coefficients $\alpha_1,c_1>0$ and Lyapunov exponent $\lambda_L\lesssim 2\pi/\beta$ for highly chaotic systems. The behavior of $(K_{L,R}(t,-t))_{VWWV}$ can be approximated by the time derivative of $-(K_I(t,-t))_{VWWV}$. So $(K_{L,R}(t,-t))_{VWWV}$ is negative at early time, becomes positive at later time, and vanishes finally. The classical approximation here agrees with the results in JT gravity (\ref{Energyu-Channel}).

Now we consider the case of $\lambda\gtrapprox1$. The general behaviors of $(K_\gamma(t,-t))_{VWWV}$ for $\gamma=L,R,I$ do not change much. The main difference between $(K_L(t,-t))_{VWWV}$ and $(K_R(t,-t))_{VWWV}$ is in their timescales. The dissipation time and the scrambling time of system $R$ are brought forward due to the $\lambda$ in the temporal argument of Eq.~(\ref{PerturbationA}). So $(K_R(t,-t))_{VWWV}$ becomes positive earlier than $(K_L(t,-t))_{VWWV}$, as shown in Fig.~\ref{FigKL2R2_1p1}. These shorter time scales advance the dissipation of fluctuation in system $R$. Recall that $(\Delta E_I)_{VWWV}$ must return to zero finally. So the integral effect is that system $R$ obtains energy while system $L$ loses energy.

\section{Entropy change}\label{SectionEntropy}

\subsection{Replica trick}

We will calculate the entropy change at time $t_f$ after we turn on an interaction $H_I$ at time $t_i$.
We will work on imaginary time first, then Wick-rotate $\tau\to it$ and obtain the entropy change as a function of real time. For the simplicity of notation, we will use imaginary time evolution as follows:
\begin{align}
O(\tau)=e^{\tau H}Oe^{-\tau H}.
\end{align}

Since the bi-systems are in a pure state, the entropies of both sides are the same. So we will only calculate the entropy of system $R$. Consider the density matrix of system $R$ at time $t_f$,
\begin{align}\label{Rho}
\tilde\rho_R(t_f)=\Tr_L\ket{\tilde\beta_I(t_f)}\bra{\tilde\beta_I(t_f)}.
\end{align}
where $\ket{\tilde\beta_I(t_f)}$ is given in Eq.~(\ref{TFDInteraction}). 
The Renyi entropy and von Neumann entropy of system $R$ can be written as
\begin{align}
S_n=\frac{\ln\Tr[\tilde\rho_R(t_f)^n]}{1-n}
=\frac{\ln Z_n-n\ln Z}{1-n},\quad S=-\Tr[\tilde\rho_R(t_f)\ln\tilde\rho_R(t_f)]
=-\partial_n\ln Z_n+\ln Z|_{n\to1},
\end{align}
where $Z_n$ is the $n$-replicated partition function and the twisted boundary conditions are applied on system $R$ at time $t_f$. 
We define a twist operator $X_n$, which cyclically permutes the replica index. We can also write the Renyi entropy of system $R$ as
\begin{subequations}\begin{align}
	e^{(1-n)S_n}&=Z_n/Z^n=\Tr[\tilde\rho_R(t_f)^{\otimes n}X_n]\\
	&=\bra{\tilde\beta_I(t_f)}^{\otimes n} (\mathbb{I}\otimes X_n) \ket{\tilde\beta_I(t_f)}^{\otimes n}\\
	&=\bra\beta^{\otimes n} 
	\ke{\kd{\mathcal T \exp\kc{-i\int_{t_i}^{t_f}dt\,\tilde H_I(it)}}^\dag}^{\otimes n} 
	(\mathbb{I}\otimes X_n) 
	\ke{\mathcal T \exp\kc{-i\int_{t_i}^{t_f}dt\,\tilde H_I(it)}}^{\otimes n} \ket\beta^{\otimes n}    \\
	&=\bra\beta^{\otimes n} 
	\ke{\mathcal T_{\mathcal C} \exp\kc{-\int_{\mathcal C^+}d\tau\,\tilde H_I(\tau)}}^{\otimes n} 
	(\mathbb{I}\otimes X_n) 
	\ke{\mathcal T_{\mathcal C} \exp\kc{-\int_{\mathcal C^-}d\tau\,\tilde H_I(\tau)}}^{\otimes n} \ket\beta^{\otimes n}    \label{TFDcontour}
	\end{align}\end{subequations}
where $\mathcal T_{\mathcal C}$ is the contour ordering. Contour $\mathcal C^-$ goes from $it_i+0^-$ to $it_f$ and contour $\mathcal C^+$ goes from $it_f$ to $it_i+0^+$, namely 
\begin{subequations}\label{Contour}\begin{align}
	\mathcal C^-&=\ke{z\in \mathbb Z|z=(it_i+0^-)(1-\kappa)+ it_f\kappa,\, \kappa\in[0,1]},\\ 
	\mathcal C^+&=\ke{z\in \mathbb Z|z=it_f(1-\kappa)+ (it_i+0^+)\kappa,\, \kappa\in[0,1]},
	\end{align}\end{subequations}
whose directions in real time are opposite. In $\mathbb{I}\otimes X_n$, $\mathbb I$ acts on the $n$ copies of system $L$, denoted as systems $L^{\otimes n}$, and $X_n$ acts on systems $R^{\otimes n}$. 

\begin{figure}
	\centering
	\subfigure[]{
		\includegraphics[height=0.25\linewidth]{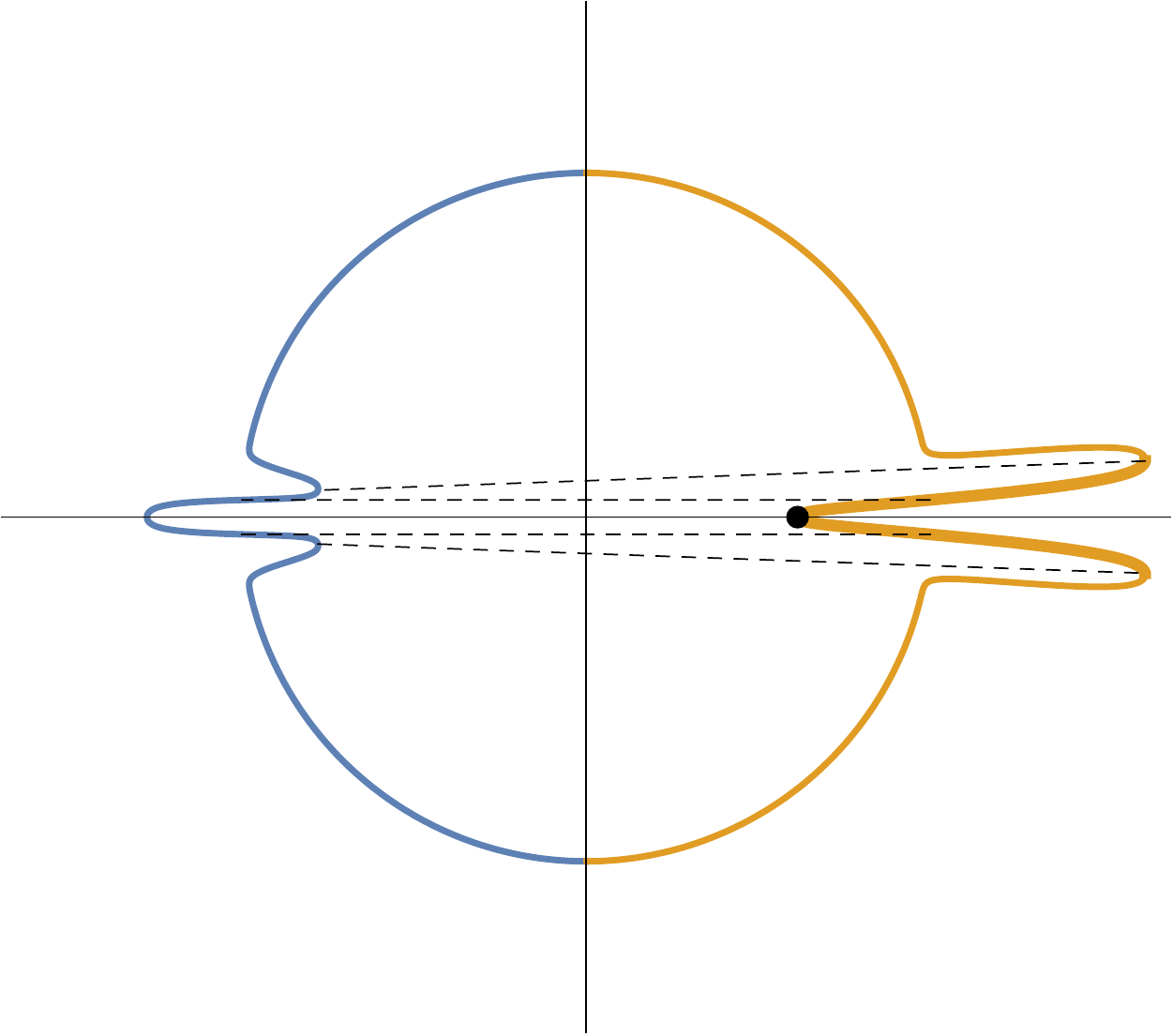}
		\label{FigReplicaA}    }
	\subfigure[]{
		\includegraphics[height=0.25\linewidth]{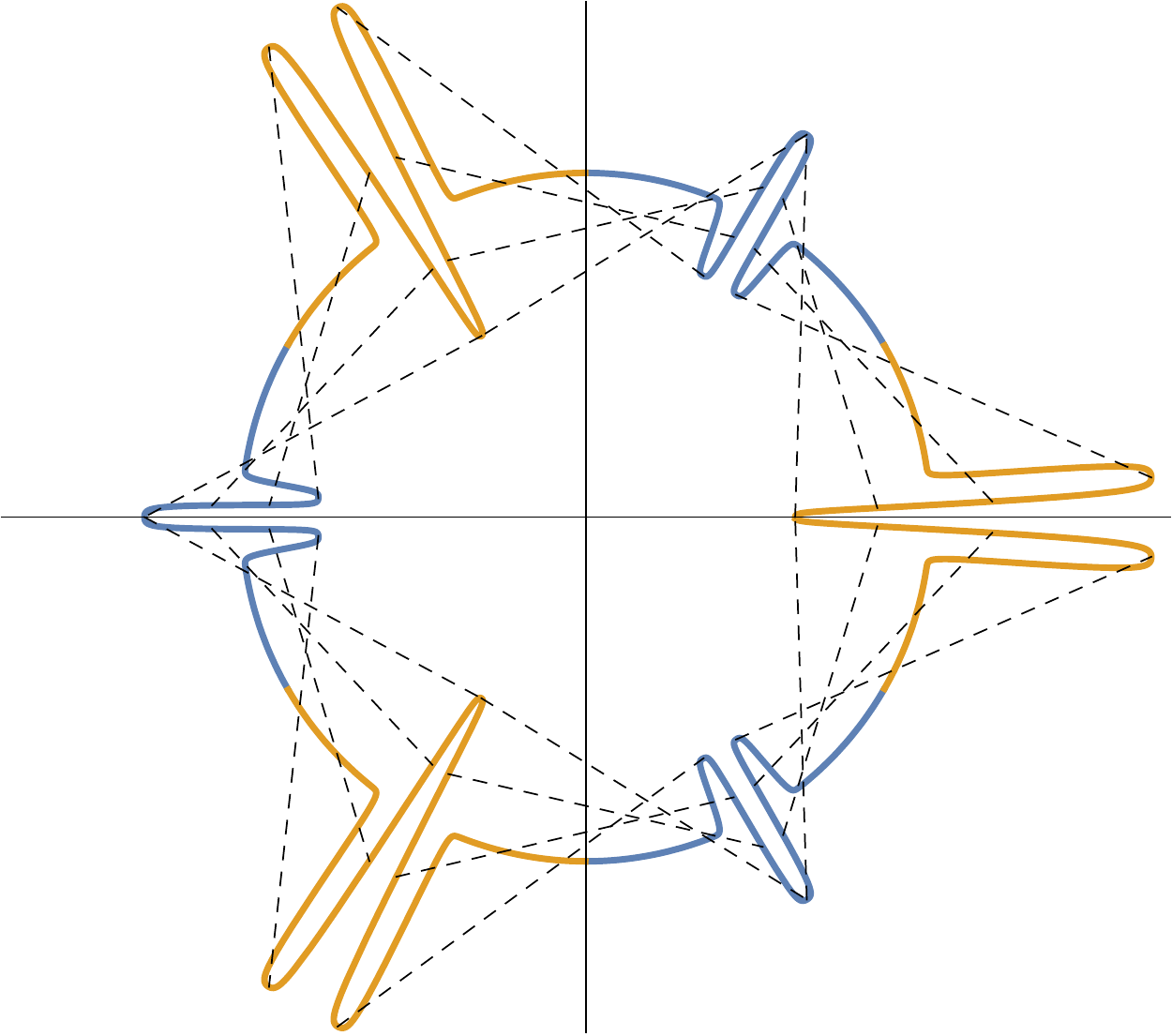}
		\label{FigReplicaB} }
	\caption{(a) The time contour $\mathcal C$ in the complex plane $z=\exp(i2\pi t_C/\beta)$, where the complex time $t_C=\tau+it$ is measured by evolution $\exp(-t_C H)$. The blue (orange) interval indicates the evolution of system $L$ ($R$). The black point indicates the twisted operator $X_n$. The dashed lines indicate the coupling between the two sides during $[t_i,t_f]$. The thick interval in the upper (lower) half plane indicates contour $\mathcal C^+$ ($\mathcal C^-$). We have chosen $-t_i=t_f$ and $\lambda=2$ in this figure. So the real-time contour of system $R$ is longer than the one of system $L$.
		(b) We unfold the time contour $\mathcal C_n$ for $n=3$ replicas in the complex plane $z^{1/n}=\exp(i2\pi t_C/n\beta)$.}
	\label{FigReplica}
\end{figure}

Equation~(\ref{TFDcontour}) can help us to write down the path integral on closed time contours. Recall that an operator of system $R$ ($L$) acting on the TFD state can be transformed into a corresponding operator of system $L$ ($R$) acting on the TFD state. Those operators on the left-hand side of $\mathbb{I}\otimes X_n$ can go through the channels of systems $L^{\otimes n}$, reach the right-hand side of $X_n$, and finally become the operators of systems $R^{\otimes n}$. Taking $H_I= V_L W_R$ as an example, we have
\begin{subequations}\label{PathIntegralVW}\begin{align}
	&\bra\beta^{\otimes n} 
	\ke{\mathcal T_{\mathcal C}
		\exp\kd{-\sum_{a=0}^{n-1}\int_{\mathcal C^+}d\tau V_{L,a}(\tau)W_{R,a}(\lambda\tau)}} 
	(\mathbb{I}\otimes X_n) 
	\ke{\mathcal T_{\mathcal C} 
		\exp\kd{-\sum_{a=0}^{n-1}\int_{\mathcal C^-}d\tau V_{L,a}(\tau)W_{R,a}(\lambda\tau)}}
	\ket\beta^{\otimes n}    \\
	=&\Tr\kd{\rho^{\otimes n} X_n\mathcal T_{\mathcal C}\exp\ke{
			-\sum_{a=0}^{n-1}\kd{\int_{\mathcal C^-}d\tau W_a(\lambda\tau)V_a(-\frac\beta2-\tau)
				+\int_{\mathcal C^+}d\tau V_a(-\frac\beta2-\tau)W_a(-\beta+\lambda\tau)}}}\\
	=&\avg{\mathcal T_{\mathcal C}\exp\ke{
			-\sum_{a=0}^{n-1}\kd{\int_{\mathcal C^-}d\tau W_a(\beta+\lambda\tau)V_a(\frac\beta2-\tau)
				+\int_{\mathcal C^+}d\tau V_a(\frac\beta2-\tau)W_a(\lambda\tau)}}X_n }_{\otimes n}, \label{PathIntegralVWfold}    \\
	=&\avg{\mathcal T_{\mathcal C}\exp\ke{
			-\sum_{a=0}^{n-1}\int_{\mathcal C}d\tau_1d\tau_2 \sigma_+(\tau_1,\tau_2) V_a(\tau_1)W_a(\tau_2)}X_n }_{\otimes n},     \\
	=&\avg{\mathcal T_{\mathcal C}\exp\kc{-\Delta I_n}}_{\otimes n},
	\end{align}\end{subequations}
where $\avg{...}_{\otimes n}=\Tr[\rho^{\otimes n}]$ and 
\begin{align}
\sigma_+(\tau_1,\tau_2)=\int_{\mathcal C^-} d\tau 
\delta(\tau_1-(\frac\beta2-\tau))\delta(\tau_2-(\beta+\lambda\tau))
+\int_{\mathcal C^+} d\tau
\delta(\tau_1-(\frac\beta2-\tau))\delta(\tau_2-\lambda\tau)    \label{gfunction}
\end{align}
We let the replica index $a$ begins at $0$ for later convenience. The deformation $\Delta I_n$ is nonlocal. The action is evaluated on time contour $\mathcal C$ where the imaginary time goes from $0$ to $\beta$, as shown in Fig.~\ref{FigReplicaA}, which is covered by an $n$-sheet manifold. Due to the contour ordering, the integral along $\mathcal C^-$ is always on the left hand side of the integral along $\mathcal C^+$. The twist operator $X_n$ imposes the following twisted boundary condition:
\begin{align}
V_a(0^-)=V_{a+1}(0^+),\quad W_a(0^-)=W_{a+1}(0^+), \quad a\sim a+n.
\end{align}
When $n=1$, we have $W(\beta+\tau)=W(\tau)$ and $\Delta I_n=0$, which is in agreement with the unitary evolution of the TFD state. One can further unfold the $n$-sheet manifold in Eq.~(\ref{PathIntegralVW}) and write the path integral into
\begin{align}
\avg{\mathcal T_{\mathcal C_n}\exp\ke{
		-\sum_{a=0}^{n-1}\kd{\int_{\mathcal C^-}d\tau W((a+1)\beta+\lambda\tau)V((a+\frac12)\beta-\tau)
			+\int_{\mathcal C^+}d\tau V((a+\frac12)\beta-\tau)W(a\beta+\lambda\tau)}}}_{n\beta},    \label{PathIntegralVWunfold}    
\end{align}
where $\avg{...}_{n\beta}=Z^{-n}\Tr[e^{-n\beta H}...]$ and
\begin{align}
V(a\beta+\tau)=V_a(\tau),\quad W(a\beta+\tau)=W_a(\tau).
\end{align}
The action is evaluated on the unfolded contour $C_n$, where the imaginary time goes from $0$ to $n\beta$, as shown in Fig.~\ref{FigReplicaB}. 
Similar approaches appeared in Refs.~\cite{Gu:2017njx,Goel:2018ubv,Chen:2019qqe}.

\subsection{Entropy at equilibrium}
We will first calculate the entropy at equilibrium by using the replica trick. For the Euclidean parametrizations $\ke{\varphi_a(\tau)}$ with $n$-sheet in the hyperbolic dish, the twisted boundary conditions are
\begin{align}\label{TwistBC}
\varphi_a(\beta)=\varphi_{a+1}(0^-) , \quad a\sim a+n.
\end{align}
It is convenient to introduce a global parametrization $f(\tau)$ which satisfies
\begin{align}\label{GlobalParameterization}
f(\tau+a\beta)=\varphi_a(\tau).
\end{align}  
Without $H_I$, the original replicated effective action $I_{n,0}$ can be written as a function of $f(\tau)$,
\begin{align}\label{ReplicaAction}
I_{n,0}=-C\sum_{a=0}^{n-1} \int_0^\beta d\tau \ke{\tan\frac{\varphi_a}2,\tau}=-C\int_0^{n\beta} d\tau \ke{\tan\frac f2,\tau}.
\end{align}
The saddle-point solution is 
\begin{align}\label{SaddlePoint}
f_c=\frac{2\pi\tau}{n\beta}.
\end{align}
Then we have 
\begin{align}
\ln Z_n=S_0+\frac{2\pi^2C}{n\beta},\quad S_n=S_0+\frac{2 \pi ^2 C(n+1)}{n\beta},\quad S=S_0+\frac{4\pi ^2C}{\beta},
\end{align} 
for the nearly extremal case, where $S_0=\phi_0/4G$ is the extremal entropy contributed by the topological term in Eq.~(\ref{DilatonGravity}) \cite{Azeyanagi:2007bj}. So $C/S_0=\bar\phi/2\pi\phi_0=\alpha_\epsilon\epsilon/2\pi$.

\subsection{The entanglement changed by extracting work}

Things become complicated when the interaction $H_I$ is turned on. The change in Renyi entropy $\Delta S_n$ is proportional to the change in replicated generating functional $\Delta\ln Z_n$, since $\Delta\ln Z=0$ under unitary evolution. In perturbation theory, it is reduced to the Feynman diagrams in the expansion of deformation $\Delta I_n$, like (\ref{PathIntegralVWunfold}), on the saddle point background (\ref{SaddlePoint}). We have
\begin{align}\label{DeltaSn}
\Delta S_n
=\frac{\Delta\ln Z_n}{1-n}
=\frac{\ln\avg{e^{-\Delta I_n}}}{1-n}
=\frac{\avg{-\Delta I_n}}{1-n}+\cdots
\end{align}
Recall that we have extracted work from the TFD state by applying the interaction $H_I=gO_LO_R$ in Sec.~\ref{SectionWork}.
As a warm-up, we will derive $\Delta S_n$ at $O(g)$. From Eq.~(\ref{DeltaSn}), we have
\begin{align}
(\Delta S_n)^{(1)}
=\frac {ign}{1-n}\int_{t_i}^{t_f} dt \kd{G_{n\beta}\kc{\frac\beta2+i(t+\lambda t)}
	-G_{n\beta}\kc{\frac\beta2-i(t+\lambda t)} },\\
G_{n\beta}(\tau)=\Tr[O(\tau)O\rho^{n}]
=2\left(\frac{n\beta}{\pi }\sin \frac{\pi\tau }{n\beta }\right)^{-2 \varDelta}+O(C^{-1}),\quad 0<\tau<n\beta \label{Gnbeta}
\end{align}
where $G_{n\beta}(\tau)$ is the Euclidean two-point function at inverse temperature $n\beta$. Then the change in von Neumann entropy is
\begin{align}
\Delta S^{(1)}= \left.\frac{2g\beta}{\lambda +1}\left(\frac\beta\pi  \cosh \left(\frac{\pi  (\lambda +1) t}{\beta }\right)\right)^{-2 \varDelta }\right|^{t_f}_{t_i}+O(C^{-1})
\end{align}
which agrees with Eq.~(\ref{EnergyRelation}) derived from the first law of entropy. We justify the consumption of entanglement in Eq.~(\ref{MI2Work}) when extracting work.

\subsection{The entanglement changed by anomalous heat flow}

\subsubsection{Renyi entropy}

We will calculate the entropy change due to the interaction (\ref{HIVW-WV}). The change at the first order vanishes, and we will consider the second-order perturbation. The insertions of two $H_I$'s on the $n$-sheet time contour happen in various ways. Both time-order and out-of-time-order correlation functions will appear.
Similarly to what we did in Eq.~(\ref{PathIntegralVW}), we can write down the deformation $\Delta I_n$ on the original replicated effective action (\ref{ReplicaAction}) with the interaction $H_I=g(V_LW_R-W_LV_R)/\sqrt2$,
\begin{subequations}\label{AddActionWormhole}\begin{align}
	-\Delta I_n
	=&-\frac{g}{\sqrt2}\sum_{a=0}^{n-1} \int_{\mathcal C} d\tau_1d\tau_2\, \sigma_+(\tau_1,\tau_2) \kd{V_a(\tau_1) W_a(\tau_2)-W_a(\tau_1) V_a(\tau_2)}-I_{M,a} \\
	=&\frac{g^2}2\sum_{a,b=0}^{n-1} \int_{\mathcal C} d\tau_1d\tau_2d\tau_3d\tau_4\, \sigma_+(\tau_1,\tau_2)\sigma_+(\tau_3,\tau_4)\kd{G^\varphi_{ab}(\tau_1,\tau_3)G^\varphi_{ab}(\tau_2,\tau_4)-G^\varphi_{ab}(\tau_1,\tau_4)G^\varphi_{ab}(\tau_2,\tau_3)}\label{GvarphiGvarphi} \\
	=&\frac{g^2}2\sum_{a,b=0}^{n-1}\int_{\mathcal C} d\tau_1d\tau_2d\tau_3d\tau_4\,\sigma_+(\tau_1,\tau_2)\sigma_+(\tau_3,\tau_4)\nn\\
	&\kd{G^f(a\beta+\tau_1,b\beta+\tau_3)G^f(a\beta+\tau_2,b\beta+\tau_4)-G^f(a\beta+\tau_1,b\beta+\tau_4)G^f(a\beta+\tau_2,b\beta+\tau_3)}
	\label{GfGf},
	\end{align}\end{subequations}
where $\sigma_+(\tau_1,\tau_2)$ is defined in Eq.~(\ref{gfunction}) and
\begin{align}
G^\varphi_{ab}(\tau_1,\tau_2)=\avg{\mathcal T_E V_a(\tau_1)V_b(\tau_2)}=\avg{\mathcal T_E W_a(\tau_1)W_b(\tau_2)}=G^f(a\beta+\tau_1,b\beta+\tau_2)    \label{Gvarphi}, \\
G^f(\tau_1,\tau_2)
=2 \left[\frac{f'(\tau_1) f'(\tau_2)}{\kc{2 \sin\frac{f(\tau_1)-f(\tau_2}2)}^2}\right]^{\varDelta } \label{Gf}.
\end{align}
$I_{M,a}$ is the action of the matter fields on the $a$-th sheet.
In Eq.~(\ref{GvarphiGvarphi}), we further integrate out the matter fields in $I_{M,a}$ which gives the correlation function $G_{ab}^\varphi$. 
Similarly to the situation of energy change at $O(g^2)$, the first (second) term in Eq.~(\ref{GvarphiGvarphi}) comes from the diagonal (cross) terms in $H_I^2$.
In Eq.~(\ref{GfGf}), we use the global parametrization (\ref{GlobalParameterization}).
According to Eq.~(\ref{ReplicaAction}), the reparametrization modes around the saddle point solution $f_c=2\pi\tau/n\beta$ are the same as the original modes in Eq.~(\ref{SchwarzianAction}) except that the inverse temperature is $n\beta$ now. We can integrate them out and obtain the expression of the four-point function $\avg{G^f(\tau_1,\tau_2)G^f(\tau_3,\tau_4)}$ up to $O(C^{-1})$, which is the same as the original one derived from the generating functional (\ref{GeneratingFunctional}) except that the inverse temperature is $n\beta$. After integrating out the four times $\ke{\tau_1,\tau_2,\tau_3,\tau_4}$ with those delta functions, we obtain the change in Renyi entropy with the form
\begin{align}\label{EntropyKernel}
\Delta S_n
=\frac{\avg{-\Delta I_n}}{1-n}+O(g^4)
\approx \int_{t_i}^{t_f} dt\int_{t_i}^{t} dt' \,Y_n(t,t').
\end{align}
We have used the perturbation theory at $O(g^2)$ and large $C$ limit. The double integrals on real time come from the integral along the contours $\mathcal C^\pm$ in Eq.~(\ref{gfunction}). The integral kernel $Y_n(t,t')$ is real and symmetric, $Y_n(t,t')=Y_n(t',t)$.
It contains two kinds of four-point functions $\avg{V(\tau_1)V(\tau_2)W(\tau_3)W(\tau_4)}_{n\beta}$ $(\tau_1>\tau_2>\tau_3>\tau_4)$ and 
$\avg{V(\tau_1)W(\tau_3)V(\tau_2)W(\tau_4)}_{n\beta}$ $(\tau_1>\tau_3>\tau_2>\tau_4)$. We can group those terms in $Y_n(t,t')$ into two parts accordingly, denoted as $(Y_n(t,t'))_{VVWW}$ and $(Y_n(t,t'))_{VWVW}$.
The former suffers from UV divergence when $\tau_1\to\tau_2$ or $\tau_3\to\tau_4$. So we will introduce a small separation $\epsilon$ to regularized it, such as $\avg{V(\tau_1+\epsilon)V(\tau_2)W(\tau_3+\epsilon)W(\tau_4)}$, while the latter is free from UV divergence since the situation of $\tau_1\to\tau_2$ or $\tau_3\to\tau_4$ never happens in Eq.~(\ref{AddActionWormhole}).

\begin{figure}
	\newcommand{\figheight}{0.2\linewidth}
	\centering
	\includegraphics[height=\figheight]{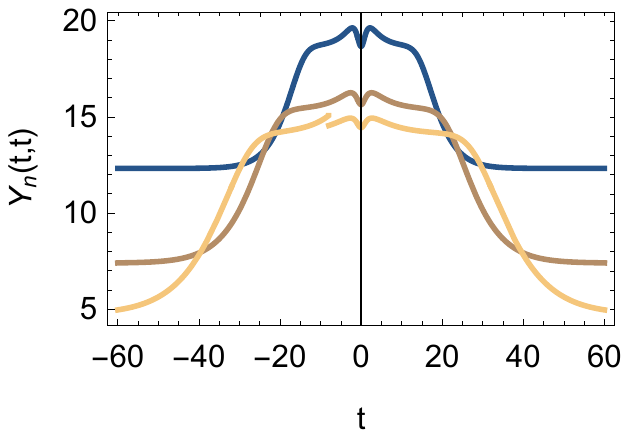}~~~~~~
	\includegraphics[height=\figheight]{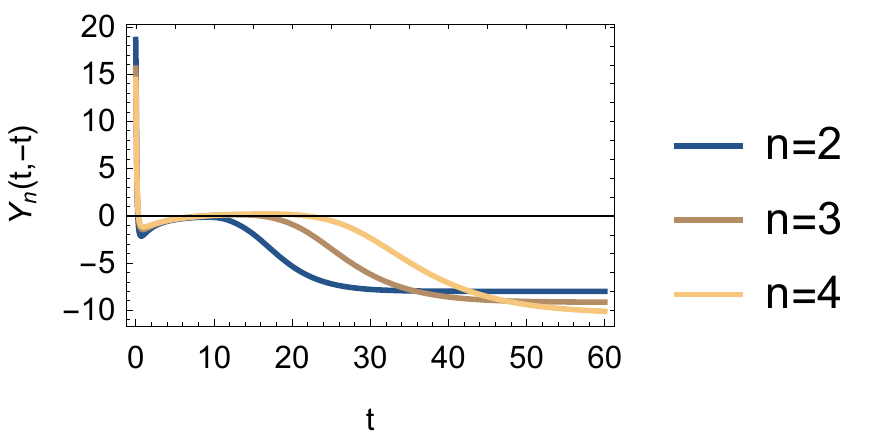}
	\caption{$Y_n(t,t)$ and $Y_n(t,-t)$ as functions of $t$ for $n=2,3,4, \,g=1,\,\beta=2\pi,\,\varDelta=1/6,\,C=10^5,\,\epsilon=0.2$, and $\lambda=1$.}
	\label{FigReKNEikonal}
\end{figure}

\begin{figure}%[!htb]
	\newcommand{\minipagewidth}{0.48\linewidth}
	\newcommand{\figwidth}{110pt}
	\centering
	\begin{minipage}[t]{\minipagewidth}
		\centering
		\includegraphics[height=0.40\linewidth]{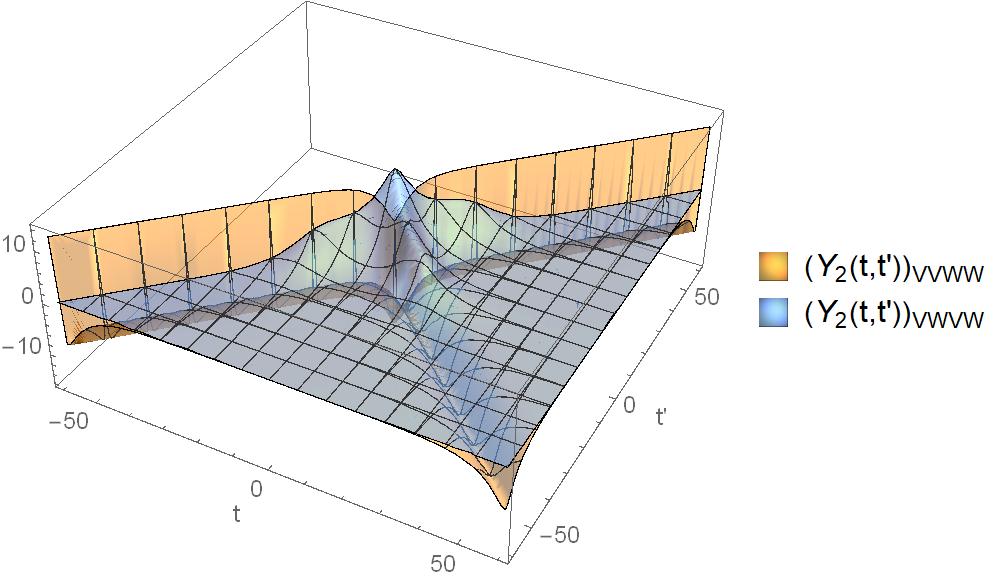}
		\caption{$(Y_2(t,t'))_{VVWW}$ (upper, orange) and $(Y_2(t,t'))_{VWVW}$ (lower, blue), where $g=1,\,\beta=2\pi,\,\varDelta=1/6,\,C=10^5,\,\epsilon=0.2$, and $\lambda=1$.}
		\label{FigReK2VVWWVWVW}
	\end{minipage}
	\hfill
	\begin{minipage}[t]{\minipagewidth}
	\includegraphics[height=0.40\linewidth]{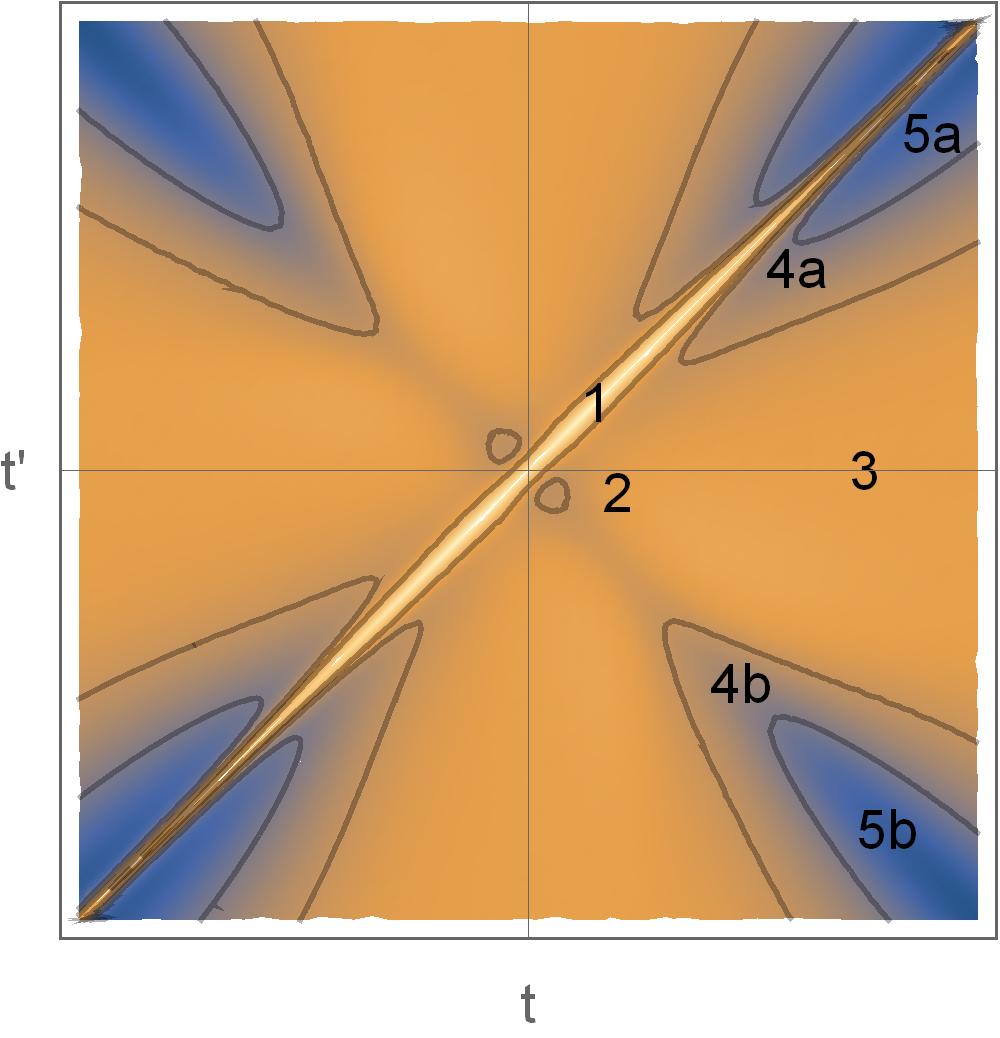}
	\caption{The illustration of the five regions in the $(t,t')$ plane. (1) UV-sensitive region. (2) Conformal region. (3) Dissipation region. (4a) and (4b) Early-time chaos region. (5a) and (5b) Later-time chaos region.}
	\label{FigRegions}
\end{minipage}
\end{figure}

\begin{figure}%[!htb]
	\centering
	\subfigure[]{
		\includegraphics[height=0.25\linewidth]{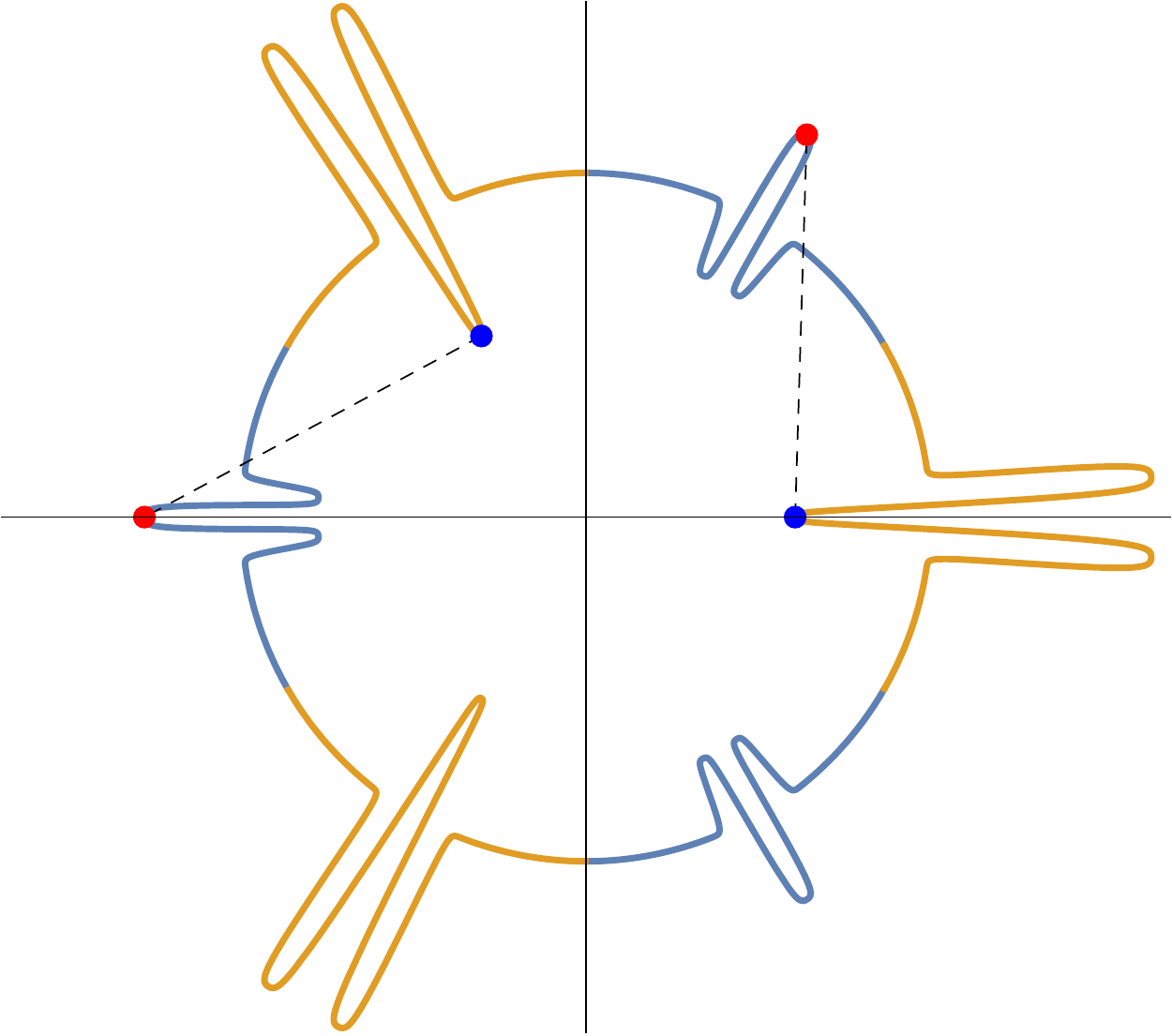}
		\label{FigReplicaOTOC1}    }
	\subfigure[]{
		\includegraphics[height=0.25\linewidth]{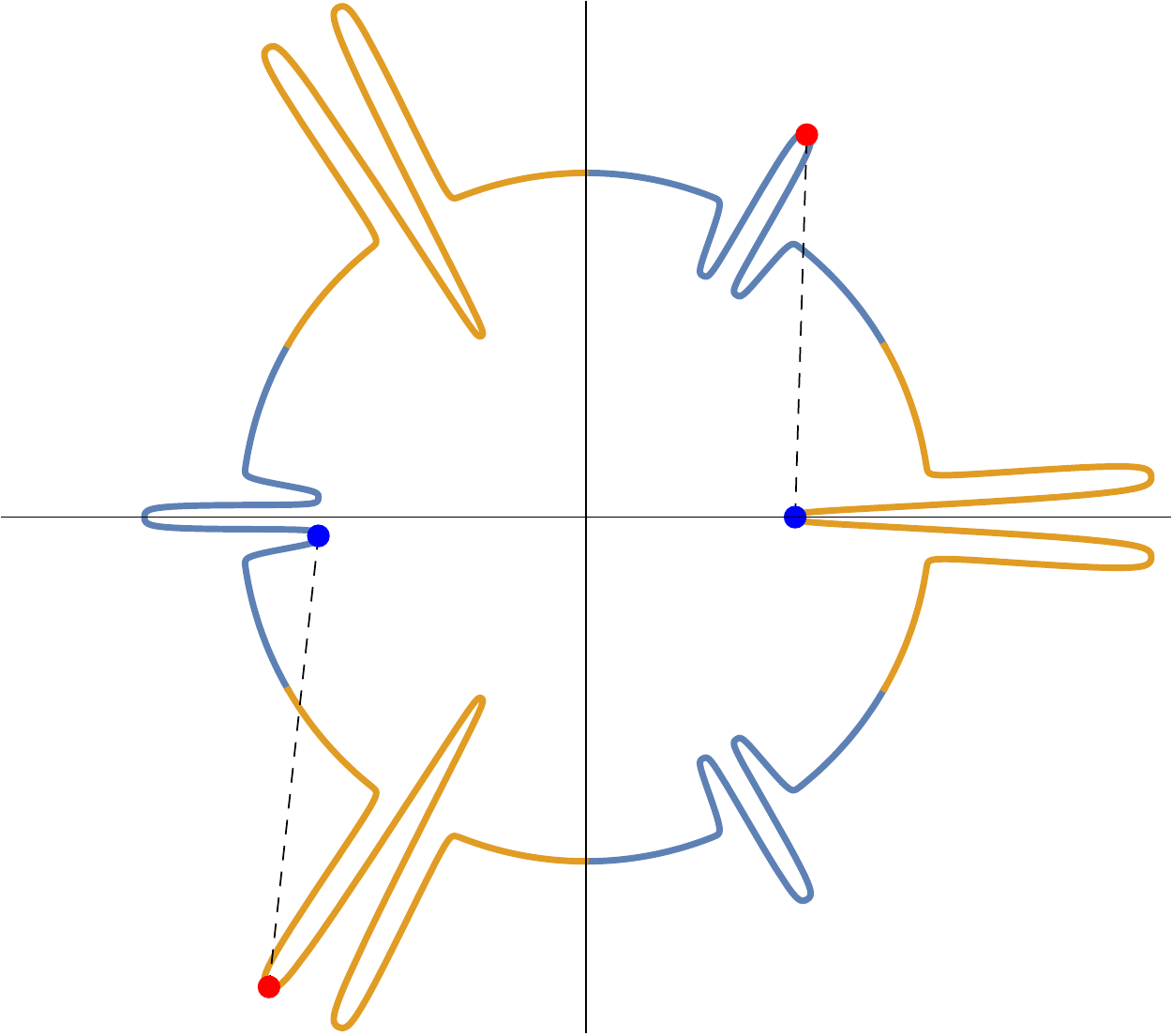}
		\label{FigReplicaOTOC2}    }
	\caption{Two OTOCs in (a) $Y_3(t,t)$ and (b) $Y_3(t,-t)$ shown in the $(n=3)$-sheet time contour. The red points and the blue points represent the operators on different types.}
	\label{FigReplicaOTOC}
\end{figure}

\begin{figure}%[!htb]
	\centering
	\includegraphics[height=0.20\linewidth]{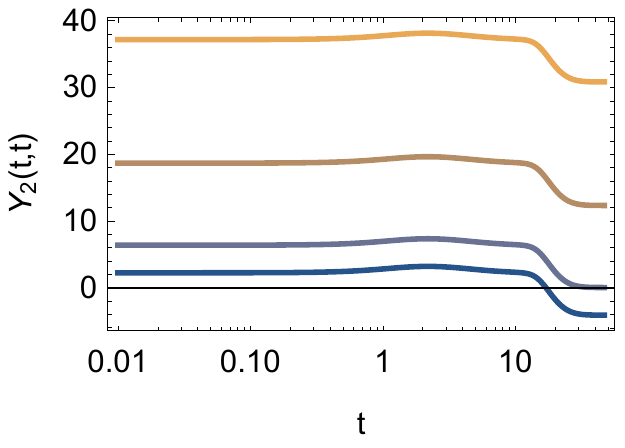}~~~~~~
	\includegraphics[height=0.20\linewidth]{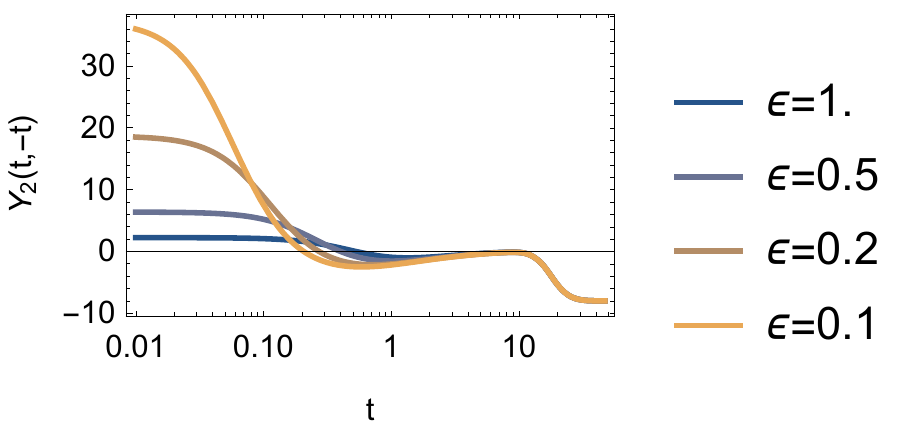}
	\caption{$Y_2(t,t)$ and $Y_2(t,-t)$ as functions of $t$, where $g=1,\,\beta=2\pi,\,\varDelta=1/6,\,C=10^5,\,\epsilon=1,0.5,0.2,0.1$, and $\lambda=1$.}
	\label{FigReKe}
\end{figure}

\begin{figure}%[!htb]
	\centering
	\newcommand{\figheight}{0.2\linewidth}
		\includegraphics[height=\figheight]{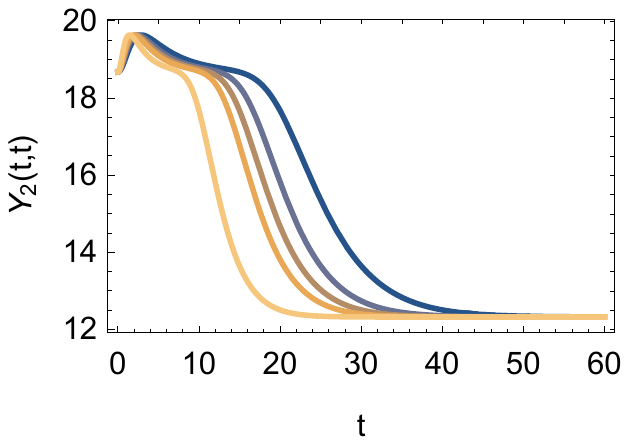}~~~~~~
		\includegraphics[height=\figheight]{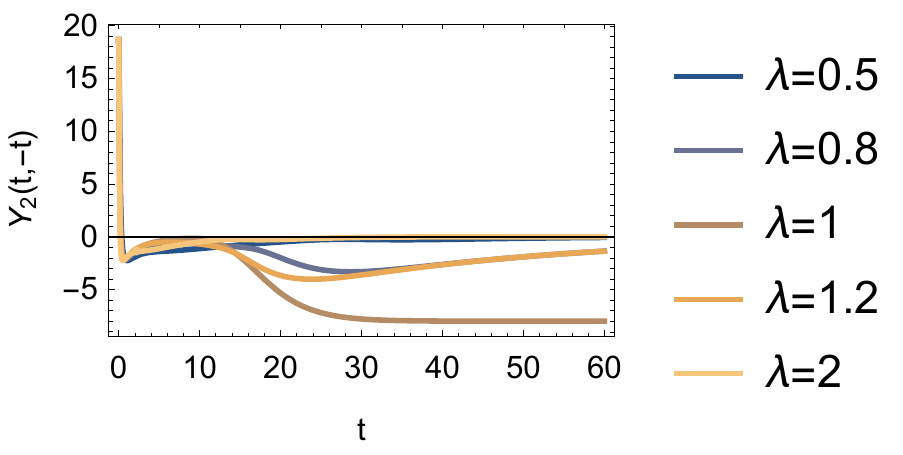}
	\caption{$Y_2(t,t)$ and $Y_2(t,-t)$ from eikonal approximation (\ref{EikonalCorrect}) for $\lambda=0.5,0.8,1,1.2,2,\,g=1,\,\beta=2\pi,\,\varDelta=1/6,\,C=10^5,\,\epsilon=0.2$.}
	\label{FigReKLtEikonal}
\end{figure}

\begin{figure}%[!htb]
	\centering
	\subfigure[]{
		\includegraphics[height=0.20\linewidth]{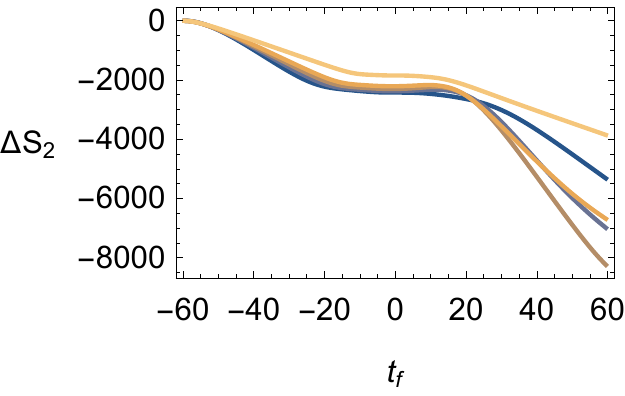}
		\label{FigRe2ti-60}
	}
	\subfigure[]{
		\includegraphics[height=0.20\linewidth]{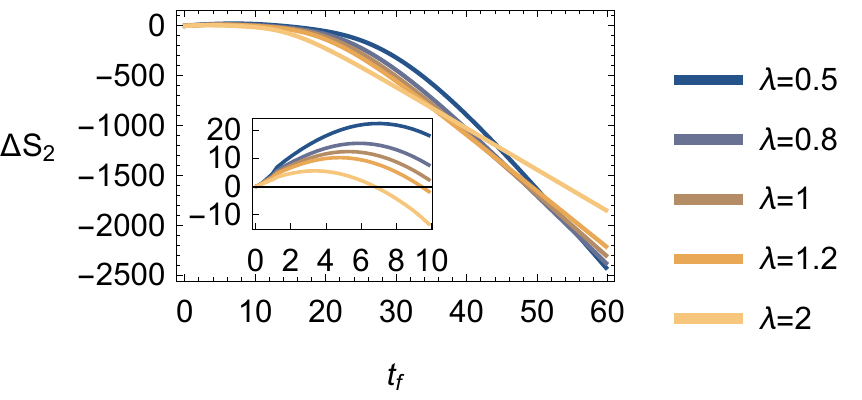}
		\label{FigRe2ti0}
	}
	\caption{$\Delta S_2$ as a function of $t_f$, where $\lambda=0.5,0.8,1,1.2,2,\,g=1,\,\beta=2\pi,\,\varDelta=1/6,\,C=10^5,\,\epsilon=0.2$, (a) $t_i=-60$ or (b) $t_i=0$.}
	\label{FigRe2}
\end{figure}

The four-point functions will be calculated in JT gravity. We will calculate $\avg{VVWW}$ by integrating out the reparametrization modes at $O(C^{-1})$ and calculate $\avg{VWVW}$ by using the eikonal approximation. One can alternatively calculate $\avg{VWVW}$ by integrating out the reparametrization modes at $O(C^{-1})$. The result is closed to the result from the eikonal approximation at early time while it becomes unreliable at late time. For $\avg{VWVW}$, we have the property
\begin{align}
\avg{V(it_1)W(it_3)V(it_2)W(it_4)}_{n\beta} =\avg{W(it_4+n\beta)V(it_1)W(it_3)V(it_2)}_{n\beta} =\avg{V(it_4+n\beta)W(it_1)V(it_3)W(it_2)}_{n\beta},
\end{align}
where we have exchanged operators $V$ and $W$ according to the $SO(M)$ symmetry of Eq.~(\ref{DilatonGravity}), while the eikonal approximation on $\avg{V(t_1)W(t_3)V(t_2)W(t_4)}$ in (6.57) of \cite{Maldacena:2016upp} does not maintain the above property. So we adjust the formula accordingly as follows
\begin{align}\label{EikonalCorrect}
\frac{\avg{V(it_1)W(it_3)V(it_2)W(it_4)}_{n\beta}}{\avg{V(it_1)V(it_2)}_{n\beta}\avg{W(it_3)W(it_4)}_{n\beta}}=\xi^{2\varDelta}U\kc{2\varDelta,1,\xi},\quad
\frac1\xi=\frac{i}{8C}\frac{2\sinh\frac{\pi(t_3+t_4-t_1-t_2)}{n\beta}}{\sinh\frac{\pi(t_1-t_2)}{n\beta}\sinh\frac{\pi(t_3-t_4)}{n\beta}}
\end{align}
where we have replaced the exponent by $2\sinh$ which is valid for both $\Re[t_3+t_4-t_1-t_2]>0$ and $\Re[t_3+t_4-t_1-t_2]<0$.

We first look at $\Delta S_n$ for integer $n\geq2$. We first analyze the case of $\lambda=1$. The configurations of $Y_n(t,t')$ is shown in Fig.~\ref{FigReKNEikonal}. We also separately draw the configuration of $(Y_n(t,t'))_{VVWW}$ and $(Y_n(t,t'))_{VVWW}$ in Fig.~\ref{FigReK2VVWWVWVW}. We can find that finally $(Y_n(t,t'))_{VVWW}$ goes to zero due to the quasinormal decay of OTOCs at late time. The characteristic time scales are UV cutoff $\epsilon$, dissipation time $t_d=\frac{\beta}{2\pi}$ and scrambling time $t_*=\frac{\beta}{2\pi}\ln\frac{2\pi C}{\beta}$. We divides the $(t,t')$ plane into several regions according to these time scales, as illustrated in Fig.~\ref{FigRegions}. 
\begin{description}
	\item[(1) UV-sensitive region, $|t-t'|\ll\epsilon$ and $|t+t'|\ll t_*$] ~\\
	The kernel regularized by the separation $\epsilon$ is UV sensitive. In Fig.~\ref{FigReKe}, we show the dependence on the choice of the separation $\epsilon$. We find $Y_n(t,t')>0$ when $t'\to t$ after such regularization. Such positivity in the UV time scale also appears even when the wormhole is absent as in Appendix \ref{SectionProductBH} and in the two-site SYK model as in Appendix \ref{SectionSYK}, where the latter has a known UV completion. During the interaction between product states, the Renyi entropy keeps increasing. So the Renyi mutual information is always positive, although the positivity of Renyi mutual information does not always hold for general states \cite{Teixeira:2012cre}. In Fig.~\ref{FigReKe}, we further find that when $\epsilon$ becomes smaller, the value of $Y_2(t,t')$ in the UV time scale has a positive shift.
	
	\item[(2) Conformal region, $\epsilon\ll|t-t'|\ll nt_d$ and $|t+t'|\ll t_*$]~\\
	 The four-point functions are factorized into conformal two-point functions. So $Y_n(t,t')\sim-|t-t'|^{-4\varDelta}<0$.
	
	\item[(3) Dissipation region, $nt_d\ll|t-t'|$ and $|t t'|\sim 0$]~\\ 
	These four-point functions are factorized into two-point functions at finite temperature, such that the kernel is dominated by dissipation, $Y_n(t,0)\sim -e^{-2\varDelta |t|/nt_d}<0$ and $Y_n(0,t')\sim -e^{-2\varDelta |t'|/nt_d}<0$. 
	
	\item[(4a) Early-time chaos region A, $nt_d\ll|t|< nt_*$ and $t\sim t'$]~\\ 
	 The kernel contains important terms $\sim-C^{-1}\beta^{1-4\varDelta}\exp\frac{4\pi t}{n\beta}$, which is contributed by the OTOCs in the first term of Eq.~(\ref{GfGf}), as illustrated in Fig.~\ref{FigReplicaOTOC1}. So the Renyi entropy $S_n$ will exponentially decrease with an exponent $2\pi t/n\beta$. Such behavior appears during general interactions between the two sides of the TFD state, as long as the interaction contains a term like $V_LW_R$ or $O_LO_R$. 
	
	\item[(4b) Early-time chaos region B, $nt_d\ll|t|< nt_*$ and $t\sim -t'$]~\\ 
	The kernel is dominated by terms $\sim-C^{-1}\beta^{1-4\varDelta}\exp\frac{4\pi t}{n\beta}$ and the Renyi entropy $S_n$ exponentially decreases with an exponent $2\pi t/n\beta$ as well. It is contributed by the OTOCs in the second term of Eq.~(\ref{GfGf}), as illustrated in Fig.~\ref{FigReplicaOTOC2}. So it will not appear if we adopt $H_I=gV_LW_R$. Such decrease in the entropy kernel corresponds to the anomalous heat flow, since both of them appear near $t\sim -t'$ and source from the cross terms in $H_I^2$.
	
	\item[(5a) Late-time chaos region A, $|t|> nt_*$ with $t\sim t'$]~\\ 
	The kernel saturates and goes to a constant, which is contributed by both $\epsilon$-regularized conformal two-point functions and OTOCs, as shown in Fig.~\ref{FigReKe}.
	
	\item[(5b) Late-time chaos region B, $|t|> nt_*$ with $t\sim -t'$]~\\
	Due to the late time decay of OTOCs, the kernel goes to a negative number
	\footnote{To obtain $\lim_{t\to\infty}Y_1(t,-t)$, we take the $n\to1$ limit first and then take the $t'\to-t$ limit. We find $\lim_{t\to\infty}\lim_{t'\to-t}\lim_{n\to1}Y_n(t,t')=4$ at the large $C$ limit.}
	\begin{align}\label{SaturatedConstant}\begin{split}
	\lim_{t\to\infty}Y_n(t,-t)
	&=-\frac{n }{n-1}\sum_{a=1}^{n-1}\avg{W((a+\frac{1}{2}) \beta +i t) V(a \beta -i t) V(\frac{\beta }{2}-i t) W(i t)}_{n\beta}+\text{conjugate}\\
	&=-\frac{2}{(n-1) n} \csc ^2\frac{\pi }{2 n} + O(C^{-1}) 
	\end{split}\end{align}
	without UV sensitivity, as shown in Fig.~\ref{FigReKe}.
\end{description}

The above two kinds of exponential decrease at early time rely on the existence of the wormhole. In Appendix \ref{SectionProductBH}, we replace the eternal black hole by two disconnected black holes, where the wormhole is absent. We calculate the change in Renyi entropy after turning on the interaction, where chaos regions (4) and (5) are replaced by dissipation region (3).

We discuss the case of general $\lambda$. $Y_n(t,t')$ for different $\lambda$ are shown in Fig.~\ref{FigReKLtEikonal}. $Y_n(t,t)$ reaches a constant at late time which is independent from $\lambda$; $Y_n(t,-t)$ approaches zero from below at late time once $\lambda\neq1$, since those four-point functions contributing to Eq.~(\ref{SaturatedConstant}) finally die out due to the mismatch between $t$ and $\lambda t$. We integrate the kernel and calculate $\Delta S_2$ as a function of time, as shown in Fig.~\ref{FigRe2}. In the case of $t_i\leq 0$, $\Delta S_2$ momentarily increases near $t_f=0$ due to the positive contribution in UV sensitive Region (1). However in other period of time, it generally decreases. Especially, in the case of $-nt_*< t_i\leq 0$, the negative contribution mainly comes from chaos regions (4a) and (5a). In the case of $t_i< -nt_*$, the negative contribution mainly come from chaos regions (4) and (5). We also check the Renyi entropy for higher $n$ and similar behaviors are observed. 

Recall that the anomalous heat flow in Sec.~\ref{SectionEnergyCorrelation} appears when $t_f\sim t_*$ in the case of $t_i< -t_*$ and $\lambda>1$, basically due to $(K_L(t,t')-K_R(t,t'))_{VWWV}<0$ near $t\sim-t'\sim t_*$. Such agreements between the kernels of energy and entropy on time scale and sign support our expectation that anomalous heat flow appears with the consumption of entanglement.
Furthermore, when $\lambda$ increases, the magnitude of the anomalous heat flow become smaller as shown in Fig.~\ref{FigKL2R2}. It agrees with the suppression of $|Y_n(t,t')|$ in chaos regions (4b) and (5b) and the suppression of $|\Delta S_2|$ near $t_f\sim t_*$, as shown in Fig.~\ref{FigReKLtEikonal} and Fig.~\ref{FigRe2ti-60}.

In Appendix \ref{SectionSYK}, we calculate the entropy changes for the initial TFD state in the two-site SYK model at early time. Regions (1), (2), (3), (4a), and (4b) also appear. In region (1), at the large-$q$ limit, we can get rid of the cutoff dependence. In region (4b), the kernel exponentially decreases as well, which agrees with the presence of the anomalous heat flow in the SYK model. But, in region (4a), the kernel exponentially increases, which is different from the case of JT gravity. 

\subsubsection{Von Neumann entropy}

To obtain the change in von Neumann entropy $\Delta S$, we calculate the summation $\sum_{a,b=0}^{n-1}$ in Eq.~(\ref{AddActionWormhole}) via contour integration and take the $n\to1$ limit. The details are given in Appendix \ref{SectionContourIntegral}. We work on a relatively simple case: large $C$ and $\varDelta=1/2$. Then $\avg{VVWW}$ is approximately factorized and $\avg{VWVW}$ is in the forms of eikonal approximation (\ref{EikonalCorrect}). The summations of factorized four-point functions have analytical forms.

\begin{figure}
	\centering
	\includegraphics[height=0.25\linewidth]{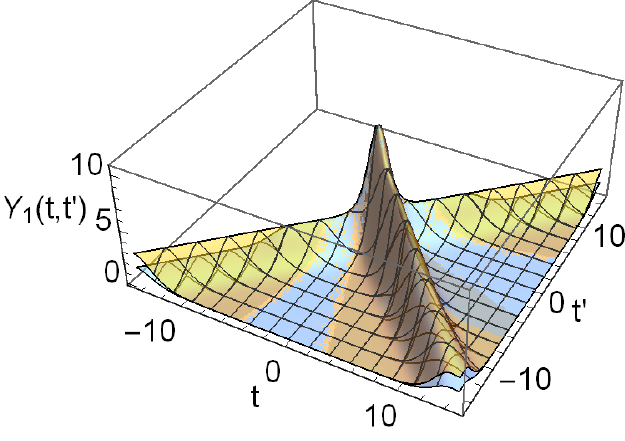}~~~~
	\includegraphics[height=0.25\linewidth]{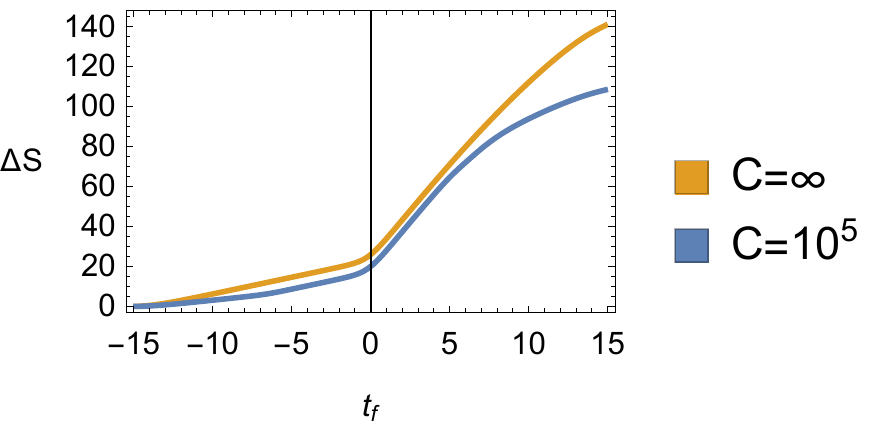}
	\caption{The changes in the von Neumann entropy $\Delta S$ and its kernel $Y_1(t,t')$, where  $g=1,\,\beta=2\pi,\,\varDelta=1/2,\,\epsilon=0,\,\lambda=1.1,\,t_i=-15$, $\,C=\infty$ (upper, orange) or $C=10^5$ (lower, blue).}
	\label{FigEEK}
\end{figure}

For the kernel of von Neumann entropy change $Y_1(t,t')$, its behavior also depends on the regions (1)-(5) introduced before. There are some differences between the $Y_{1}(t,t')$ and $Y_{n>1}(t,t')$ when $\Delta=1/2$. We can set the separation $\epsilon\to0^+$, but keeping $Y_1(t,t)$ finite. In regions (2), (3), and (4), the kernel $Y_1(t,t')$ is positive. In Region (5), the kernel $Y_1(t,t')$ goes to positive numbers when $\lambda=1$ and zero when $\lambda>1$. By comparing the kernel for finite $C$ with the kernel for infinite $C$, we find that the gravitational scattering (finite $C$ effect) lowers down the entanglement in chaos regions (4) and (5), as shown in Fig.~\ref{FigEEK}, which agrees with the time scale of anomalous heat flow.

The change in von Neumann entropy $\Delta S$ now becomes positive, which is different from the change in Renyi entropy. But the gravitational scattering still lowers $\Delta S$. Recall that the gravitational scattering mainly contributes to the anomalous heat flow as well. So we may identify it as the entanglement consumption related to the anomalous heat flow.

\section{Conclusion and outlook}

Quantum entanglement can be taken as the resource to extract work or transfer energy.
In this paper, we extract the work and study the thermodynamic arrow of time in the presence of wormholes in the AdS/CFT correspondence and the SYK model, where the field theory description is strongly coupled and highly chaotic. The initial state is taken to be the eternal black hole in JT gravity or the TFD state in the two-site SYK model. Such kind of states has two faces: it is highly entangled and local thermal. When some proper interactions between the two sides are turned on, we discover the anomalous heat flow, which transfers the energy from the colder side to the hotter side and further investigate its concomitant consumption of entanglement. Due to the property of the local thermal state, besides the anomalous heat flow, other channels allow the energy to diffuse from the hotter side to the colder side. The former effect reverses the thermodynamic arrow of time while the latter effect obeys it. Finally, within the setup in this paper, the latter effect wins the game at the perturbation theory. So far, we do not need to generalize the laws of black hole thermodynamics in the presence of wormholes.

The following issues deserve further investigation in the future.
\begin{itemize}
	\item The anomalous heat flow in the bulk picture. The energy-momentum tensor will carry the energy transfer in the bulk, which deserves to be solved after the interaction is turned on. This might reveal the specific role of the wormhole in the presence of the anomalous heat flow.

	\item Go beyond perturbation theory. Since $I_{n,0}\sim N$ and $\Delta I_n\sim g^2$ in the main text, we have taken the large-$N$ limit and the small $g$ limit to support the perturbation theory. If we change the scale of $\Delta I_n$ by replacing $g^2\to Ng^2$, then $I_{n,0}$ and $\Delta I_n$ will have the same scale. One should solve the full backreaction, which is similar to the eternal traversable wormhole \cite{Maldacena:2018lmt} and the geometrical interpretation of the twist operators in the generalized SYK model \cite{Gu:2017njx}. However, it is difficult to handle the highly non-local term in $\Delta I_n$ here. One may have to consider other interactions to maintain the solvability and the anomalous heat flow simultaneously.
	
	\item The possibility of reversing the total thermodynamic arrow of time. We have scanned the reasonable parameter space satisfying $\epsilon<\beta<C$ in JT gravity but do not observe the reversion of the total thermodynamic arrow of time. Those disconnected diagrams universally appear for general interactions as long as they contain the term $V_LW_R$. This reflects the fact that the TFD state is a local thermal state on each side, and thermal diffusion may be unavoidable. 
	
	\item About black hole evaporation near the Page time. 
	Black hole evaporation is a non-equilibrium process, where energy transfers from a black hole of finite temperature to empty space of zero temperature \cite{Page:2013dx}. At the Page time, the Hawking radiation and the remainder black hole are maximally entangled. It is argued that, after some quantum manipulations on the Hawking radiation, the total system can be transformed into a TFD state \cite{Maldacena:2013xja}. However, even at this time, it does not means that the two local systems have the same temperature. So the TFD state with unbalanced Hamiltonian considered here captures both of the entanglement and the temperature gradient between the remaining black hole and the Hawking radiation after the manipulations, where the former corresponds to the system $R$ and the latter corresponds to system $L$. Interactions between them can trigger subsequent evaporation. The calculation in this paper offers a framework to estimate the changes in energy and entropy of the evaporating black hole near the Page time.
\end{itemize}

\begin{acknowledgments}
We are grateful to Yu-Sen An, Rong-Gen Cai, Yiming Chen, Johanna Erdmenger, Song He,  Shao-Kai Jian, Sang Pyo Kim, Hong Lu, Li Li, Yue-Zhou Li, Yi Ling, Zi-Wen Liu, Chen-Te Ma, Ren\'{e} Meyer, Aavishkar A. Patel, Yu Tian, Xiao-Ning Wu and Jiaju Zhang for helpful discussions and correspondence. We also thank the referee for helpful comments on the analytic continuation of the Renyi entropy. 
This work is supported in part by the Natural Science Foundation of China under Grants No. 11875053, No. 11847229 (Z.-Y.X.), and No. 11675081 (L.Z.). Z.-Y.X. also acknowledges the support from China Postdoctoral Science Foundation under the National Postdoctoral Program for Innovative Talents No. BX20180318.
	
\end{acknowledgments}

\appendix

\section{Define work and heat}\label{SectionFirstLaw}
We will review the definitions of work and heat given in Ref.~\cite{Alipour:2016}, where the two local systems are treated equally.
One can divided the state of bi-systems into a non-correlated part and correlated part at time $t$,
\begin{align}
\rho(t)=\rho_L(t)\otimes \rho_R(t)+\chi(t).
\end{align}
The infinitesimal change in local energy can be divided into work and heat
\begin{align}
d E_\gamma(t)=d W_\gamma(t)+d Q_\gamma(t),\quad \gamma=L,R.
\end{align}
By introducing two auxiliary parameters $\alpha_L$ and $\alpha_R=1-\alpha_L$, one can rearrange the total Hamiltonians as
\begin{align}
H_{tot}=&H_L^{eff}(t)+H_R^{eff}(t)+H_I^{eff}(t),\\
H_\gamma^{eff}(t)=&H_\gamma+\Tr_{\bar\gamma}[\rho_{\bar\gamma}(t)H_I]-\alpha_\gamma\Tr[\rho_L(t)\otimes\rho_R(t) H_I].
\end{align}
One can define the so-called binding energy according to the correlation and the interaction
\begin{align}
U_I=\Tr[\chi(t)H_I^{eff}(t)].
\end{align}
Work and heat are defined as
\begin{subequations}\label{WorkHeat}\begin{align}
	dW_L(t)=&-dW_R(t)=-\alpha_L\Tr[d\rho_L(t)\otimes \rho_R(t) H_I]+\alpha_R\Tr[\rho_L(t)\otimes d\rho_R(t) H_I],\\
	dQ_\gamma(t)=&-i\Tr\kd{\chi(t)[H_\gamma^{eff}(t),H_I]}dt,
	\end{align}\end{subequations}
which satisfy
\begin{align}
d W_L(t)+d W_R(t)=0,\quad
d Q_L(t)+d Q_R(t)+d U_I(t)=0.
\end{align}

The two parameters $\alpha_L,\alpha_R$ should be determined by the relation between the pseudo-temperature and the equilibrium temperature in Ref.~\cite{Alipour:2016}. However, in the models of this paper, the term $\Tr[\rho_L(t)\otimes\rho_R(t') H_I]$ vanishes at the first- and second-order perturbation. So, the effect of the two parameters and the work done on the local systems vanish.

\section{Interaction between two disconnected black holes}\label{SectionProductBH}

We consider the unbalanced Hamiltonian $\tilde H_0$ and prepare a product of thermal state $\rho=\tau(\beta)\otimes\tau(\beta/\lambda)$ in $CFT_L\otimes CFT_R$, which is dual to two disconnected black holes living in two disconnected AdS$_2$ spaces \cite{Product}. Similarly, we consider that in each space there is dilaton-gravity with the JT term and some free matter fields
\begin{align}\begin{split}
I_0=&I_L+I_R,\\
I_\gamma=&
-\frac1{16\pi G_N}\kd{\int d^2x \sqrt{g_\gamma}\phi_\gamma (R_\gamma+2)+2\int_\partial dx \sqrt{h_\gamma}\phi_{\gamma,\partial} K_\gamma}- \int d^2x \sqrt{g_\gamma}\kd{(\partial\chi_{\gamma})^2+m^2\partial\chi_{\gamma}^2},\quad \gamma=L,R,
\end{split}\end{align}
where the topology terms are omitted. As shown in the main text, the dynamic of the system $R$ in unbalanced $\tilde H_0$ can be obtained by redefining the time of the system $R$ in balanced $H_0$. The boundary conductions are
\begin{align}\begin{split}
ds^2_{\partial,L}&=\frac1{\epsilon^2}d\tau^2,\quad \phi_{\partial,L}=\frac{\bar\phi}{\epsilon},\\
ds^2_{\partial,R}&=\frac{\lambda^2}{\epsilon^2}d\tau^2,\quad \phi_{\partial,R}=\frac{\bar\phi/\lambda}{\epsilon/\lambda}.
\end{split}\end{align}
Furthermore, at the low-energy limit, each $I_\gamma$ with scalar source $J_{\gamma,i}$ is reduced to a Schwarzian action plus the generating functional
\begin{align}\begin{split}
I_{eff}=
&-C\int d \tau_1 \ke{\mu_L, \tau_1}
-\int d\tau_1 d\tau_2 J_{L}(\tau_1)J_{L}(\tau_2)
\kd{\frac{\mu_L'(\tau_1)\mu_L'(\tau_2)}{(\mu_L(\tau_1)-\mu_L(\tau_2))^2}}^{\varDelta},\\
&-\frac{C}{\lambda}\int d \tau_1 \ke{\mu_R, \tau_1}
-\int d\tau_1 d\tau_2 J_{R}(\tau_1)J_{R}(\tau_2)
\kd{\frac{\mu_R'(\lambda\tau_1)\mu_R'(\lambda\tau_2)}{(\mu_R(\lambda\tau_1)-\mu_R(\lambda\tau_2))^2}}^{\varDelta}.
\end{split}\end{align}
When $J_{\gamma}=0$, the on-shell solutions with different temperatures are
\begin{align}
\mu_L=\tan\frac{\pi\tau}{\beta},\quad \mu_R=\tan\frac{\pi\lambda\tau}{\beta}.
\end{align}
Real time is obtained by Wick rotation $\mu\to i\nu$ and $\tau\to it$. So far, the $LL$ and $RR$ correlations here are the same as those of the TFD state in the main text.

After turning on the interaction $H_I=gO_LO_R$, we find that the $O(g^1)$ contribution vanishes because of $\avg{O_LO_R}=0$. At $O(g^2)$, since the four-point function is factorized $\avg{O_LO_RO_LO_R}=\avg{O_LO_L}\avg{O_RO_R}$, the energy changes are contributed by the diagrams given in Figs.~\ref{Fig4Pt_d1} and \ref{Fig4Pt_d2}, whose kernels are
\begin{align}\label{EnergyProduct}
K_L(t,t')=-G_{\beta }(-i \varDelta t \lambda ) G_{\beta }'(-i \varDelta t)-G_{\beta }(i \varDelta t \lambda ) G_{\beta }'(i \varDelta t),\\
K_R(t,t')=-G_{\beta }(-i \varDelta t) G_{\beta }'(-i \varDelta t \lambda )-G_{\beta }(i \varDelta t) G_{\beta }'(i \varDelta t \lambda ).
\end{align}
Our calculations in the main text already tell us that $\Delta E_L+\Delta E_R\geq 0$ and $\Delta E_L-\Delta E_R\geq 0$.
The same things happen if we alternatively consider interaction $H_I=gV_LW_R$ or $H_I=g(V_LW_R-W_LV_R)/\sqrt2$. Since $\avg{VW}=0$, in both cases, we find $\avg{H_IH_I}=g^2\avg{V_LV_L}\avg{W_RW_R}$ and only the diagrams given in Figs.~\ref{Fig4Pt_d1} and \ref{Fig4Pt_d2} are left.

\begin{figure}
	\includegraphics[height=0.25\linewidth]{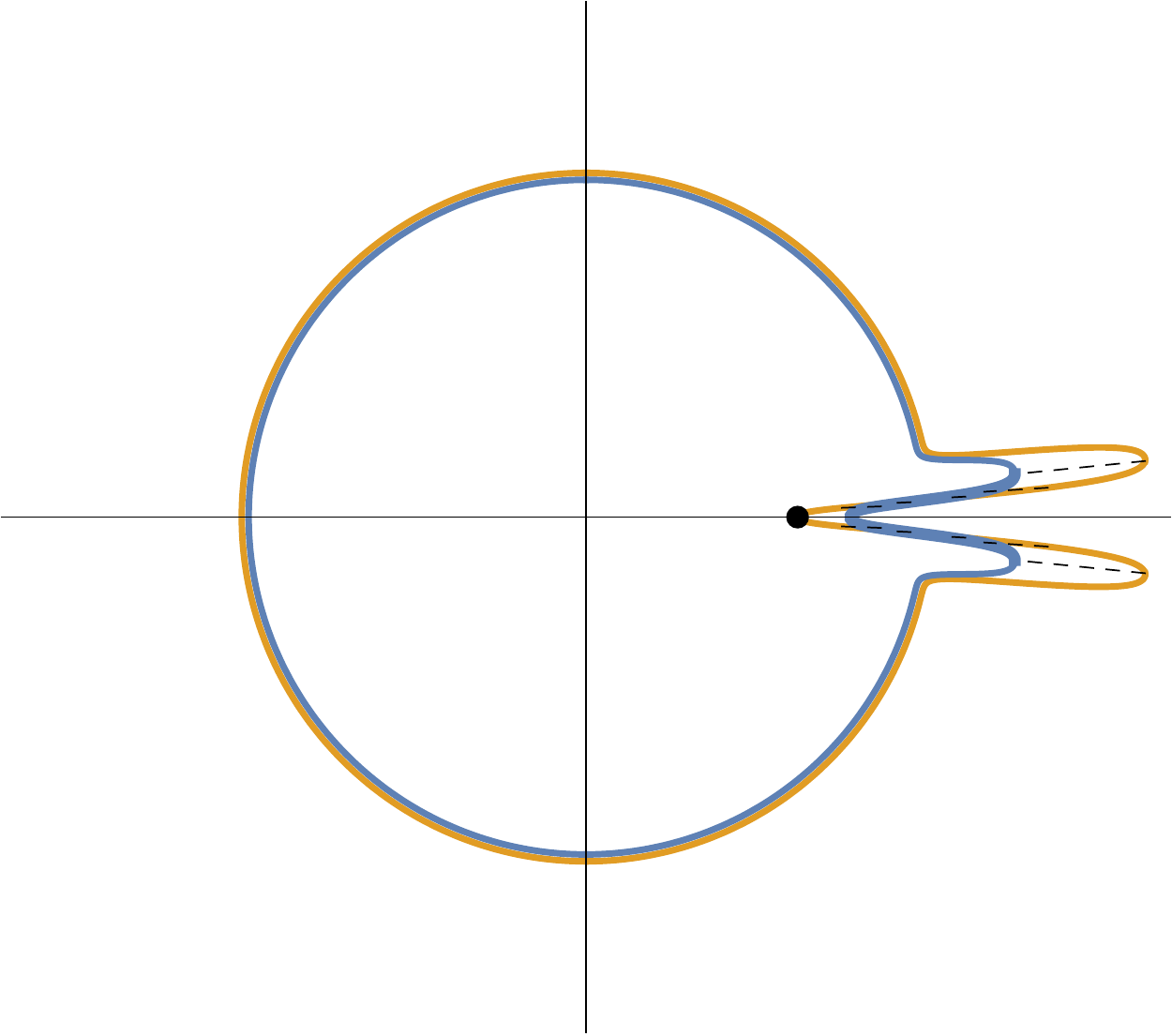}
	\caption{The time contour $\mathcal C$ in the complex plane $z=\exp(i2\pi t_C/\beta)$ for initial product state, where the complex time $t_C=\tau+it$ is measured by evolution $\exp(-t_C H)$. The blue (orange) contour indicates the evolution of system $L$ ($R$). The black point indicates the twisted operator $X_n$. The dashed lines indicate the coupling between the two sides during $[t_i,t_f]$, where $t_i<0<t_f$. Thick interval in the upper (lower) half plane indicates contour $\mathcal C^+$ ($\mathcal C^-$). We choose $\lambda=2$ here.}
	\label{FigReplicaP}
\end{figure}

In the following, we will calculate the entropy changes by using replica trick. Now the entropy change of system $L$ may not be the same as the one of system $R$ since the total system is not a pure state anymore.
We will use the parametrization $\mu_{\gamma,a}=\tan\frac{\varphi_{\gamma,a}}2$ and $\varphi_{\gamma,a}\sim \varphi_{\gamma,a}+2\pi$.
Similarly to the method in Sec.~\ref{SectionEntropy}, Renyi entropy $S_{n,R}$ of system $R$ can be written as
\begin{subequations}\begin{align}
	e^{(1-n)S_{n,R}}
	=&\Tr\kd{
		\ke{\mathcal T_{\mathcal C} \exp\kc{-\int_{\mathcal C^-}d\tau\,\tilde H_I(\tau)}}^{\otimes n}
		(\rho\otimes\rho)^{\otimes n}
		\ke{\mathcal T_{\mathcal C} \exp\kc{-\int_{\mathcal C^+}d\tau\,\tilde H_I(\tau)}}^{\otimes n} 
		(\mathbb{I}\otimes X_n) } \\
	=&\Tr\kd{(\rho\otimes\rho)^{\otimes n}
		\ke{\mathcal T_{\mathcal C} \exp\kc{-\int_{\mathcal C^-}d\tau\,\tilde H_I(\tau+\beta)
				-\int_{\mathcal C^+}d\tau\,\tilde H_I(\tau)}}^{\otimes n} 
		(\mathbb{I}\otimes X_n) } \\
	=&\Tr\kd{(\rho\otimes\rho)^{\otimes n} {\mathcal T_{\mathcal C}} \exp\kc{-\sum_a\int_{\mathcal C} d\tau_1d\tau_2\, \sigma(\tau_1,\tau_2) V_{L,a}(\tau_1)W_{R,a}(\tau_2)}(\mathbb I\otimes X_n)}
	\end{align}\end{subequations}
where, at the last step, we consider $H_I=g V_L W_R$ and
\begin{align}
\sigma(\tau_1,\tau_2)=\int_{\mathcal C^-}d\tau\delta(\tau_1-(\beta+\tau))\delta(\tau_2-(\beta+\lambda\tau))+\int_{\mathcal C^+} d\tau\delta(\tau_1-\tau)\delta(\tau_2-\lambda\tau),
\end{align}
where contours $\mathcal C_\pm$ are defined in Eq.~(\ref{Contour}) and contour $C$ goes from $0$ to $\beta$ containing $\mathcal C_\pm$, as shown in Fig.~\ref{FigReplicaP}.
The twisted boundary conditions are only applied on system $R$, leading to boundary conditions $O_{L,a}(\beta)=O_{L,a}(0^-)$ and $O_{R,a}(\beta)=O_{R,a+1}(0^-)$ for any scalar operator $O$ in the path integral of action $I_n=I_{n,0}+\Delta I_n$. The undeformed replicated effective action $I_{n,0}$ can be written as
\begin{subequations}\label{SchwarzianProduct}\begin{align}
	I_{n,0}=&-C\sum_{a=0}^{n-1} \int_0^\beta d\tau \ke{\tan\frac{\varphi_{L,a}}2,\tau}-C\sum_{a=0}^{n-1} \int_0^\beta d\tau \ke{\tan\frac{\varphi_{R,a}}2,\tau}\\
	=&-C\sum_{a=0}^{n-1} \int_0^\beta d\tau \ke{\tan\frac{\varphi_{L,a}}2,\tau}-C\int_0^{n\beta} d\tau \ke{\tan\frac {f_R} 2,\tau},
	\end{align}\end{subequations}
where $f_R(\tau+a\beta)=\varphi_{R,a}(\tau)$.
The saddle-point solution is $\phi_{L,a,c}=2\pi\tau/\beta,\,f_{R,c}=2\pi\tau/n\beta$.
The deformation on effective action up to $O(g^2)$ is
\begin{subequations}\label{AddActionProduct}\begin{align}
	-\Delta I_n
	=&-g\sum_{a=0}^{n-1} \int_{\mathcal C} d\tau_1d\tau_2\, \sigma(\tau_1,\tau_2) V_{L,a}(\tau_1) W_{R,a}(\tau_2)-I_{M,L,a}-I_{M,R,a} \\
	=&\frac{g^2}2\sum_{a=0}^{n-1} \int_{\mathcal C} d\tau_1d\tau_2d\tau_3d\tau_4\, \sigma(\tau_1,\tau_2)\sigma(\tau_3,\tau_4) G^{\varphi_{L,a}}_{aa}(\tau_1,\tau_3)G^{\varphi_{R,a}}_{aa}(\tau_2,\tau_4) \label{ReplicaDiagonal} \\
	=&\frac{g^2}2\sum_{a=0}^{n-1}\int_{\mathcal C} d\tau_1d\tau_2d\tau_3d\tau_4\,\sigma(\tau_1,\tau_2)\sigma(\tau_3,\tau_4)
	G^{\varphi_{L,a}}_{aa}(\tau_1,\tau_3)G^{f_R}(a\beta+\tau_2,a\beta+\tau_4)\\
	=&\frac{ng^2}2\int_{\mathcal C} d\tau_1d\tau_2d\tau_3d\tau_4\,\sigma(\tau_1,\tau_2)\sigma(\tau_3,\tau_4) 
	G^{\varphi_{L}}(\tau_1,\tau_3)G^{f_R}(\tau_2,\tau_4) \label{GfGvarphi},
	\end{align}\end{subequations}
where $G^{\varphi_{L,a}}_{aa},\, G^{\varphi_{R,a}}_{aa},\, G^{f_R}$ are defined in Eqs.~(\ref{Gvarphi}) and (\ref{Gf}). $I_{M,\gamma,a}$ is the action of the matter fields in the $\gamma$ $AdS_2$ space on the $a$-th sheet. In Eq.~(\ref{ReplicaDiagonal}), we have used $G^{\varphi_{L,a}}_{ab}(\tau_1,\tau_2)=\delta_{ab}G^{\varphi_{L,a}}_{aa}(\tau_1,\tau_2)$, since the different sheets of system $L$ are disconnected. In Eq.~(\ref{GfGvarphi}), we have used the translational symmetry on $\tau$ and let $G^{\varphi_{L,a}}_{aa}(\tau_1,\tau_2)=G^{\varphi_L}(\tau_1,\tau_2)$. For the interaction $H_I=g(V_LW_R-W_LV_R)/\sqrt2$, we have
\begin{subequations}\begin{align}
	-\Delta I_n
	=&\frac g{\sqrt2}\sum_{a=0}^{n-1} \int_{\mathcal C} d\tau_1d\tau_2\, \sigma(\tau_1,\tau_2) (V_{L,a}(\tau_1) W_{R,a}(\tau_2)-W_{L,a}(\tau_1) V_{R,a}(\tau_2))-I_{M,L,a}-I_{M,R,a} \\
	=&\frac{g^2}2\sum_{a=0}^{n-1} \int_{\mathcal C} d\tau_1d\tau_2d\tau_3d\tau_4\, \sigma(\tau_1,\tau_2)\sigma(\tau_3,\tau_4) G^{\varphi_{L,a}}_{aa}(\tau_1,\tau_3)G^{\varphi_{R,a}}_{aa}(\tau_2,\tau_4)
	\end{align}\end{subequations}
and the final result is the same. The calculation of Renyi entropy $S_{n,L}$ of system $L$ is paralleled to the above.

\begin{figure}
	\includegraphics[height=0.25\linewidth]{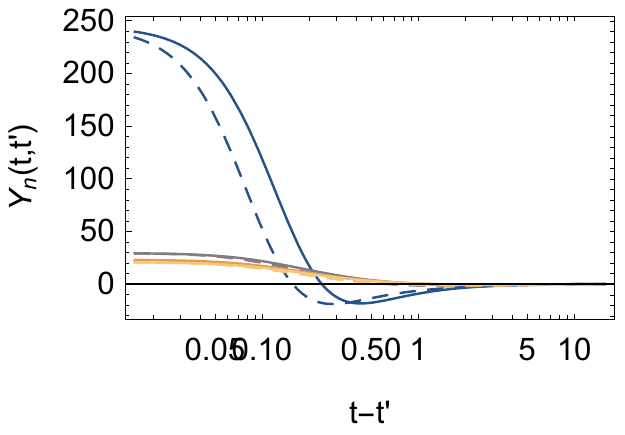}
	\includegraphics[height=0.25\linewidth]{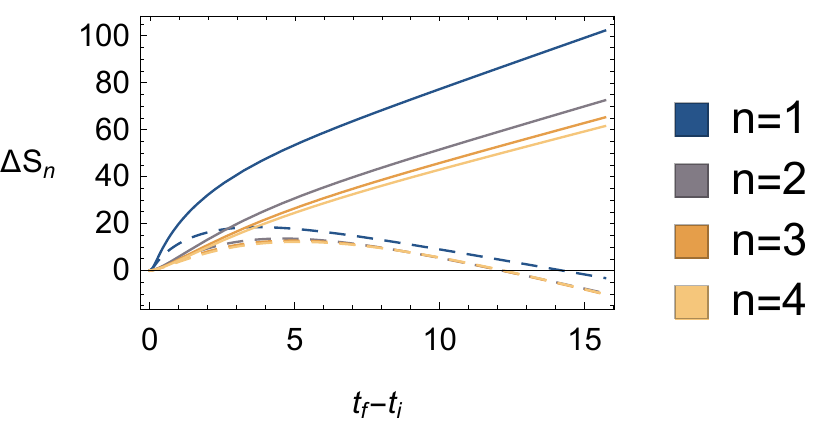}
	\caption{$Y_n(t,t')$ as a function of $t-t'$ and $\Delta S_n$ as a function of $t_f-t_i$ for the product state in JT gravity, where $n=1,2,3,4, \,g=1,\,\beta=2\pi,\,\varDelta=1/6,\,C=10^5,\,\epsilon=0.2$ and $\lambda=2$. Solid (dashed) lines correspond to system $L$ ($R$). The entropy changes for the product state in the two-site SYK model are the same.}
	\label{FigReKP}
\end{figure}

Finally, the change in Renyi entropy at $O(g^2)$ is a double integral (\ref{EntropyKernel}) as well. Since the  $G^{\varphi_L}$ and $G^{f_R}$ in Eq.~(\ref{GfGvarphi}) are uncorrelated with each other, the integral kernel $Y_{n,\gamma}(t,t')$ is factorized into
\begin{align}
Y_{n,L}(t,t')&=\frac{n}{n-1} \kc{G_{n\beta}(\epsilon+i\Delta t)G_{\beta}(\epsilon+i\lambda\Delta t)
	-G_{n\beta}(\beta-\epsilon+i\Delta t)G_{\beta}(\epsilon-i\lambda\Delta t)} + (\Delta t\leftrightarrow -\Delta t),\\\
Y_{n,R}(t,t')&=\frac{n}{n-1} \kc{G_{\beta }(\epsilon +i \Delta t) G_{n\beta }(\epsilon +i \lambda\Delta t  )-G_{\beta }(\epsilon -i \Delta t) G_{n\beta}(\beta-\epsilon +i \lambda\Delta t )} + (\Delta t\leftrightarrow -\Delta t),\\
\quad \Delta t=t-t', \nn
\end{align}
where $G_{n\beta}$ is given in Eq.~(\ref{Gnbeta}). $Y_{n,\gamma}(t,t')$ has similar UV sensitivities to those of the kernel of the eternal black hole in Sec.~\ref{SectionEntropy}. The formula above is analytical in $n$. So we can take the $n\to1$ limit and obtain the change in the von Neumann entropy, whose kernels are
\begin{align}\label{EntropyProduct}
Y_{1,L}(t,t')=-\beta  \left(G_{\beta }(-i \Delta t \lambda ) G_{\beta }'(-i \Delta t)+G_{\beta }(i \Delta t \lambda ) G_{\beta }'(i \Delta t)\right),\\
Y_{1,R}(t,t')=-\beta  \left(G_{\beta }(-i \Delta t) G_{\beta }'(-i \Delta t \lambda )+G_{\beta }(i \Delta t) G_{\beta }'(i \Delta t \lambda )\right).
\end{align}
The configurations of kernel $Y_n(t,t')$ and $\Delta S_n$ are shown in Fig.~\ref{FigReKP}. By comparing Eq.~(\ref{EnergyProduct}) with Eq.~(\ref{EntropyProduct}) or comparing Fig.~\ref{FigE_tree} with Fig.~\ref{FigReKP}, we find that the energy-entropy inequalities (\ref{EnergyEntropy}) are saturated at $O(g^2)$ for the initial product state.

\section{Energy and entropy in the two-site SYK model}\label{SectionSYK}
\subsection{TFD state in the two-site SYK model}
We will investigate similar phenomena in the two-site SYK model in parallel. The local Hamiltonian of the SYK model is a random $q$-body interaction between $N$ Majorana fermions \cite{SYK}
\begin{align}
H=i^{q/2}\sum_{1\leq j_1<j_2<...<j_q\leq N} J_{j_1j_2...j_q}\psi^{j_1}\psi^{j_2}...\psi^{j_q},\\
\avg{J_{j_1j_2...j_q}^2}=\frac{2^{q-1}J^2(q-1)!}{qN^{q-1}} \quad \text{(no sum)},
\end{align}
where $q\geq4$ and $N$ is even.
The effective action is \cite{Maldacena:2016hyu}
\begin{align}
\frac {I_0} N=\frac12 \ln\det(\partial_\tau-\Sigma)-\frac12\iint d\tau_1 d\tau_2 \kd{ \Sigma(\tau_1,\tau_2)G(\tau_1,\tau_2)-\frac{ J^2}{q}G(\tau_1,\tau_2)^q},
\end{align}
where $G(\tau_1,\tau_2)=\frac1N\sum_i\avg{\mathcal T_E\psi^i(\tau_1)\psi^i(\tau_2)}$. At the limit $N\gg\beta J\gg1$, the saddle point solution is $G_c(\tau_1,\tau_2)=b\left(\frac{J\beta}{\pi }\sin \frac{\pi(\tau_1-\tau_2)}{\beta }\right)^{-2 \varDelta}$ where scaling dimension $\varDelta=[\psi^i]=1/q$ and coefficient $b^q\pi=(\frac12-\varDelta)\tan(\pi\varDelta)$. The extremal entropy is $S_0=\alpha_0 N$ where the constant $\alpha_0$ is a function of $q$ \cite{Maldacena:2016hyu}. The leading UV correction is reparametrization $G(\tau_1,\tau_2)= [f'(\tau_1)f'(\tau_2)]^\varDelta G_c(f(\tau_1),f(\tau_2))$ controlled by the Schwarzian derivative
\begin{align}\label{SchwarzianSYK}
I_{eff}=-\frac{N\alpha_S}{\mathcal J} \int d\tau \ke{f,\tau},
\end{align}
where $\mathcal J=\sqrt q  J 2^{(1-q)/2}$ and the constant $\alpha_S$ is determined by numerical calculation \cite{Maldacena:2016hyu,Kitaev:2017awl}. The $SL(2,R)$ symmetry of the ground state is $f\to \frac{af_L+b}{cf_L+d}$ with $ad-bc=1$.
We can give a correspondence between the parameters in JT gravity and the SYK model by matching their effective actions and the extremal entropy
\begin{align}\label{JTSYK}
\frac1q=\varDelta,\quad 
\frac {N\alpha_S} {\mathcal J}=C,\quad
\frac{\alpha_S}{\alpha_0\mathcal J}=\frac{\alpha_\epsilon\epsilon}{2\pi}.
\end{align}

We consider two-site SYK$_L$ and SYK$_R$ which consist of two copies of $N$ Majorana fermions $\ke{\psi_L^i}$ and $\ke{\psi_R^i}$.
We first consider the balanced Hamiltonian  \cite{Gu:2017njx,Maldacena:2018lmt}
\begin{align}
H_0=H_L+H_R,    \quad
H_L=H\otimes 1,\quad H_R=1\otimes H^T,
\end{align}
where the transpose is taken on a basis $\ke{\ket{\Psi}}$. Those coupling $J_{j_1j_2...j_q}$ in $H_L$ and $H_R$ are the same. So the disordering of SYK$_L$ and SYK$_R$ are correlated. Given such a basis $\ke{\ket{\Psi}}$, we construe a state 
\begin{align}
\ket I=\sum_\Psi \ket{\Psi}_L\ket{\Psi}_R,
\end{align}
which satisfies $(H_L-H_R)\ket I=0$ automatically.
We specify the basis by imposing \cite{Gu:2017njx}
\begin{align}\label{ISYK}
(\psi_L^j+i\psi_R^j)\ket I=0,\quad j=1,2,...,N.
\end{align}
One can find $H^T=H$ in this basis.
The TFD state at $t=0$ is obtained from the Euclidean evolution of $\ket I$
\begin{align}
\ket{\beta}=Z_\beta^{-1/2}e^{-\beta(H_L+H_R)/4} \ket I.
\end{align}
By using Eq.~(\ref{ISYK}), the two-point function evaluated on the TFD state can be transformed into
$
\bra\beta \psi_L^j(t_1)\psi_R^k(t_2) \ket\beta=i \Tr[y\psi^j(-t_1)y\psi^k(t_2)]
$,
where $y=Z^{-1/2}e^{-\beta H/2}$ and $\Tr[O]\equiv\Tr_R[O_R]$.

Similarly to what we did in the main text, we further consider the unbalanced Hamiltonian
\begin{align}
\tilde H_0=H_L+\tilde H_R,\quad 
\tilde H_R=\lambda H_R,
\end{align}
on the two-site SYK model. The TFD state which evolves with $\tilde H_0$ is $\ket{\tilde\beta(t)}=e^{-i\tilde H_0t}\ket\beta$.
It is equivalent to redefining the time of SYK$_R$. 

\subsection{Anomalous heat flow}

To construe the anomalous heat flow in the two-site SYK model, we turn on the following interaction after $t_i$,
\begin{align}\label{HISYK}
H_I=i\sum_{j,k=1}^N g_{jk}\psi_L^j\psi_R^k,\\
g_{jk}=g_{kj},\quad \avg{g_{jk}g_{lm}}=N^{-2}g^2(\delta_{jl}\delta_{km}+\delta_{jm}\delta_{kl}).
\end{align}
At the large-$N$ limit, the energy changes at $t_f$ are
\begin{subequations}\label{EnergySYK}\begin{align}
	\Delta E_L
	=&(-i)^3g^2N^{-2}\sum_{jk} \int_{t_i}^{t_f} dt \int_{t_i}^{t} dt' \avg{[\dot\psi_L^j(t)\psi_R^k(\lambda t),\psi_L^j(t')\psi_R^k(\lambda t')]}
	+\avg{[\dot\psi_L^j(t)\psi_R^k(\lambda t),\psi_L^k(t')\psi_R^j(\lambda t')]}    \nn \\
	=&2g^2N^{-2}\Im\sum_{jk} \int_{t_i}^{t_f} dt \int_{t_i}^{t} dt' \Tr[\psi^j(-t')\dot\psi^j(-t)y\psi^k(\lambda t)\psi^k(\lambda t')y]
	+\Tr[\psi^k(-t)\dot\psi^j(-t_f)y\psi^k(\lambda t_f)\psi^j(\lambda t)y],\\
	\Delta E_R
	=&(-i)^3g^2N^{-2}\lambda\sum_{jk} \int_{t_i}^{t_f} dt \int_{t_i}^{t} dt' \avg{[\psi_L^j(t)\dot\psi_R^k(\lambda t),\psi_L^j(t')\psi_R^k(\lambda t')]}
	+\avg{[\psi_L^j(t)\dot\psi_R^k(\lambda t),\psi_L^k(t')\psi_R^j(\lambda t')]}    \nn \\
	=&2g^2N^{-2}\lambda\Im\sum_{jk} \int_{t_i}^{t_f} dt \int_{t_i}^{t} dt' \Tr[\psi^j(-t')\psi^j(-t)y\dot\psi^k(\lambda t)\psi^k(\lambda t')y]
	+\Tr[\psi^k(-t')\psi^j(-t)y\dot\psi^k(\lambda t)\psi^j(\lambda t')y],\\
	\Delta E_I
	=&-i^3g^2N^{-2}\sum_{jk} \int_{t_i}^{t_f} dt \avg{[\psi_L^j(t_f)\psi_R^k(\lambda t_f),\psi_L^j(t)\psi_R^k(\lambda t)]}
	+\avg{[\psi_L^j(t_f)\psi_R^k(\lambda t_f),\psi_L^k(t)\psi_R^j(\lambda t)]}    \nn \\
	=&2g^2N^{-2}\Im\sum_{jk} \int_{t_i}^{t_f} dt 
	\Tr[\psi^j(-t)\psi^j(-t_f)y\psi^k(\lambda t_f)\psi^k(\lambda t)y]
	+\Tr[\psi^k(-t)\psi^j(-t_f)y\psi^k(\lambda t_f)\psi^j(\lambda t)y],
	\end{align}\end{subequations}
where $\psi_\gamma^j(t)=e^{itH_\gamma}\psi_\gamma^j e^{-itH_\gamma}\,(\gamma=L,R)$, $\psi^j(t)=e^{itH}\psi^j e^{-itH}$. The above equations have the same form as Eq.~(\ref{PerturbationA}) \footnote{Exchanging the two fermions in the second term will give a minus sign.}. The disconnected four-point functions are determined by the conformal symmetry. The leading contribution of the connected four-point functions at early time are determined by the Schwarzian action (\ref{SchwarzianSYK}). So the energy changes at early time here are the same as those in JT gravity. The anomalous heat flow in the SYK model is contributed by the second term in Eq.~(\ref{EnergySYK}).

If we consider an initial product state in the two-site SYK model and turn on the same interaction (\ref{HISYK}), similarly to what we did in Appendix~\ref{SectionProductBH}, only the first term in Eq.~(\ref{EnergySYK}) is left.

\begin{figure}
	\includegraphics[height=0.3\linewidth]{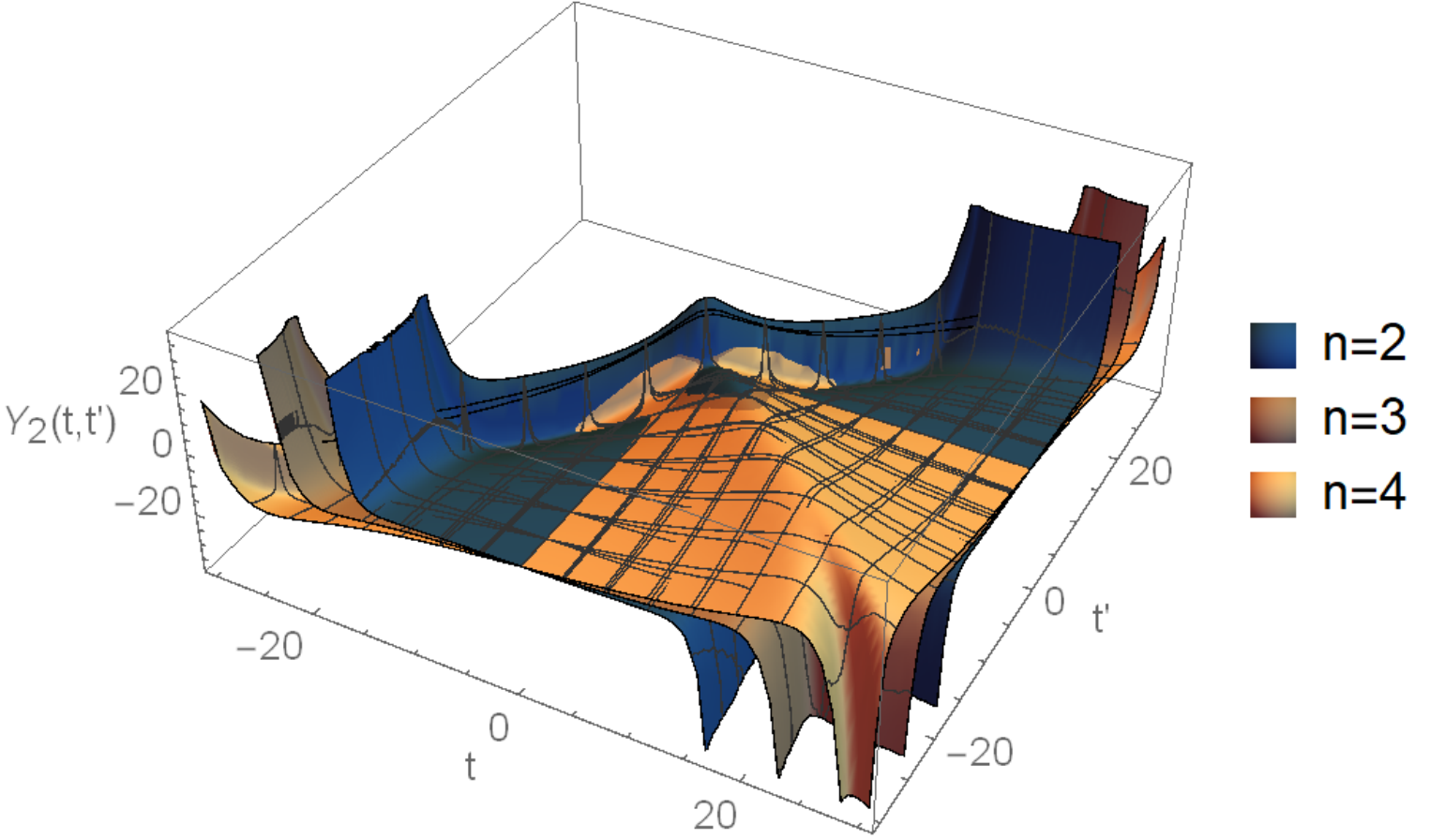}
	\caption{$Y_n(t,t')$ for an initial TFD state in the two-site SYK model. Parameters are $g=1,\,\beta=2\pi,\,\varDelta=1/6,\,C=10^5,\,\epsilon=0.2,\,\lambda=1$, and $n=2,3,4$ (from top to bottom near $t=t'$), which can be translated into the parameters of the SYK model $q,J,N$ according to relation (\ref{JTSYK}).}
	\label{FigReKSYK}
\end{figure}

To calculate the change in Renyi entropy $\Delta S_n$, we the apply replica trick on system $R$. We will alternately use imaginary time $\psi(\tau)=e^{\tau H}\psi e^{-\tau H}$ below. Similarly to the case in Eq.~(\ref{PathIntegralVW}), for the interaction (\ref{HISYK}), we have
\begin{subequations}\begin{align}
	&e^{(1-n)S_n}    \nn\\
	=&\bra\beta^{\otimes n} 
	\ke{\mathcal T_{\mathcal C}
		\exp\kd{-i\sum_{a,jk}\int_{\mathcal C^+}d\tau g_{jk}\psi_{L,a}^j(\tau)\psi_{R,a}^k(\lambda\tau)}} 
	(\mathbb{I}\otimes X_n) 
	\ke{\mathcal T_{\mathcal C} 
		\exp\kd{-i\sum_{a,jk}\int_{\mathcal C^-}d\tau g_{jk}\psi_{L,a}^j(\tau)\psi_{R,a}^k(\lambda\tau)}}
	\ket\beta^{\otimes n}    \\
	=&\Tr\kd{\rho^{\otimes n} X_n\mathcal T_{\mathcal C}\exp\ke{
			\sum_{a,jk}\kd{\int_{\mathcal C^-}d\tau g_{jk}\psi_a^k(\lambda\tau)\psi_a^j(-\frac\beta2-\tau)
				+\int_{\mathcal C^+}d\tau g_{jk} \psi_a^j(-\frac\beta2-\tau)\psi_a^k(-\beta+\lambda\tau)}}}\\
	=&\avg{\mathcal T_{\mathcal C}\exp\ke{-
			\sum_{a,jk}\kd{\int_{\mathcal C^-}d\tau g_{jk}\psi_a^j(\frac\beta2-\tau)\psi_a^k(\beta+\lambda\tau)
				-\int_{\mathcal C^+}d\tau g_{jk}\psi_a^j(\frac\beta2-\tau)\psi_a^k(\lambda\tau)}}X_n }_{\otimes n} \\
	=&\avg{\mathcal T_{\mathcal C}\exp\ke{
			-\sum_{a,jk}\int_{\mathcal C}d\tau_1d\tau_2 \sigma_-(\tau_1,\tau_2) g_{jk}\psi_a^j(\tau_1)\psi_a^k(\tau_2)}X_n }_{\otimes n}, 
	\end{align}\end{subequations}
where
\begin{align}
\sigma_-(\tau_1,\tau_2)=\int_{\mathcal C^-} d\tau 
\delta(\tau_1-(\frac\beta2-\tau))\delta(\tau_2-(\beta+\lambda\tau))
-\int_{\mathcal C^+} d\tau
\delta(\tau_1-(\frac\beta2-\tau))\delta(\tau_2-\lambda\tau).
\end{align}
In the final expression, twist operator $X_n$ is located at a time $\tau_*$ which is infinitesimally less than $0$. It imposes the relations
\begin{align}\label{TwistRelation}
\psi_a^j(\tau_*^-)=\psi_{a+1}^j(\tau_*^+),\quad \forall a=0,1,...,n-1,\quad (\psi_n^j=\psi_0^j).
\end{align}
We write down the replicated action, integrate out the disorder, and introduce the bi-local field $G_{ab}(\tau_1,\tau_2)=N^{-1}\sum_j\avg{\mathcal T_E\psi_a^j(\tau_1)\psi_b^j(\tau_2)}$ and Legendre multiplier $\Sigma_{ab}(\tau_1,\tau_2)$,
\begin{align}
Z_n=\int \prod_{j,a}\mathcal D\psi^j_a e^{-I_n[\psi]}=\int  \prod_{ab}\mathcal DG_{ab} \mathcal D\Sigma_{ab} e^{-I_n[G,\Sigma]},
\end{align}
where
\begin{subequations}\label{DeformationSYKfold}\begin{align}
	&-I_n    \nn\\
	=&\sum_{a=0}^{n-1}\int_{\mathcal C}d\tau \kd{-\frac12 \sum_j\psi_a^j(\tau)\partial_\tau\psi_a^j(\tau)
		+ i^{q/2}\sum_{j_1<j_2<...<j_q} J_{j_1j_2...j_q}\psi_a^{j_1}(\tau)\psi_a^{j_2}(\tau)...\psi_a^{j_q}(\tau)} \nn\\
	&    - \sum_{a,jk}\int_{\mathcal C} d\tau_1d\tau_2 \sigma_-(\tau_1,\tau_2) g_{jk} \psi_a^j(\tau_1)\psi_a^k(\tau_2) \\
	=&-\frac12\sum_{a,j}\int_{\mathcal C}d\tau \psi_a^j(\tau)\partial_\tau\psi_a^j(\tau)
	+\frac{J^2}{2qN^{q-1}} \sum_{ab,j_1<j_2<...<j_q}\int_{\mathcal C}d\tau_1 d\tau_2  \psi_a^{j_1}(\tau_1)\psi_b^{j_1}(\tau_2)\psi_a^{j_2}(\tau_1)\psi_b^{j_2}(\tau_2)...\psi_a^{j_q}(\tau_1)\psi_b^{j_q}(\tau_2)    \nn    \\
	&+ \frac{g^2}{2N^2}\sum_{ab,jk}\int_{\mathcal C} d\tau_1d\tau_2d\tau_3d\tau_4 \,\sigma_-(\tau_1,\tau_2)\sigma_-(\tau_3,\tau_4) \kd{\psi_a^j(\tau_1)\psi_a^k(\tau_2)\psi_b^j(\tau_3)\psi_b^k(\tau_4) +\psi_a^j(\tau_1)\psi_a^k(\tau_2)\psi_b^k(\tau_3)\psi_b^j(\tau_4)}    \\
	=&\frac N2\ln\det(\partial_\tau\delta_{ab}-\Sigma_{ab})    -\frac N2\sum_{ab}\int_{\mathcal C} d\tau_1 d\tau_2  \kd{\Sigma_{ab}(\tau_1,\tau_2)G_{ab}(\tau_1,\tau_2)-\frac {J^2}{q}G_{ab}(\tau_1,\tau_2)^q}        \nn    \\
	&+ \frac{g^2}{2}\sum_{ab}\int_{\mathcal C} d\tau_1d\tau_2d\tau_3d\tau_4 \,\sigma_-(\tau_1,\tau_2)\sigma_-(\tau_3,\tau_4) \kd{-G_{ab}(\tau_1,\tau_3)G_{ab}(\tau_2,\tau_4) +G_{ab}(\tau_1,\tau_4)G_{ab}(\tau_2,\tau_3)}    \\
	=& -I_{n,0}-\Delta I_n,
	\end{align}\end{subequations}
where $\Delta I_n$ is a highly non-local deformation on the original bi-local action $I_{n,0}$. 
Relation (\ref{TwistRelation}) leads to the twisted boundary conditions in the path integral
\begin{subequations}\label{TwistSYK}\begin{align}
	&\psi_a^j(\beta)=\psi_{a+1}^j(0^-),\quad \psi_{n-1}^j(\beta)=-\psi_0^j(0^-) \quad \text{for}\quad a=0,1,...,n-2,\\
	&G_{ab}(\beta,\tau)=G_{(a+1)b}(0^-,\tau), \quad  G_{(n-1)b}(\beta,\tau)=-G_{0b}(0^-,\tau) \quad \text{for}\quad a=0,1,...,n-2,\quad \forall b,\\
	&G_{ab}(\tau,\beta)=G_{a(b+1)}(\tau,0^-),\quad G_{a(n-1)}(\tau,\beta)=G_{a0}(\tau,0^-) \quad \text{for}\quad b=0,1,...,n-2,\quad  \forall a .
	\end{align}\end{subequations}
Similar conditions arise in Refs.~\cite{Liu:2017kfa,Gu:2017njx}.
As we did in Eq.~(\ref{Gvarphi}), we introduce a global fermion $\psi^j(\tau)$ and global bi-local fields $G(\tau_1,\tau_2),\Sigma(\tau_1,\tau_2)$ by
\begin{align}
\psi_a^j(\tau)=\psi^j(a\beta+\tau),\quad
G_{ab}(\tau_1,\tau_2)=G(a\beta+\tau_1,b\beta+\tau_2),\quad \Sigma_{ab}(\tau_1,\tau_2)=\Sigma(a\beta+\tau_1,b\beta+\tau_2).
\end{align}
According to Eq.~(\ref{TwistSYK}), these global fields are continuous in $\tau\in(0,n\beta)$ and anti-periodic
\begin{align}
\psi^j(n\beta)=-\psi^j(0^-),\quad
G(n\beta,\tau)=-G(0^-,\tau),\quad     G(\tau,n\beta)=-G(\tau,0^-),\quad     
\Sigma(n\beta,\tau)=-\Sigma(0^-,\tau),\quad     \Sigma(\tau,n\beta)=-\Sigma(\tau,0^-).
\end{align}
Then we have
\begin{subequations}\label{UnfoldSYK}\begin{align}
-I_{n,0}=&\frac N2\ln\det(\partial_\tau-\Sigma)-\frac N2\int_{\mathcal C_n} d\tau_1 d\tau_2  \kd{\Sigma(\tau_1,\tau_2)G(\tau_1,\tau_2)-\frac {J^2}{q}G(\tau_1,\tau_2)^q},    \label{UnfoldSYK0} \\
-\Delta I_n=&\frac{g^2}{2}\sum_{ab}\int_{\mathcal C} d\tau_1d\tau_2d\tau_3d\tau_4 \,\sigma_-(\tau_1,\tau_2)\sigma_-(\tau_3,\tau_4)    \nn\\ &\kd{-G(a\beta+\tau_1,b\beta+\tau_3)G(a\beta+\tau_2,b\beta+\tau_4) +G(a\beta+\tau_1,b\beta+\tau_4)G(a\beta+\tau_2,b\beta+\tau_3)},  \label{DeformationSYKunfold}
\end{align}\end{subequations}
where $\mathcal C_n$ is the unfolded time contour in Fig.~\ref{FigReplica} and the $\tau$ in $\int_{\mathcal C_n}d\tau$ goes from $0$ to $n\beta$. Equation~(\ref{UnfoldSYK0}) is just the action of the SYK model at the inverse temperature $n\beta$, with the saddle point solution and effective action in the same forms when $\beta J\gg1$. Considering the reparametrization modes in Eq.~(\ref{UnfoldSYK0}), we will obtain the Schwarzian derivative as in the case of JT gravity in Eq.~(\ref{ReplicaAction}). If we integrate out the four times in Eq.~(\ref{DeformationSYKunfold}), expand the product into eight terms and sort the time ordering of fermions, we will find that the deformation $\Delta I$ in the SYK model is different from Eq.~(\ref{AddActionWormhole}) in JT gravity, where some terms have inverse signs. 
Perturbation theory (\ref{DeltaSn}) at $O(g^2)$ is applicable at the large-$N$ limit. The kernel $Y_n(t,t')$ consists of four-point functions $\sum_{jk}\avg{\psi^j\psi^j\psi^k\psi^k}$ and $\sum_{jk}\avg{\psi^j\psi^k\psi^j\psi^k}$, which are controlled by effective action (\ref{SchwarzianSYK}) as well. So, at early time, we will apply the four-point functions $\avg{VVWW}$ and $\avg{VWVW}$ from JT gravity but take care of their signs when evaluating $Y_n(t,t')$. The parameters of the SYK model and JT gravity can be related by using relation (\ref{JTSYK}). The configuration of kernel $Y_n(t,t')$ at early time is shown in Fig.~\ref{FigReKSYK}. The kernel $Y_n(t,t')$ exponentially increase at early-time chaos region A $(t\sim t')$, contributed by the first term in Eq.~(\ref{DeformationSYKunfold}). The kernel $Y_n(t,t')$ exponentially decrease at early-time chaos region B $(t\sim -t')$, contributed by the second term in Eq.~(\ref{DeformationSYKunfold}). Its decrease agrees with the fact that the anomalous heat flow in the SYK model is also contributed by the second term in Eq.~(\ref{EnergySYK}).

For an initial product state of the two-site SYK model, we calculate the entropy changes in the way of Appendix \ref{SectionProductBH} and obtain
\begin{subequations}\begin{align}
	e^{(1-n)S_{n,R}}
	=&\Tr\kd{(\rho\otimes\rho)^{\otimes n} {\mathcal T_{\mathcal C}} \exp\kc{-i\sum_{a,jk}\int_{\mathcal C} d\tau_1d\tau_2\, \sigma(\tau_1,\tau_2)g_{jk} \psi_{L,a}^j(\tau_1)\psi_{R,a}^k(\tau_2)}(\mathbb I\otimes X_n)}\\
	=&\int \prod_{\gamma,a,j}\mathcal D\psi_{\gamma,a}^j \exp\kc{-I_{n,0}-\Delta I_n},
	\end{align}\end{subequations}
with boundary conditions $\psi_{L,a}(\beta)=-\psi_{L,a}(0^-)$ for $(a=0,1,...,n-1)$, $\psi_{R,a}(\beta)=\psi_{R,a+1}(0^-)$ for $(a=0,1,...,n-2)$, and $\psi_{R,n-1}(\beta)=-\psi_{R,0}(0^-)$.
We can transform $I_{n,0}$ into the effective action (\ref{SchwarzianProduct}) and find
\begin{align}
&-\Delta I_n    \nn\\
=&-i\sum_{a,jk}\int_{\mathcal C} d\tau_1d\tau_2\, \sigma(\tau_1,\tau_2)g_{jk} \psi_{L,a}^j(\tau_1)\psi_{R,a}^k(\tau_2)\\
=&-\frac{g^2}{2N^2}\sum_{ab,jk}\int_{\mathcal C} d\tau_1d\tau_2d\tau_3d\tau_4\, \sigma(\tau_1,\tau_2)\sigma(\tau_3,\tau_4)
\kd{\psi_{L,a}^j(\tau_1)\psi_{R,a}^k(\tau_2)\psi_{L,b}^j(\tau_3)\psi_{R,b}^k(\tau_4)
	+\psi_{L,a}^j(\tau_1)\psi_{R,a}^k(\tau_2)\psi_{L,b}^k(\tau_3)\psi_{R,b}^j(\tau_4)}\\
=&\frac{g^2}{2}\sum_{a}\int_{\mathcal C} d\tau_1d\tau_2d\tau_3d\tau_4\, \sigma(\tau_1,\tau_2)\sigma(\tau_3,\tau_4)
G_{L,aa}(\tau_1,\tau_2)G_{R,aa}(\tau_2,\tau_4).
\end{align}
The final result is the same as deformation (\ref{AddActionProduct}) in JT gravity. So the energy-entropy inequalities (\ref{EnergyEntropy}) are saturated as well.

\subsection{Large-$q$ limit and UV-sensitive region}
In the main part of this paper, we only use the correlators with nearly conformal symmetry, which is only valid in the IR. 
Due to the lack of the UV part of the boundary theory which is dual to JT gravity in the bulk, we are unable to give a detailed description on the change in energy and entropy in the UV-sensitive region. Thanks to the fact that the UV fixed point of the SYK model is known as free fermions, we can calculate the UV behaviors. Here, we will consider the large-$q$ limit. The Euclidean Green's function with $1/q$ expansion is \cite{Maldacena:2016hyu}
\begin{align}\label{GLargeq}
G(\tau)=\frac12 \sgn(\tau)\kd{1+\frac2q \ln\frac{\cos\frac{\pi\nu(\beta)}2}{\cos\kd{\pi\nu(\beta)\kc{\frac12-\frac{|\tau|}\beta}}}+\cdots},
\end{align}
where the logarithm is required to be of order $1$ and $\nu(\beta)$ is determined by
\begin{align}
\beta\mathcal J=\frac{\pi\nu(\beta)}{\cos\frac{\pi\nu(\beta)}{2}}.
\end{align}
In (\ref{GLargeq}), the second term, as a correction, should be smaller than the first term. So we should keep the real time small and focus on the UV-sensitive region. Then we can assume that four-point functions are factorized into the product of two-point functions.
We turn on the interaction (\ref{HISYK}) at time $t_i$ and calculate the changes in energy and entropy at time $t_f$.

\begin{figure}
	\includegraphics[height=0.25\linewidth]{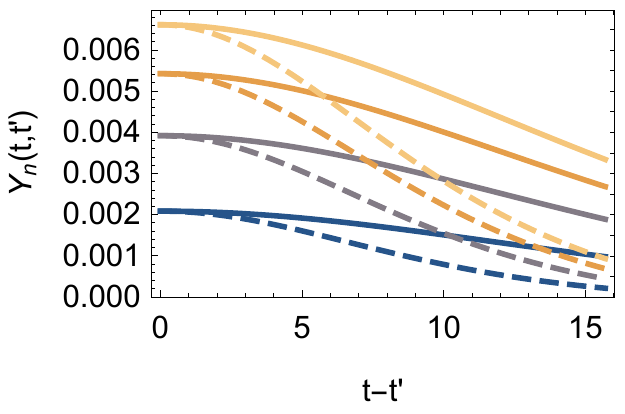}~~~~
	\includegraphics[height=0.25\linewidth]{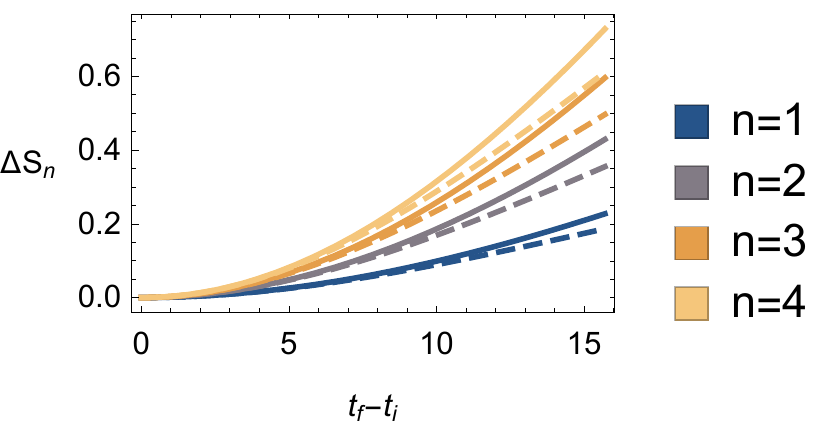}
	\caption{$Y_n(t,t')$ as a function of $t-t'$ and $\Delta S_n$ as a function of $t_f-t_i$ for an initial product state in the two-site SYK model at the large-$q$ limit, where $n=1,2,3,4, \,g=1,\,\beta=2\pi,\,\mathcal J=30,\,q=24$, and $\lambda=2$. Solid (dashed) lines correspond to system $L$ ($R$).}
	\label{FigReKPSYKLargeq}
\end{figure}

For an initial product state in the two-side SYK model, we use the method of Appendix~\ref{SectionProductBH}. The energy-entropy inequalities (\ref{EnergyEntropy}) are saturated. We illustrate the entropy change in Fig.~\ref{FigReKPSYKLargeq}. The kernels of both entropies and energies are positive within the period where approximation (\ref{GLargeq}) is valid. When $\lambda=1$, the increase in energy agrees with the fact that the product of thermal states is a passive state; the increase in entropy agrees with the fact that interactions increase entanglement for an initial product state. When $\lambda>1$, as shown in Fig.~\ref{FigReKPSYKLargeq}, we find that the entropies obtained in the way of the large-$q$ limit increase as well. This agrees with the behaviors of the Renyi entropy in the UV-sensitive region obtained in the way of UV regularization in Fig.~\ref{FigReKP}.

\begin{figure}
	\includegraphics[height=0.25\linewidth]{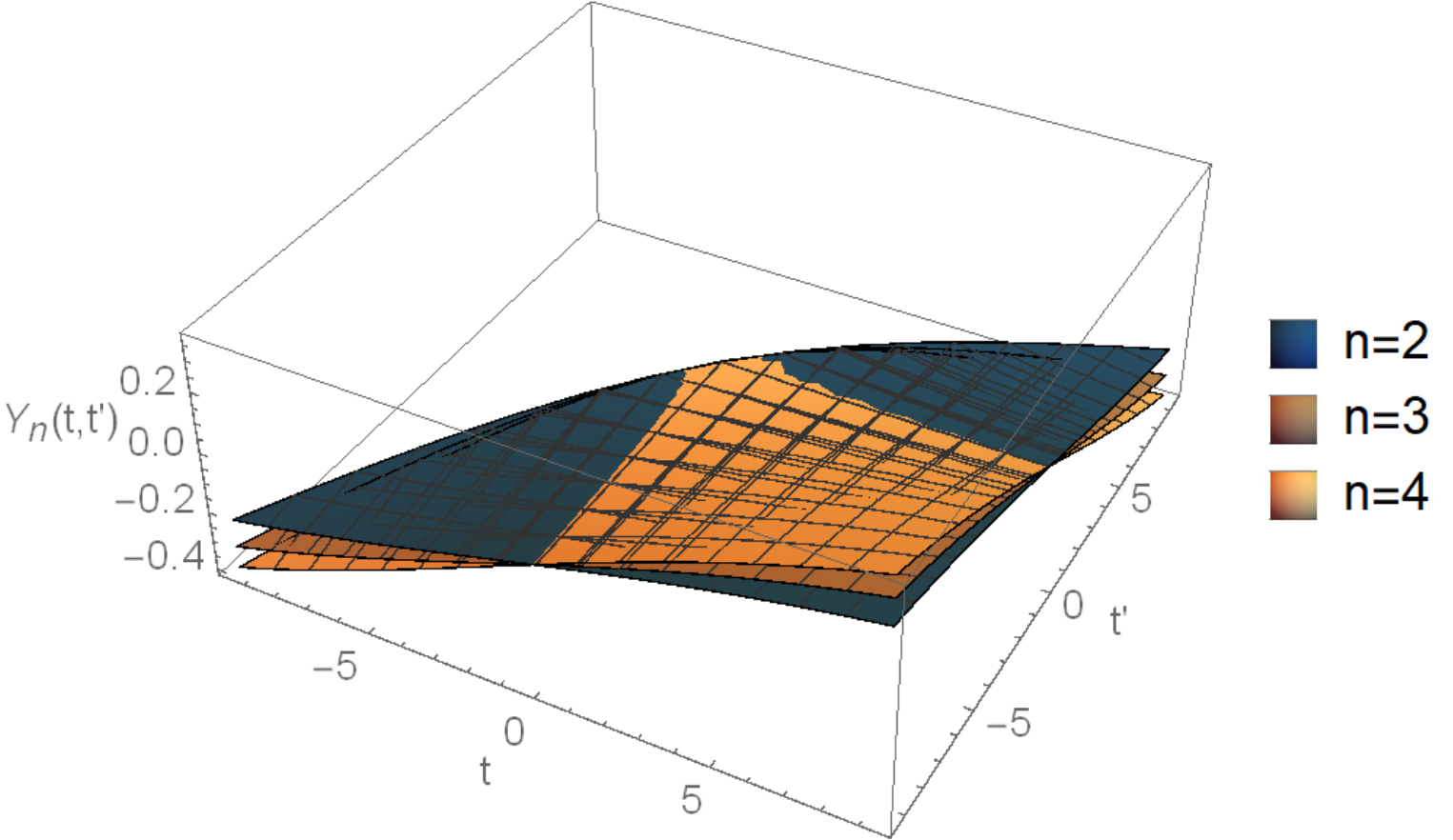}
	\includegraphics[height=0.25\linewidth]{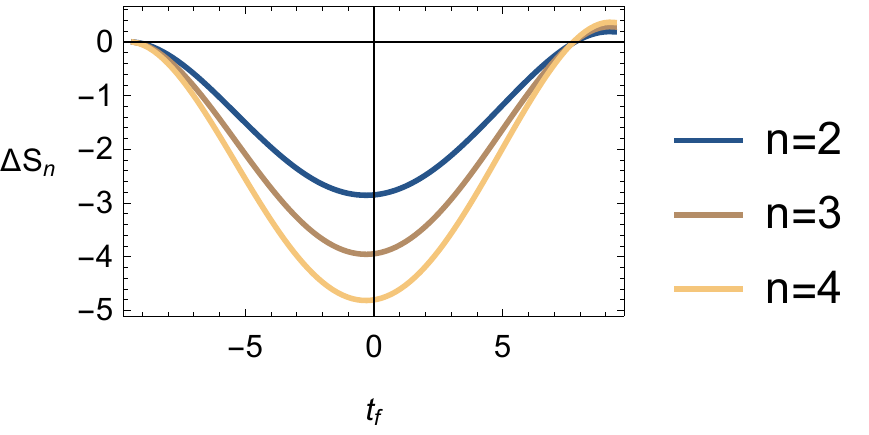}
	\caption{$Y_n(t,t')$ as a function of $\kc{t,t'}$ and $\Delta S_n$ as a function of $t_f$ for an initial TFD state in the two-site SYK model, where $g=1,\,\beta=2\pi,\,\mathcal J=30,\,q=24,\,\lambda=2,\,t_i=-3\pi$, and $n=2,3,4$ (from top to bottom near $t=t'$).}
	\label{FigReKSYKLargeq}
\end{figure}

For an initial TFD state in the two-site SYK model, we use Eqs.~(\ref{EnergySYK}), (\ref{UnfoldSYK}), and (\ref{EntropyKernel}) to calculate the changes in energy and entropy. The energy changes are the same as the case of the product state because of the factorization of the four-point functions, while the entropy changes are different because of the initial entanglement. So the energy-entropy inequalities (\ref{EnergyEntropy}) may not be saturated any more. We illustrate the result in Fig.~\ref{FigReKSYKLargeq}. 

\section{Classical model}\label{SectionClassical}

We will try to discuss the phenomenon analogous to anomalous heat flow in the classical model of \cite{Maldacena:2017axo}. Consider a local Hamiltonian
\begin{align}
H(\ke{x_i},\ke{p_i})=\frac12 \sum_i^N p_i^2+\sum_a^N \kc{\sum_{ij}^N x_i J_{aij} x_j}^2,
\end{align}
where $J_{aij}$ are random couplings satisfying $\avg{J_{aij}^2}=N^{-1}$. Such model is non-integrable and expected to exhibit classical chaos.

Consider two copies of phase space $\kc{x_i^L,p_i^L}$ and $\kc{x_i^R,p_i^R}$ where $i=1,2,...,N$ and their unbalanced Hamiltonian
\begin{align}
H_0=H_L+H_R=H(\ke{x_i^L},\ke{p_i^L})+\lambda H(\ke{x_i^R},\ke{p_i^R}).
\end{align}
Analogous to the TFD state, we prepare an ensemble at $t=0$ each state of which satisfies
\begin{align}
x_i^L=x_i^R,\quad p_i^L=-p_i^R,
\end{align}
and distributes according to the Gibb distribution $\rho(\ke{x_i},\ke{p_i})\sim e^{-\beta H(\ke{x_i},\ke{p_i})}$.
Each state evolved by $H_0$ will satisfy $x_i^L(\lambda t)=x_i^R(-t),\, p_i^L(\lambda t)=-p_i^R(-t)$.

To trigger heat flow, at time $t_i$, we turn on the interaction
\begin{align}
H_I=\sum_{ij} x_i^L g_{ij} x^R_j,
\end{align}
where $g_{ij}$ is random coupling satisfying $g_{ij}=-g_{ji}$ and $\avg{g_{ij}g_{kl}}=N^{-2}g^2(\delta_{ik}\delta_{jl}-\delta_{il}\delta_{jk})$ for $i\neq j, k\neq l$.

Similarly, by using the classical interaction picture, we can estimate the energy change at time $t_f$ with second-order perturbation
\begin{subequations}\begin{align}
	\Delta E_I=&\int_{t_i}^{t_f}dt\,\avg{\avg{\ke{H_I(t_f),H_I(t)}}}=\int_{t_i}^{t_f} dt\, K_I(t_f,t), \\
	K_I(t,t')=&=\frac{g^2}{N^2}\sum_{i\neq j} \avg{\avg{\ke{x_i^L(t)x_j^R(\lambda t),x_i^L(t')x_j^R(\lambda t')}}}- \avg{\avg{\ke{x_i^L(t)x_j^R(\lambda t),x_j^L(t')x_i^R(\lambda t')}}}\\
	=&(K_I(t,t'))_{ijij}+(K_I(t,t'))_{ijji},
	\end{align}\end{subequations}
whose form is the same as Eq.~(\ref{PerturbationAI}). When $t=-t'$ and $\lambda=1$, $(K_I(t,-t))_{ijji}$ can be written as
\begin{align}
(K_I(t,-t))_{ijji}=-\frac{g^2}{2N^2}\sum_{i\neq j}\avg{\avg{\ke{x_i(t)^2,x_j(-t)^2}_P}},
\end{align}
which agrees with the classical limit (\ref{ClassicalLimit}) in JT gravity.

\section{Von Neumann entropy from contour integration}\label{SectionContourIntegral}

According to Eq.~(\ref{AddActionWormhole}), the kernel in Eq.~(\ref{EntropyKernel}) is a summation on the two replicated indexes $\ke{a,b}$. Due to the translational invariance, only the difference $b-a$ is important,
\begin{align}\label{SubKernal}
Y_n(t,t')=\sum_{a,b=0}^{n-1} y^{b-a}_n(t,t')=n\sum_{d=0}^{n-1}y_n^d(t,t'),
\end{align}
where we have used the rotational symmetry of trace
\begin{align}\label{RotationalSymmetry}
y_n^{d}(t,t')=y_n^{d-n}(t,t').
\end{align}
The time orderings of the terms in $y_n^d(t,t')$ may depend on the index $d$. We will split the summation (\ref{SubKernal}) into two parts $y_n^0(t,t')$ and $\sum_{d=1}^{n-1}y_n^d(t,t')$, since the time orderings of the terms in each part are found to be independent from $d$. Then we apply the factorized four-point function to $\avg{VVWW}$ and the eikonal approximation to $\avg{VWVW}$.

The summation on $d$ can be transformed into the contour integral
\begin{align}
\sum_{d=1}^{n-1}y_n^d(t,t') = -i \int_{\mathcal C} dz y_n^{-iz}(t,t') f(z) = -i \int_{\mathcal C'} dz y_n^{-iz}(t,t') f(z),
\end{align}
where $f(z)=1/\kc{e^{2\pi z}-1}$, the contour $\mathcal C$ runs anticlockwise around the poles $z=i,2i,...,(n-1)i$, and the contour $\mathcal C'$ runs clockwise around the other poles and branch cuts in the stripe $\Im z\in(-1,n-1]$. Because of the periodicity $y_n^{-iz}(t,t') f(z)=y_n^{-i(z+in)}(t,t') f(z+in)$, we are free to choose the location of the stripe as long as its width is fixed to be $in$. One can check that these poles and branch cuts surrounded by $\mathcal C'$ are all located in the narrower stripe $\Im z\in(-1,1)$. To have a relatively simple contour integral, we chose $\varDelta=1/2$ in the following analysis.

For those $\avg{VVWW}$ terms, which are factorized, we have an analytical expression of the summation,
\begin{align}\label{SummationVVWW}
\sum_{d=1}^{n-1} \kc{\csc ^2\frac{\pi(d+iu)}{n} }^{1/2} \kc{\csc ^2\frac{\pi(d+iv)}{n} }^{1/2}
=\csch\frac{\pi  u}{n} \csch\frac{\pi  v}{n}+n (\coth\pi u-\coth\pi v) \csch\frac{\pi  (u-v)}{n},
\end{align}
where $|\Im u|,|\Im v|<1$. 
Due to the factor $\frac1{1-n}$ in Renyi entropy (\ref{EntropyKernel}), only the $O(n-1)$ term in the series of Eq.~(\ref{SummationVVWW}) contributes to the von Neumann entropy. Those leading $O(1)$ terms in the series cancel each other out finally.

\begin{figure}
	\centering
	\includegraphics[height=0.25\linewidth]{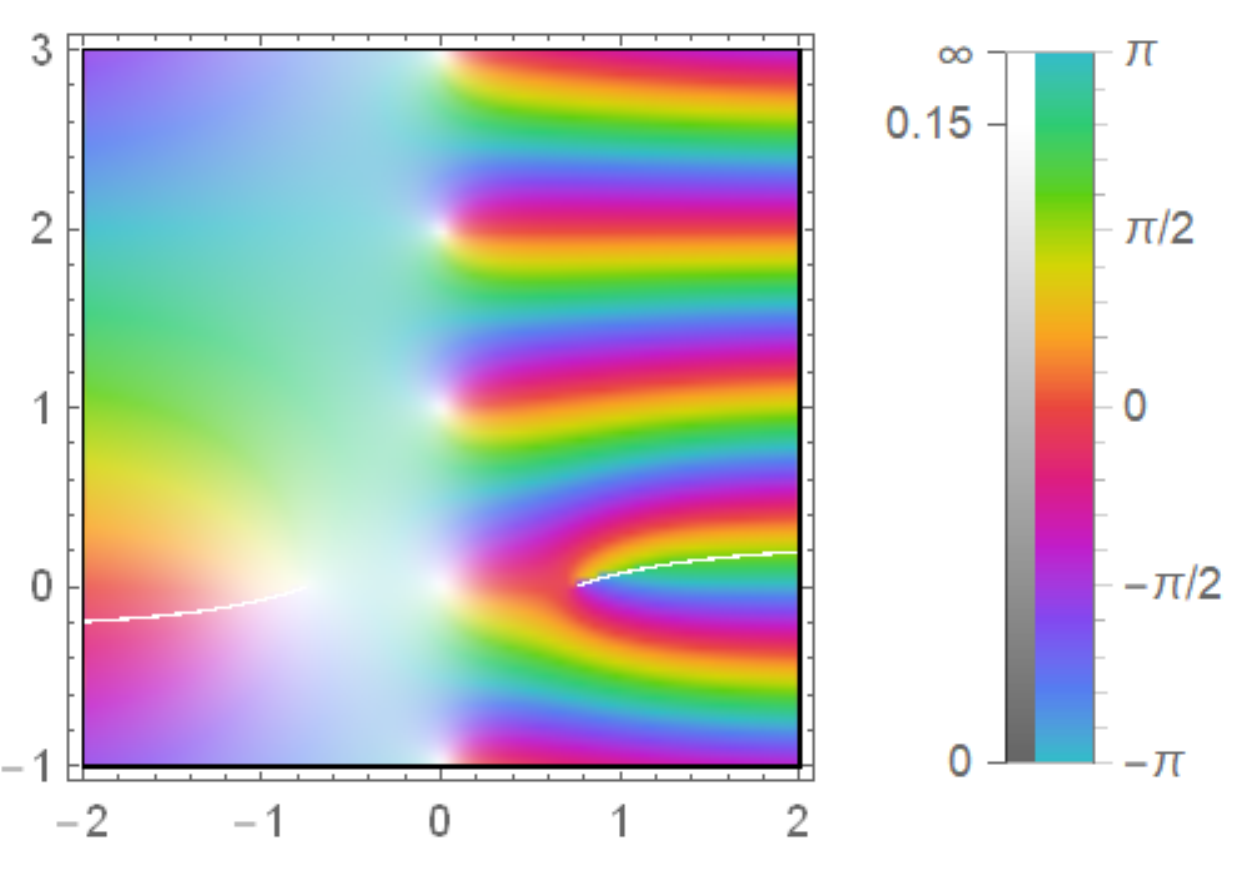}
	\caption{A term $\avg{V((\frac12-iz)\beta+i\lambda t') W(-iz\beta-it') V(\frac12\beta+i\lambda t) W(-it)}_{n\beta}f(z)$ in $y_n^{-iz}(t,t')f(z)$ on the $z$ plane, where $\varDelta=1/2,\,C=1,\,\beta=2\pi,\,\lambda=1,\,t=\pi,\,t'=\pi/2$, and $n=4$.}
	\label{FigComplexPlot}
\end{figure}

For those $\avg{VWVW}$ terms, we encounter the branch cut of the confluent hypergeometric function $U(1,1,\xi)$ on the negative axis $\xi<0$, which is mapped to two branch cuts in the $z$ plane, as illustrated in Fig.~\ref{FigComplexPlot}. The contour $\mathcal C'$ runs around these two branch cuts and the pole at the origin. We use $U(1,1,\xi+i0^+)-U(1,1,\xi+i0^-)=-2i\pi e^\xi$ for $\xi<0$. However, due to the complicated dependence in Eq.~(\ref{EikonalCorrect}), we have to calculate the integrals numerically and then take the $n\to1$ limit. Alternatively, it is more convenient to expand those four-point functions about $n=1$ first and then calculate the contour integrals. The leading $O(1)$ terms of the series have a period $i$. So the contour integrals around the branch cuts and the pole at the origin cancel. The $O(n-1)$ terms are not periodic. So the contour integrals around the branch cuts and the pole at the origin do not cancel. They contribute to the von Neumann entropy. We encounter $U(2,2,\xi)$ and its two branch cuts in the stripe $\Im z\in(-1,1)$. The treatment on the contour integrals around the branch cuts is similar.

\begin{thebibliography}{99}
	
	%%%%%%%%%%%%%%%%%%%%%%%%%%%%%%%%%%%%%%%%%%%%%%%%%%%%%%%%%%%%%    QT
	
	\bibitem{QT:book}
	J. Gemmer, M. Michel, G. Mahler, 
	``Quantum thermodynamics: emergence of thermodynamic behavior within composite quantum systems,'' 
	2nd edition, Springer, Heidelberg, New York, 2009;
	G. Mahler, 
	``Quantum Thermodynamic Processes: energy and information flow at the nanoscale,'' 
	Pan Stanford Publications, Singapore, 2015;
	%C. Gogolin, J. Eisert, Equilibration, thermalisation, and the emergence of statistical mechanics in closed quantum systems, Rep. Prog. Phys. 79 (2016) 056001.
	F. Binder, et al. 
	``Thermodynamics in the quantum regime,''
	Fundamental Theories of Physics, Springer, 2019.
	
	\bibitem{SecondLaw}
	%\bibitem{Kondepudi:1998}
	D. Kondepudi, I. Prigogine, 
	``Modern Thermodynamics: From Heat Engines to Dissipative Structures,'' 
	(Wiley, New York, 1998);
	%\bibitem{Sagawa:2012}
	T. Sagawa, 
	``Second law-like inequalities with quantum relative entropy: an introduction,'' 
	in: Lectures on Quantum Computing, Thermodynamics and Statistical Physics, WORLD SCIENTIFIC, 2012: pp. 125–190. 
	arXiv:1202.0983 [cond-mat].
	%doi:10.1142/9789814425193_0003.
	
	\bibitem{Partovi:2008}
	M. Hossein Partovi, 
	``Verschraenkung versus Stosszahlansatz: Disappearance of the Thermodynamic Arrow in a High-Correlation Environment,'' 
	Physical Review E 77, no. 2 (February 11, 2008). 
	%https://doi.org/10.1103/PhysRevE.77.021110.
	
	\bibitem{Jennings:2010}
	Jennings, David, and Terry Rudolph, 
	``Entanglement and the Thermodynamic Arrow of Time.'' 
	Physical Review E 81, no. 6 (June 23, 2010). 
	%https://doi.org/10.1103/PhysRevE.81.061130.
	
	\bibitem{Lloyd:2015}
	S.~Lloyd, Z.-W.~Liu, S.~Pirandola, V.~Chiloyan, Y.~Hu, S.~Huberman, G.~Chen, ``No energy transport without discord,'' 
	arXiv:1510.05035 [quant-ph];
	%http://arxiv.org/abs/1510.05035 (accessed June 13, 2019).
	%\cite{Henao:2018}
	%\bibitem{Henao:2018}
	I.~Henao and M.~S.~Roberto 
	``Role of Quantum Coherence in the Thermodynamics of Energy Transfer,''
	Physical Review E 97, no. 6 (June 4, 2018). 
	%https://doi.org/10.1103/PhysRevE.97.062105.
	
	\bibitem{Micadei:2017}
	K.~Micadei,  J.~P.~S.~Peterson, A.~M.~Souza, et al,
	``Reversing the direction of heat flow using quantum correlations,'' 
	Nat Commun 10, 2456 (2019) 
	%doi:10.1038/s41467-019-10333-7
	[arXiv:1711.03323 [Cond-Mat, Physics:Quant-Ph]]
	%, November 9, 2017. 
	%http://arxiv.org/abs/1711.03323.
	
	%\cite{Lesovik:2019klx}
	\bibitem{Lesovik:2019klx} 
	G.~B.~Lesovik, I.~A.~Sadovskyy, M.~V.~Suslov, A.~V.~Lebedev and V.~M.~Vinokur,
	``Arrow of time and its reversal on the IBM quantum computer,''
	Sci.\ Rep.\  {\bf 9}, no. 1, 4396 (2019).
	%doi:10.1038/s41598-019-40765-6
	%%CITATION = doi:10.1038/s41598-019-40765-6;%%
	
	%\cite{Bera:2017}
	\bibitem{Bera:2017}
	M.~N.~Bera, A.~Riera, M.~Lewenstein and A.~Winter, 
	``Generalized laws of thermodynamics in the presence of correlations,''
	Nature Communications. 8 (2017) 2180. 
	%doi:10.1038/s41467-017-02370-x.
	
	%\cite{Vitagliano:2018}
	\bibitem{Vitagliano:2018}
	G.~Vitagliano, C.~Klöckl, M.~Huber and N.~Friis, 
	``Trade-off Between Work and Correlations in Quantum Thermodynamics,''
	arXiv:1803.06884 [quant-ph].
	%(2018). http://arxiv.org/abs/1803.06884 (accessed November 3, 2018).
	
	\bibitem{PassiveState}
	%\bibitem{Haag:1974}
	R. Haag, D. Kastler, E.B. Trych-Pohlmeyer, 
	``Stability and equilibrium states'', 
	Commun.Math. Phys. 38 (1974) 173–193;
	%\bibitem{Lenard:1978}
	A. Lenard, 
	``Thermodynamical proof of the Gibbs formula for elementary quantum systems,'' 
	J Stat Phys. 19 (1978) 575–586; 
	%doi:10.1007/BF01011769.
	%\bibitem{Pusz:1978}
	W. Pusz, S.L. Woronowicz, 
	``Passive states and KMS states for general quantum systems,''
	Commun.Math. Phys. 58 (1978) 273–290. 
	%doi:10.1007/BF01614224.
	
	%\cite{Llobet:2015}
	\bibitem{Llobet:2015}
	M.~Perarnau-Llobet, K.~V.~Hovhannisyan, M.~Huber, P.~Skrzypczyk, N.~Brunner and A.~Acín, 
	``Extractable Work from Correlations,''
	Phys. Rev. X. 5 (2015) 041011. 
	%doi:10.1103/PhysRevX.5.041011.
	
	\bibitem{Brandao:2013}
	F. Brandão, M. Horodecki, N. Ng, J. Oppenheim, S. Wehner, 
	``The second laws of quantum thermodynamics,'' 
	Proceedings of the National Academy of Sciences. 112 (2015) 3275–3279. 
	%https://doi.org/10.1073/pnas.1411728112.
	arXiv:1305.5278 [quant-ph].
	
	\bibitem{Alipour:2016}
	S. Alipour, F. Benatti, F. Bakhshinezhad, M. Afsary, S. Marcantoni, A.T. Rezakhani, 
	``Correlations in quantum thermodynamics: Heat, work, and entropy production,'' 
	Sci Rep. 6 (2016) 35568. 
	%https://doi.org/10.1038/srep35568.
	arXiv:1606.08869 [quant-ph]
	
	%\cite{Huber:2015}
	\bibitem{Huber:2015}
	M.~Huber, M.~Perarnau-Llobet, K.~V.~Hovhannisyan, P.~Skrzypczyk, C.~Klöckl, N.~Brunner and A.~Acín, 
	``Thermodynamic cost of creating correlations,'' 
	New J. Phys. 17 (2015) 065008. 
	%doi:10.1088/1367-2630/17/6/065008.
	
	%\cite{Bruschi:2015}
	\bibitem{Bruschi:2015}
	D.~E.~Bruschi, M.~Perarnau-Llobet, N.~Friis, K.~V.~Hovhannisyan, M.~Huber, 
	``The thermodynamics of creating correlations: Limitations and optimal protocols,''
	Physical Review E. 91 (2015) 032118. 
	%doi:10.1103/PhysRevE.91.032118.
	
	%\cite{Friis:2016}
	\bibitem{Friis:2016}
	N.~Friis, M.~Huber, M.~Perarnau-Llobet, 
	``Energetics of correlations in interacting systems,'' 
	Phys. Rev. E. 93 (2016) 042135. 
	%doi:10.1103/PhysRevE.93.042135.
	
	
	%%%%%%%%%%%%%%%%%%%%%%%%%%%%%%%%%%%%% AdS/CFT
	
	%\cite{Susskind:1994vu}
	\bibitem{Susskind:1994vu} 
	L.~Susskind,
	``The World as a hologram,''
	J.\ Math.\ Phys.\  {\bf 36}, 6377 (1995)
	%doi:10.1063/1.531249
	[hep-th/9409089].
	%%CITATION = doi:10.1063/1.531249;%%
	%2671 citations counted in INSPIRE as of 01 Jun 2019
	
	\bibitem{AdSCFT}
	%\cite{Maldacena:1997re}
	%\bibitem{Maldacena:1997re} 
	J.~M.~Maldacena,
	``The Large N limit of superconformal field theories and supergravity,''
	Int.\ J.\ Theor.\ Phys.\  {\bf 38}, 1113 (1999)
	[Adv.\ Theor.\ Math.\ Phys.\  {\bf 2}, 231 (1998)]
	%doi:10.1023/A:1026654312961, 10.4310/ATMP.1998.v2.n2.a1
	[hep-th/9711200];
	%%CITATION = doi:10.1023/A:1026654312961, 10.4310/ATMP.1998.v2.n2.a1;%%
	%14641 citations counted in INSPIRE as of 01 Jun 2019
	%\cite{Witten:1998qj}
	%\bibitem{Witten:1998qj} 
	E.~Witten,
	``Anti-de Sitter space and holography,''
	Adv.\ Theor.\ Math.\ Phys.\  {\bf 2}, 253 (1998)
	%doi:10.4310/ATMP.1998.v2.n2.a2
	[hep-th/9802150];
	%%CITATION = doi:10.4310/ATMP.1998.v2.n2.a2;%%
	%9495 citations counted in INSPIRE as of 01 Jun 2019
	%\cite{Gubser:1998bc}
	%\bibitem{Gubser:1998bc} 
	S.~S.~Gubser, I.~R.~Klebanov and A.~M.~Polyakov,
	``Gauge theory correlators from noncritical string theory,''
	Phys.\ Lett.\ B {\bf 428}, 105 (1998)
	%doi:10.1016/S0370-2693(98)00377-3
	[hep-th/9802109].
	%%CITATION = doi:10.1016/S0370-2693(98)00377-3;%%
	%8109 citations counted in INSPIRE as of 01 Jun 2019
	
	
	%%%%%%%%%%%%%%%%%%%%%%%%%%%%%%%%%%%% EE
	
	\bibitem{HEE}
	%\cite{Ryu:2006bv}
	%\bibitem{Ryu:2006bv} 
	S.~Ryu and T.~Takayanagi,
	``Holographic derivation of entanglement entropy from AdS/CFT,''
	Phys.\ Rev.\ Lett.\  {\bf 96}, 181602 (2006)
	%doi:10.1103/PhysRevLett.96.181602
	[hep-th/0603001];
	%%CITATION = doi:10.1103/PhysRevLett.96.181602;%%
	%1924 citations counted in INSPIRE as of 11 Jun 2019
	%\cite{Faulkner:2013ana}
	%\bibitem{Faulkner:2013ana} 
	T.~Faulkner, A.~Lewkowycz and J.~Maldacena,
	``Quantum corrections to holographic entanglement entropy,''
	JHEP {\bf 1311}, 074 (2013)
	%doi:10.1007/JHEP11(2013)074
	[arXiv:1307.2892 [hep-th]];
	%%CITATION = doi:10.1007/JHEP11(2013)074;%%
	%304 citations counted in INSPIRE as of 01 Nov 2019
	%\cite{Lewkowycz:2013nqa}
	%\bibitem{Lewkowycz:2013nqa} 
	A.~Lewkowycz and J.~Maldacena,
	``Generalized gravitational entropy,''
	JHEP {\bf 1308}, 090 (2013)
	%doi:10.1007/JHEP08(2013)090
	[arXiv:1304.4926 [hep-th]].
	%%CITATION = doi:10.1007/JHEP08(2013)090;%%
	%508 citations counted in INSPIRE as of 01 Nov 2019
	
	%%%%%%%%%%%%%%%%%%%%%%%%%%%%%%%%%%%%%%%%%%%%% BH Information & ER=EPR
	
	%\cite{Polchinski:2016hrw}
	\bibitem{Polchinski:2016hrw} 
	J.~Polchinski,
	``The Black Hole Information Problem,''
	%doi:10.1142/9789813149441_0006
	arXiv:1609.04036 [hep-th].
	%%CITATION = doi:10.1142/9789813149441_0006;%%
	%49 citations counted in INSPIRE as of 18 Oct 2019
	
	%\cite{Harlow:2014yka}
	\bibitem{Harlow:2014yka} 
	D.~Harlow,
	``Jerusalem Lectures on Black Holes and Quantum Information,''
	Rev.\ Mod.\ Phys.\  {\bf 88}, 015002 (2016)
	%doi:10.1103/RevModPhys.88.015002
	[arXiv:1409.1231 [hep-th]].
	%%CITATION = doi:10.1103/RevModPhys.88.015002;%%
	%140 citations counted in INSPIRE as of 01 Jun 2019
	
	%\cite{Page:1993df}
	\bibitem{Page:1993df} 
	D.~N.~Page,
	``Average entropy of a subsystem,''
	Phys.\ Rev.\ Lett.\  {\bf 71}, 1291 (1993)
	%doi:10.1103/PhysRevLett.71.1291
	[gr-qc/9305007].
	%%CITATION = doi:10.1103/PhysRevLett.71.1291;%%
	%413 citations counted in INSPIRE as of 11 Oct 2019
	
	%\cite{Page:2013dx}
	\bibitem{Page:2013dx} 
	D.~N.~Page,
	``Time Dependence of Hawking Radiation Entropy,''
	JCAP {\bf 1309}, 028 (2013)
	%doi:10.1088/1475-7516/2013/09/028
	[arXiv:1301.4995 [hep-th]].
	%%CITATION = doi:10.1088/1475-7516/2013/09/028;%%
	%67 citations counted in INSPIRE as of 11 Oct 2019
	
	%\cite{Israel:1976ur}
	\bibitem{Israel:1976ur} 
	W.~Israel,
	``Thermo field dynamics of black holes,''
	Phys.\ Lett.\ A {\bf 57}, 107 (1976).
	%doi:10.1016/0375-9601(76)90178-X
	%%CITATION = doi:10.1016/0375-9601(76)90178-X;%%
	%433 citations counted in INSPIRE as of 31 Oct 2019
	
	%\cite{Susskind:1993if}
	\bibitem{Susskind:1993if} 
	L.~Susskind, L.~Thorlacius and J.~Uglum,
	``The Stretched horizon and black hole complementarity,''
	Phys.\ Rev.\ D {\bf 48}, 3743 (1993)
	%doi:10.1103/PhysRevD.48.3743
	[hep-th/9306069].
	%%CITATION = doi:10.1103/PhysRevD.48.3743;%%
	%847 citations counted in INSPIRE as of 18 Oct 2019
	
	%\cite{Maldacena:2001kr}
	\bibitem{Maldacena:2001kr} 
	J.~M.~Maldacena,
	``Eternal black holes in anti-de Sitter,''
	JHEP {\bf 0304}, 021 (2003)
	%doi:10.1088/1126-6708/2003/04/021
	[hep-th/0106112].
	%%CITATION = doi:10.1088/1126-6708/2003/04/021;%%
	%753 citations counted in INSPIRE as of 01 Apr 2019
	
	\bibitem{VanRaamsdonk}
	%\cite{VanRaamsdonk:2010pw}
	%\bibitem{VanRaamsdonk:2010pw} 
	M.~Van Raamsdonk,
	``Building up spacetime with quantum entanglement,''
	Gen.\ Rel.\ Grav.\  {\bf 42}, 2323 (2010)
	[Int.\ J.\ Mod.\ Phys.\ D {\bf 19}, 2429 (2010)]
	%doi:10.1007/s10714-010-1034-0, 10.1142/S0218271810018529
	[arXiv:1005.3035 [hep-th]];
	%%CITATION = doi:10.1007/s10714-010-1034-0, 10.1142/S0218271810018529;%%
	%536 citations counted in INSPIRE as of 31 Oct 2019
	%\cite{VanRaamsdonk:2016exw}
	%\bibitem{VanRaamsdonk:2016exw} 
	M.~Van Raamsdonk,
	``Lectures on Gravity and Entanglement,''
	%doi:10.1142/9789813149441_0005
	arXiv:1609.00026 [hep-th].
	%%CITATION = doi:10.1142/9789813149441_0005;%%
	%48 citations counted in INSPIRE as of 31 Oct 2019
	
	\bibitem{Firewall}
	%\cite{Almheiri:2012rt}
	%\bibitem{Almheiri:2012rt} 
	A.~Almheiri, D.~Marolf, J.~Polchinski and J.~Sully,
	``Black Holes: Complementarity or Firewalls?,''
	JHEP {\bf 1302}, 062 (2013)
	%doi:10.1007/JHEP02(2013)062
	[arXiv:1207.3123 [hep-th]];
	%%CITATION = doi:10.1007/JHEP02(2013)062;%%
	%920 citations counted in INSPIRE as of 18 Oct 2019
	%\cite{Almheiri:2013hfa}
	%\bibitem{Almheiri:2013hfa} 
	A.~Almheiri, D.~Marolf, J.~Polchinski, D.~Stanford and J.~Sully,
	``An Apologia for Firewalls,''
	JHEP {\bf 1309}, 018 (2013)
	%doi:10.1007/JHEP09(2013)018
	[arXiv:1304.6483 [hep-th]].
	%%CITATION = doi:10.1007/JHEP09(2013)018;%%
	%328 citations counted in INSPIRE as of 18 Oct 2019
	
	%\cite{Maldacena:2013xja}
	\bibitem{Maldacena:2013xja} 
	J.~Maldacena and L.~Susskind,
	``Cool horizons for entangled black holes,''
	Fortsch.\ Phys.\  {\bf 61}, 781 (2013)
	%doi:10.1002/prop.201300020
	[arXiv:1306.0533 [hep-th]].
	%%CITATION = doi:10.1002/prop.201300020;%%
	%577 citations counted in INSPIRE as of 11 Jun 2019
	
	%%%%%%%%%%%%%%%%%%%%%%%%%%%%%%%%%%%%%%%%%%%%%%%%%% Chaos
	
	\bibitem{Chaos}
	%\cite{Sfetsos:1994xa}
	%\bibitem{Sfetsos:1994xa} 
	K.~Sfetsos,
	``On gravitational shock waves in curved space-times,''
	Nucl.\ Phys.\ B {\bf 436}, 721 (1995)
	%doi:10.1016/0550-3213(94)00573-W
	[hep-th/9408169];
	%%CITATION = doi:10.1016/0550-3213(94)00573-W;%%
	%100 citations counted in INSPIRE as of 01 Nov 2019
	%\cite{Sekino:2008he}
	%\bibitem{Sekino:2008he} 
	Y.~Sekino and L.~Susskind,
	``Fast Scramblers,''
	JHEP {\bf 0810}, 065 (2008)
	%doi:10.1088/1126-6708/2008/10/065
	[arXiv:0808.2096 [hep-th]];
	%%CITATION = doi:10.1088/1126-6708/2008/10/065;%%
	%460 citations counted in INSPIRE as of 21 Oct 2019
	%\cite{Shenker:2013pqa}
	%\bibitem{Shenker:2013pqa} 
	S.~H.~Shenker and D.~Stanford,
	``Black holes and the butterfly effect,''
	JHEP {\bf 1403}, 067 (2014)
	%doi:10.1007/JHEP03(2014)067
	[arXiv:1306.0622 [hep-th]].
	%%CITATION = doi:10.1007/JHEP03(2014)067;%%
	%523 citations counted in INSPIRE as of 21 Oct 2019
	%\cite{Shenker:2013yza}
	%\bibitem{Shenker:2013yza} 
	S.~H.~Shenker and D.~Stanford,
	``Multiple Shocks,''
	JHEP {\bf 1412}, 046 (2014)
	%doi:10.1007/JHEP12(2014)046
	[arXiv:1312.3296 [hep-th]];
	%%CITATION = doi:10.1007/JHEP12(2014)046;%%
	%231 citations counted in INSPIRE as of 01 Nov 2019
	%\cite{Roberts:2014ifa}
	%\bibitem{Roberts:2014ifa} 
	D.~A.~Roberts and D.~Stanford,
	``Two-dimensional conformal field theory and the butterfly effect,''
	Phys.\ Rev.\ Lett.\  {\bf 115}, no. 13, 131603 (2015)
	%doi:10.1103/PhysRevLett.115.131603
	[arXiv:1412.5123 [hep-th]];
	%%CITATION = doi:10.1103/PhysRevLett.115.131603;%%
	%192 citations counted in INSPIRE as of 01 Nov 2019
	%\bibitem{Kitaev:2015hch}
	A.~Kitaev. 
	``Hidden correlations in the Hawking radiation and thermal noise,''
	2015. KITP seminar, Feb. 12. 
	http://online.kitp.ucsb.edu/online/joint98/kitaev;
	%\cite{Maldacena:2015waa}
	%\bibitem{Maldacena:2015waa} 
	J.~Maldacena, S.~H.~Shenker and D.~Stanford,
	``A bound on chaos,''
	JHEP {\bf 1608}, 106 (2016)
	%doi:10.1007/JHEP08(2016)106
	[arXiv:1503.01409 [hep-th]].
	%%CITATION = doi:10.1007/JHEP08(2016)106;%%
	%673 citations counted in INSPIRE as of 21 Oct 2019
	
	%\cite{Shenker:2014cwa}
	\bibitem{Shenker:2014cwa} 
	S.~H.~Shenker and D.~Stanford,
	``Stringy effects in scrambling,''
	JHEP {\bf 1505}, 132 (2015)
	%doi:10.1007/JHEP05(2015)132
	[arXiv:1412.6087 [hep-th]].
	%%CITATION = doi:10.1007/JHEP05(2015)132;%%
	%235 citations counted in INSPIRE as of 16 Oct 2019
	
	%%%%%%%%%%%%%%%%%%%%%%%%%%%%%%%%%%%%%%%%%%% wormhole
	
	%\cite{Hayden:2007cs}
	\bibitem{Hayden:2007cs} 
	P.~Hayden and J.~Preskill,
	``Black holes as mirrors: Quantum information in random subsystems,''
	JHEP {\bf 0709}, 120 (2007)
	%doi:10.1088/1126-6708/2007/09/120
	[arXiv:0708.4025 [hep-th]].
	%%CITATION = doi:10.1088/1126-6708/2007/09/120;%%
	%412 citations counted in INSPIRE as of 21 Oct 2019
	
	%\cite{Gao:2016bin}
	\bibitem{Gao:2016bin} 
	P.~Gao, D.~L.~Jafferis and A.~Wall,
	``Traversable Wormholes via a Double Trace Deformation,''
	JHEP {\bf 1712}, 151 (2017)
	%doi:10.1007/JHEP12(2017)151
	[arXiv:1608.05687 [hep-th]].
	%%CITATION = doi:10.1007/JHEP12(2017)151;%%
	%82 citations counted in INSPIRE as of 01 Apr 2019
	
	%\cite{Maldacena:2017axo}
	\bibitem{Maldacena:2017axo} 
	J.~Maldacena, D.~Stanford and Z.~Yang,
	``Diving into traversable wormholes,''
	Fortsch.\ Phys.\  {\bf 65}, no. 5, 1700034 (2017)
	%doi:10.1002/prop.201700034
	[arXiv:1704.05333 [hep-th]].
	%%CITATION = doi:10.1002/prop.201700034;%%
	%74 citations counted in INSPIRE as of 01 Apr 2019
	
	%\cite{Gao:2018yzk}
	\bibitem{Gao:2018yzk} 
	P.~Gao and H.~Liu,
	``Regenesis and quantum traversable wormholes,''
	JHEP {\bf 1910}, 048 (2019)
	%doi:10.1007/JHEP10(2019)048
	[arXiv:1810.01444 [hep-th]].
	%%CITATION = doi:10.1007/JHEP10(2019)048;%%
	%10 citations counted in INSPIRE as of 24 Oct 2019
	
	%\cite{Susskind:2017nto}
	\bibitem{Susskind:2017nto} 
	L.~Susskind and Y.~Zhao,
	``Teleportation through the wormhole,''
	Phys.\ Rev.\ D {\bf 98}, no. 4, 046016 (2018)
	%doi:10.1103/PhysRevD.98.046016
	[arXiv:1707.04354 [hep-th]].
	%%CITATION = doi:10.1103/PhysRevD.98.046016;%%
	%18 citations counted in INSPIRE as of 21 Oct 2019
	
	%\cite{Maldacena:2018lmt}
	\bibitem{Maldacena:2018lmt} 
	J.~Maldacena and X.~L.~Qi,
	``Eternal traversable wormhole,''
	arXiv:1804.00491 [hep-th].
	%%CITATION = ARXIV:1804.00491;%%
	%40 citations counted in INSPIRE as of 01 Apr 2019
	
	%\cite{Garcia-Garcia:2019poj}
	\bibitem{Garcia-Garcia:2019poj} 
	A.~M.~García-García, T.~Nosaka, D.~Rosa and J.~J.~M.~Verbaarschot,
	``Quantum chaos transition in a two-site Sachdev-Ye-Kitaev model dual to an eternal traversable wormhole,''
	Phys.\ Rev.\ D {\bf 100}, no. 2, 026002 (2019)
	%doi:10.1103/PhysRevD.100.026002
	[arXiv:1901.06031 [hep-th]].
	%%CITATION = doi:10.1103/PhysRevD.100.026002;%%
	%15 citations counted in INSPIRE as of 23 Oct 2019
	
	%\cite{Chen:2019qqe}
	\bibitem{Chen:2019qqe} 
	Y.~Chen and P.~Zhang,
	``Entanglement Entropy of Two Coupled SYK Models and Eternal Traversable Wormhole,''
	JHEP {\bf 1907}, 033 (2019)
	%doi:10.1007/JHEP07(2019)033
	[arXiv:1903.10532 [hep-th]].
	%%CITATION = doi:10.1007/JHEP07(2019)033;%%
	%3 citations counted in INSPIRE as of 21 Oct 2019
	
	%%%%%%%%%%%%%%%%%%%%%%%%%%%%%%%%%%%%%%%%%%%%%%%%%%%%%%% SYK AdS2
	
	\bibitem{SYK}
	%\bibitem{kitaevtalk2015} 
	A. Kitaev, ``A simple model of quantum holography",
	Talks at KITP on April 7, 2015 and May 27, 2015.
	http://online.kitp.ucsb.edu/online/entangled15/kitaev/,
	http://online.kitp.ucsb.edu/online/entangled15/kitaev2;
	%\cite{Sachdev:1992fk}
	%\bibitem{Sachdev:1992fk}
	S.~Sachdev and J.~Ye,
	``Gapless spin fluid ground state in a random, quantum Heisenberg magnet,''
	Phys.\ Rev.\ Lett.\  {\bf 70}, 3339 (1993)
	%doi:10.1103/PhysRevLett.70.3339
	[cond-mat/9212030];
	%%CITATION = %doi:10.1103/PhysRevLett.70.3339;%%
	%172 citations counted in INSPIRE as of 25 Aug 2017
	%gu_SYK chain
	%\cite{Sachdev:2015efa}
	%\bibitem{Sachdev:2015efa} 
	S.~Sachdev,
	``Bekenstein-Hawking Entropy and Strange Metals,''
	Phys.\ Rev.\ X {\bf 5}, no. 4, 041025 (2015)
	%doi:10.1103/PhysRevX.5.041025
	[arXiv:1506.05111 [hep-th]];
	%%CITATION = doi:10.1103/PhysRevX.5.041025;%%
	%200 citations counted in INSPIRE as of 19 Sep 2019
	
	
	%\cite{Maldacena:2016hyu}
	\bibitem{Maldacena:2016hyu} 
	J.~Maldacena and D.~Stanford,
	``Remarks on the Sachdev-Ye-Kitaev model,''
	Phys.\ Rev.\ D {\bf 94}, no. 10, 106002 (2016)
	%doi:10.1103/PhysRevD.94.106002
	[arXiv:1604.07818 [hep-th]].
	%%CITATION = doi:10.1103/PhysRevD.94.106002;%%
	%568 citations counted in INSPIRE as of 17 Sep 2019
	
	%\cite{Kitaev:2017awl}
	\bibitem{Kitaev:2017awl} 
	A.~Kitaev and S.~J.~Suh,
	``The soft mode in the Sachdev-Ye-Kitaev model and its gravity dual,''
	JHEP {\bf 1805}, 183 (2018)
	%doi:10.1007/JHEP05(2018)183
	[arXiv:1711.08467 [hep-th]].
	%%CITATION = doi:10.1007/JHEP05(2018)183;%%
	%137 citations counted in INSPIRE as of 19 Sep 2019
	
	%\cite{Gu:2017njx}
	\bibitem{Gu:2017njx} 
	Y.~Gu, A.~Lucas and X.~L.~Qi,
	``Spread of entanglement in a Sachdev-Ye-Kitaev chain,''
	JHEP {\bf 1709}, 120 (2017)
	%doi:10.1007/JHEP09(2017)120
	[arXiv:1708.00871 [hep-th]].
	%%CITATION = doi:10.1007/JHEP09(2017)120;%%
	%23 citations counted in INSPIRE as of 05 Apr 2019
	
	%\cite{Goel:2018ubv}
	\bibitem{Goel:2018ubv} 
	A.~Goel, H.~T.~Lam, G.~J.~Turiaci and H.~Verlinde,
	``Expanding the Black Hole Interior: Partially Entangled Thermal States in SYK,''
	JHEP {\bf 1902}, 156 (2019)
	%doi:10.1007/JHEP02(2019)156
	[arXiv:1807.03916 [hep-th]].
	%%CITATION = doi:10.1007/JHEP02(2019)156;%%
	%17 citations counted in INSPIRE as of 07 Nov 2019
	
	%\cite{Cottrell:2018ash}
	\bibitem{Cottrell:2018ash} 
	W.~Cottrell, B.~Freivogel, D.~M.~Hofman and S.~F.~Lokhande,
	``How to Build the Thermofield Double State,''
	JHEP {\bf 1902}, 058 (2019)
	%doi:10.1007/JHEP02(2019)058
	[arXiv:1811.11528 [hep-th]].
	%%CITATION = doi:10.1007/JHEP02(2019)058;%%
	%13 citations counted in INSPIRE as of 24 Oct 2019
	
	%%%%%%%%%%%%%%%%%%%%%%%%%%%% JT
	
	\bibitem{JT}
	%\cite{Almheiri:2014cka}
	%\bibitem{Almheiri:2014cka} 
	A.~Almheiri and J.~Polchinski,
	``Models of AdS$_{2}$ backreaction and holography,''
	JHEP {\bf 1511}, 014 (2015)
	%doi:10.1007/JHEP11(2015)014
	[arXiv:1402.6334 [hep-th]];
	%%CITATION = doi:10.1007/JHEP11(2015)014;%%
	%217 citations counted in INSPIRE as of 24 Oct 2019
	%\cite{Jensen:2016pah}
	%\bibitem{Jensen:2016pah} 
	K.~Jensen,
	``Chaos in AdS$_2$ Holography,''
	Phys.\ Rev.\ Lett.\  {\bf 117}, no. 11, 111601 (2016)
	%doi:10.1103/PhysRevLett.117.111601
	[arXiv:1605.06098 [hep-th]];
	%%CITATION = doi:10.1103/PhysRevLett.117.111601;%%
	%266 citations counted in INSPIRE as of 24 Oct 2019
	%\cite{Engelsoy:2016xyb}
	%\bibitem{Engelsoy:2016xyb} 
	J.~Engelsöy, T.~G.~Mertens and H.~Verlinde,
	``An investigation of AdS$_{2}$ backreaction and holography,''
	JHEP {\bf 1607}, 139 (2016)
	%doi:10.1007/JHEP07(2016)139
	[arXiv:1606.03438 [hep-th]];
	%%CITATION = doi:10.1007/JHEP07(2016)139;%%
	%205 citations counted in INSPIRE as of 24 Oct 2019
	
	%\cite{Maldacena:2016upp}
	\bibitem{Maldacena:2016upp} 
	J.~Maldacena, D.~Stanford and Z.~Yang,
	``Conformal symmetry and its breaking in two dimensional Nearly Anti-de-Sitter space,''
	PTEP {\bf 2016}, no. 12, 12C104 (2016)
	%doi:10.1093/ptep/ptw124
	[arXiv:1606.01857 [hep-th]].
	%%CITATION = doi:10.1093/ptep/ptw124;%%
	%257 citations counted in INSPIRE as of 07 Apr 2019
	
	
	\bibitem{Product}
	%\cite{Kiritsis:2006hy}
	%\bibitem{Kiritsis:2006hy} 
	E.~Kiritsis,
	``Product CFTs, gravitational cloning, massive gravitons and the space of gravitational duals,''
	JHEP {\bf 0611}, 049 (2006)
	%doi:10.1088/1126-6708/2006/11/049
	[hep-th/0608088];
	%%CITATION = doi:10.1088/1126-6708/2006/11/049;%%
	%58 citations counted in INSPIRE as of 01 Apr 2019
	%\cite{Jian:2017tzg}
	%\bibitem{Jian:2017tzg} 
	S.~K.~Jian, Z.~Y.~Xian and H.~Yao,
	``Quantum criticality and duality in the Sachdev-Ye-Kitaev/AdS$_2$ chain,''
	Phys.\ Rev.\ B {\bf 97}, no. 20, 205141 (2018)
	%doi:10.1103/PhysRevB.97.205141
	[arXiv:1709.02810 [hep-th]].
	%%CITATION = doi:10.1103/PhysRevB.97.205141;%%
	%16 citations counted in INSPIRE as of 01 Apr 2019
	
	%\cite{Alfinito:2007zz}
	\bibitem{Alfinito:2007zz} 
	E.~Alfinito and G.~Vitiello,
	``Double universe and the arrow of time,''
	J.\ Phys.\ Conf.\ Ser.\  {\bf 67}, 012010 (2007).
	%doi:10.1088/1742-6596/67/1/012010
	%%CITATION = doi:10.1088/1742-6596/67/1/012010;%%
	%1 citations counted in INSPIRE as of 07 Nov 2019
	
	%\cite{Tsuji:2016kep}
	\bibitem{Tsuji:2016kep} 
	N.~Tsuji, T.~Shitara and M.~Ueda,
	``Out-of-time-order fluctuation-dissipation theorem,''
	Phys.\ Rev.\ E {\bf 97}, no. 1, 012101 (2018)
	%doi:10.1103/PhysRevE.97.012101
	[arXiv:1612.08781 [cond-mat.stat-mech]].
	%%CITATION = doi:10.1103/PhysRevE.97.012101;%%
	%14 citations counted in INSPIRE as of 06 Apr 2019
	
	%\cite{Azeyanagi:2007bj}
	\bibitem{Azeyanagi:2007bj} 
	T.~Azeyanagi, T.~Nishioka and T.~Takayanagi,
	``Near Extremal Black Hole Entropy as Entanglement Entropy via AdS(2)/CFT(1),''
	Phys.\ Rev.\ D {\bf 77}, 064005 (2008)
	%doi:10.1103/PhysRevD.77.064005
	[arXiv:0710.2956 [hep-th]].
	%%CITATION = doi:10.1103/PhysRevD.77.064005;%%
	%131 citations counted in INSPIRE as of 30 Sep 2019
	
	%\cite{Teixeira:2012cre}
	\bibitem{Teixeira:2012cre}
	A.~Teixeira, A.~Matos, L.~Antunes, 
	``Conditional Renyi Entropies,'' 
	IEEE Trans. Inform. Theory. 58 (2012) 4273–4277. %doi:10.1109/TIT.2012.2192713.
	
	%\cite{Liu:2017kfa}
	\bibitem{Liu:2017kfa} 
	C.~Liu, X.~Chen and L.~Balents,
	``Quantum Entanglement of the Sachdev-Ye-Kitaev Models,''
	Phys.\ Rev.\ B {\bf 97}, no. 24, 245126 (2018)
	%doi:10.1103/PhysRevB.97.245126
	[arXiv:1709.06259 [cond-mat.str-el]].
	%%CITATION = doi:10.1103/PhysRevB.97.245126;%%
	%16 citations counted in INSPIRE as of 09 Oct 2019
	
	
\end{thebibliography}
\end{document}